\documentclass{aa}  
\usepackage{longtable,lscape}
\usepackage{rotating}

\usepackage{graphicx}
\usepackage{txfonts}
\usepackage{orcidlink}
\begin{document}

   \title{Unveiling the nature of SN\,2022jli: The first double-peaked stripped-envelope supernova showing periodic undulations and dust emission at late times}

   \subtitle{}


   \author{R\'egis Cartier\inst{1,2,3}\orcidlink{0000-0003-4553-4033}\thanks{rgcartier@gmail.com}
          \and
          Carlos Contreras\inst{4}\orcidlink{0000-0001-6293-9062}
          \and
          Maximilian Stritzinger\inst{5}\orcidlink{0000-0002-5571-1833}
          \and
          Mario Hamuy\inst{6}\orcidlink{0000-0001-7981-8320}
          \and
          Pilar Ruiz-Lapuente\inst{7,8}\orcidlink{0000-0001-9046-4420}
          \and \\
          Jose L. Prieto\inst{2,9}\orcidlink{0000-0003-1072-2712}
          \and
          Joseph P. Anderson\inst{10}\orcidlink{0000-0003-0227-3451}
          \and
          Aleksandar Cikota\inst{1}\orcidlink{0000-0001-7101-9831}
          \and
          Matthias Gerlach\inst{1,2,9,11}
     }



   \institute{Gemini Observatory, NSF's National Optical-Infrared Astronomy Research Laboratory, Casilla 603, La Serena, Chile
     \and
         {Instituto de Estudios Astrof\'{\i}sicos, Facultad de Ingenier\'{\i}a y Ciencias, Universidad Diego Portales, Av. Ej\'ercito Libertador 441, Santiago, Chile}
    \and
        {Centro de Astronom\'ia (CITEVA), Universidad de Antofagasta, Avenida Angamos 601, Antofagasta, Chile}
    \and
         {Las Campanas Observatory, Carnegie Observatories, Casilla 601, La Serena, Chile}
    \and
        {Department of Physics and Astronomy, Aarhus University, Ny Munkegade 120, DK-8000 Aarhus C, Denmark}
    \and
        {Fundaci\'on Chilena de Astronom\'ia, Santiago, Chile}
    \and
        {Instituto de F\'isica Fundamental, Consejo Superior de Investigaciones Cient\'ificas, c/. Serrano 121, E-28006, Madrid, Spain}
    \and
        {Institut de Ci\`encies del Cosmos (UB-IEEC), c/. Mart\'i i Franqu\'es 1, E-080228, Barcelona, Spain}
    \and
        {Millennium Institute of Astrophysics MAS, Nuncio Monse\~nor Sotero Sanz 100, Off. 104, Providencia, Santiago, Chile}
    \and
        {European Southern Observatory, Alonso de C\'ordova 3107, Casilla 19, Santiago, Chile}
    \and
        {Instituto de Astrof\'isica, Facultad de F\'isica, Pontificia Universidad Cat\'olica de Chile, Av. Vicu\~na Mackenna 4860, Santiago, Chile}    
   }
   \titlerunning{Observations of SN\,2022jli} 
   \authorrunning{Cartier et\,al. 2024}
   
   \date{Received ; accepted}

  \abstract
      {We present optical and infrared observations from maximum light until around $+800$ days of supernova (SN)\,2022jli,
        a peculiar stripped-envelope (SE) SN showing two maxima, each one with a peak luminosity of about $3 \times 10^{42}$\,erg\,s$^{-1}$,
        separated by 50 days. The second maximum is followed by unprecedented periodic undulations with a period of $P \sim 12.5$ days.
  The spectra and the photometric evolution of the first maximum are consistent with the behaviour of a standard SE SN with an ejecta
  mass of $\sim 1.5$\,$M_{\odot}$ and a radioactive $^{56}$Ni mass of $\sim 0.12$\,$M_{\odot}$. The optical spectra after
  $+400$\,days relative to the first maximum correspond to a standard SN\,Ic event, and at late times SN\,2022jli exhibits a
  significant drop in the optical luminosity, implying that the physical phenomena that produced the secondary maximum have
  ceased to power the SN light curve. Among other potential scenarios, we discuss how the second maximum could be powered
    by a magnetar, while the light curve periodic undulations could be produced by accretion of material from a companion star
  onto the neutron star in a binary system. The near-infrared spectra shows clear first CO overtone emission from about $+190$ days after
  the first maximum, and it becomes undetected at $+400$\,days. A significant near-infrared excess from hot dust emission
  is detected at $+238$\,days, having been produced by either newly formed dust in the SN ejecta or a strong near-infrared dust echo.
  Depending on the assumptions of the dust composition, the estimated dust mass is $2-16 \times 10^{-4}$\,$M_{\odot}$.
  The potential magnetar power of the second maximum can fit into a more general picture in which magnetars are the power source of SE
    super-luminous SNe, and could explain bumps, undulations, and late-time excess emission in SE SNe. The CO detection and the dust
    emission of SN 2022jli are key to understanding the molecule and dust formation in the ejecta of SE SNe and in their environment.
  }


   \maketitle
%

\nolinenumbers

\section{Introduction}

Massive stars with a zero-age main-sequence (ZAMS) mass higher than $\approx 10~M_{\sun}$ end their
lives as core-collapse supernovae (SNe), giving birth to neutron stars and black holes.
These stellar explosions can be broadly divided
into hydrogen-rich Type~II SNe exhibiting strong and long-lasting hydrogen features in their spectra,
and hydrogen-deficient stripped-envelope (SE) SNe that have lost the majority of their hydrogen
and helium envelopes over their evolutionary lifetimes \citep[see][]{filippenko97,galyam17}.

An initial scenario proposed for SE SNe was the explosion of single
massive  Wolf-Rayet (WR) stars that have lost their envelopes through strong winds \citep{conti75}.
In this scenario, SE SNe (IIb, Ib, and Ic) follow a sequence in mass, whereby the most massive
stars suffer from stronger winds, losing a larger fraction of their envelopes before their demise as a SN event.
This predicts that the most massive stars demise as SNe\,Ic, losing all of their hydrogen and most,
if not all, of their helium envelopes.

An alternative progenitor scenario for SE SNe is massive stars in binary systems that shed much of their
envelopes through binary interaction \citep{podsiadlowski92}. At the same time, in the late stages of their evolution,
stellar winds can play a key role in peeling off the remaining hydrogen (Ib's) and helium (Ic's) in their
outermost layers \citep[see e.g.,][]{yoon17,dessart20}. This scenario allows a larger range of initial stellar
masses to become SE SNe, but still, the most massive stars lose a larger fraction of their envelopes and explode
as SNe\,Ic \citep{yoon17,dessart20}. This scenario is supported by the large fraction of young massive stars
in binary systems \citep{sana12}, and by their small ejecta masses ($M_{\mathrm{ej}}$) ranging between about
$1-5$\,$M_{\odot}$ \citep{drout11,cano13,taddia15,lyman16,prentice16,taddia18b}, which are too small to be
consistent with single massive stars. Their radioactive $^{56}$Ni mass ($M_{\mathrm{Ni}}$) is in the range
of $0.01 - 0.7$\,$M_{\odot}$ from sample studies of SE SNe \citep{drout11,cano13,taddia15,lyman16,prentice16,taddia18b,anderson19,sharon20,afsariardchi21,rodriguez23}.

Standard SE SNe display bell-shaped light curves that have a single maximum \citep[see e.g.,][]{taddia15,taddia18},\footnote{Some SE SNe exhibit an early bump a
few days after the explosion and several days before the $^{56}$Ni powered maximum \citep[eg., SN\,1993J; see][]{richmond94}.
This early excess emission is explained by shock cooling of extended material.} powered by the decay of radioactive  $^{56}$Ni.
SN\,2005bf \citep{monard05} was the first SE SN reported to show a peculiar light curve morphology composed of two maxima
separated by about 25\,days \citep{anupama05, tominaga05, folatelli06, maeda07,stritzinger18}. The canonical radioactive decay
of $^{56}$Ni distributed close to the SN centre is not able to produce bumpy or double-peaked light curves.
However, a double-peaked distribution of radioactive  $^{56}$Ni, with an enhanced  $^{56}$Ni abundance near the SN surface,
provides a reasonable alternative to produce double-peaked light curves \citep[see][for a discussion]{orellana22}. This was the first
scenario proposed to explain the light curves of SN\,2005bf \citep{tominaga05,folatelli06}; however, it faces the difficulty
of reconciling the relatively high $^{56}$Ni mass predicted with the fast decline and the faint SN luminosity at about
a year after the explosion \citep[see][]{maeda07}. Motivated by this discrepancy, \citet{maeda07} proposed that the energy injection from
a newly born magnetar could explain the maximum light morphology, the rapid decline, and the faint luminosity of SN\,2005bf at late times.
PTF11mnb \citep{taddia18} and SN\,2019cad \citep{gutierrez21} are two SE SNe that display a similar
light curve morphology to SN\,2005bf. In both cases their light curves can be modelled with a double-peaked distribution of $^{56}$Ni;
however, \citet{gutierrez21} show that the light curves of SN\,2019cad can equally be modelled with a central distribution of $^{56}$Ni
and power injection from a magnetar, which produces the second maximum. Another noteworthy event is SN\,2022xxf \citep{itagaki22}.
This SN was discovered before maximum and promptly classified as a  SN~Ic broad line \citep[Ic-BL;][]{balcon22}. It shows two
prominent maxima separated by $75$\,days.  The second maximum reached a slightly brighter luminosity and bluer colours, compared
with the first maximum \citep{kuncarayakti23}. The first maximum appears to correspond to a standard SN Ic-BL event. In turn, it has been argued that the second maximum was powered by the interaction between the SN ejecta and a massive (a few $M_{\odot}$) hydrogen-
and helium-poor circumstellar medium \citep[CSM; see][]{kuncarayakti23}. This hypothesis is based on spectral features suggesting ejecta-CSM
interaction and on the appearance of a conspicuous intermediate-width line component ($\mathrm{FWHM} \sim 2500$\,km\,s$^{-1}$) on top
of the broad emission ($\mathrm{FWHM} \sim 5000$\,km\,s$^{-1}$). The spectral evolution is slow, and after the second maximum the
light curve declines rapidly. The hypothesis proposed by \citet{kuncarayakti23} requires further investigation to confirm
such a massive hydrogen- and helium-poor CSM.

Hydrogen-deficient super-luminous SE SNe  (SLSNe-Ic) correspond to a luminous class of objects encompassing peak
luminosities from $\sim 10^{43}$\,erg\,s$^{-1}$ or $-20$\,mag \citep[see e.g.,][]{decia18,lunnan18,angus19,cartier22} to
$\sim 3 \times 10^{44}$ erg\,s$^{-1}$ or $-22.5$\,mag \citep[see e.g.,][]{decia18,lunnan18,angus19,cartier22}.
Their high peak luminosities are difficult to produce through standard SN powering mechanisms such as the canonical
decay of $^{56}$Ni synthesised in the SN explosion. The most popular scenarios to explain the light curves of SLSNe-Ic are power injection from a magnetar \citep{kasen10, woosley10, inserra13}, or strong interaction with a hydrogen- and helium-poor CSM \citep[see e.g.,][for a discussion]{chatzopoulos13,sorokina16,moriya18}. Pre-maximum bumps are often found in the light curves
of the SLSNe-Ic population \citep{leloudas12, nicholl15, smith16, nicholl16c, angus19}. These early bumps are usually explained
by the shock cooling emission from extended material surrounding the progenitor star \citep[see][]{nicholl16c, piro15, piro21}.
Subsequent bumps, humps, and light curve undulations are frequently found throughout SLSNe-Ic light curve evolution
\citep{yan15,nicholl16,yan17,hosseinzadeh22,cartier22,gutierrez22}. Some bumps are likely explained by the interaction
between the SN ejecta with a dense CSM \citep{yan15,yan17}, while in other cases they can be explained by the variable power
injection from a magnetar \citep[see][]{moriya22}.

A combination of mechanisms such as decay of radioactive nickel, power injection from a magnetar, and interaction between
the SN ejecta with a dense CSM are invoked to explain the double-peaked light curve of the luminous SN\,2019stc \citep{gomez21}.
Early bumps can also be the result of the collision between the SN ejecta and a companion star, yielding double-peaked light curves
\citep{kasen10b}, the result of magnetar-driven shock-break out \citep{kasen16}, or the collision of the newborn neutron star
with the companion star in a binary system \citep[see][]{hirai22}.

SN\,2022jli is a SE SN discovered near maximum light in the nearby galaxy NGC\,157. SN\,2022jli shows a strong secondary maximum
followed by periodic or quasi-periodic undulations \citep{moore23, chen24}. \citet{chen24} proposed that the first maximum corresponds
to a SE SN, and the second maximum and the subsequent light curve undulations are produced by the accretion from a companion star onto
a compact object at periastron encounters in a binary system \citep[see also][]{king24}. Alternatively, \citet{moore23} proposed that
the second maximum corresponds to a massive $12 \pm 6$\,$M_{\odot}$ SN ejecta powered by the radioactive decay of $^{56}$Ni, and the
periodic variability could a consequence of ejecta-CSM interaction with the CSM configured in nested shells or due to the collision
of a neutron star with the companion star, as in the model of \citet{hirai22}. During the revision of this article, accretion
models from a shock-inflated companion star into a neutron star in a binary system appeared. These models seem to explain
the second maximum and the periodic light curve undulations of SN\,2022jli \citep{hirai25,lu25}. A recent hydrodynamical
model that combines the radioactive decay of $\sim 0.15$\,$M_{\odot}$ of $^{56}$Ni and a magnetar successfully reproduces
the first and second maxima of SN\,2022jli, but does not reproduce the periodic undulations of SN\,2022jli \citep{orellana25}.

This work is structured as follows. In Section \ref{sec:observations} we present optical and infrared (IR)
observations of SN\,2022jli from maximum light to $+772$\,days. In Section \ref{sec:analysis} we present the analysis
of the observations of SN\,2022jli. In Section \ref{sec:discussion}, we discuss the super-Eddington accretion scenario
\citep{chen24, king24} and the SN ejecta-CSM interaction scenario \citep{moore23}, and propose an alternative scenario in which in addition
to the accretion of material from a companion star onto the neutron star, the power from the spin-down rotational energy from a
magnetar contributes to powering the second maximum. In Section \ref{sec:summary} we summarise our conclusions.

\section{Observations and data reduction}
\label{sec:observations}

\begin{figure}
\centering
\includegraphics[width=90mm]{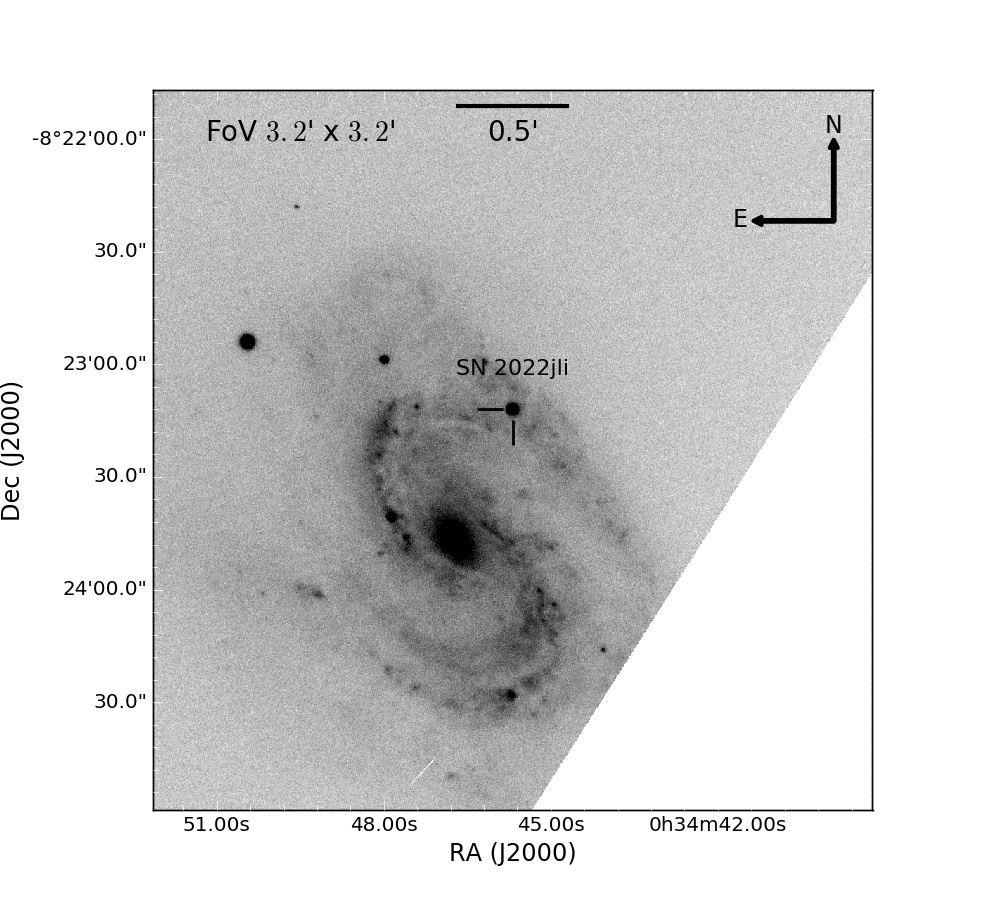}
\caption{Field of SN\,2022jli in the $i$ band observed with the LDSS-3
  mounted at the Clay telescope at Las Campanas Observatory on May 22, 2022.
  The position of the SN is indicated in the figure. North is up and east is
  to the left.
}
\label{finder_chart_fig}
\end{figure}

SN\,2022jli was discovered on 2022 May 05.17 UTC by Libert Monard \citep{monard2022} at an
unfiltered (clear filter) apparent magnitude of 14.4\,mag, which was later corrected to
$14.22 \pm 0.15$\,mag and reported to the Transient Name Server (TNS). The SN is located
in the galaxy NGC\,157 at equatorial co-ordinates of $\alpha = 00$:$34$:$45.7$ and
$\delta = -08$:$23$:$12.07$ (J2000), in the northern spiral arm of its host galaxy, $16\farcs2$ west and $35\farcs09$ north of the galaxy centre (see Fig. \ref{finder_chart_fig}).
SN\,2022jli was classified as a SN\,Ic at a phase of one to three days after maximum based on a spectrum obtained
on 2022 May 11.14 UTC by \citet{2022TNSCR1261....1G}, and later confirmed as a SN~Ic by ePESSTO+
\citep{2022TNSCR1409....1C} based on a spectrum obtained on 2022 May 24.42 UTC. Due to some confusion
about the SN co-ordinates reported to TNS, the ePESSTO+ classification was erroneously reported as
a separate object named SN\,2022jzy.

\subsection{Photometry}
\label{photometry_obs_sec}

We analysed public optical photometry from the Kleinkaroo Observatory reported by Monard to the TNS,
from the Gaia mission\footnote{\url{http://gsaweb.ast.cam.ac.uk/alerts/alert/Gaia22cbu/}} \citep{gaia16},
from the ATLAS survey \citep{tonry18, smith20}, from the Zwicky Transient Facility \citep[ZTF;][]{bellm19},
and from All-Sky Automated Survey for Supernovae \citep[ASAS-SN;][]{shappee14}. We also present optical
photometry obtained with the LDSS-3 instrument mounted to the 6.5-m Magellan Clay telescope, the EFOSC2
instrument with the 3.58-m New Technology Telescope (NTT), the Goodman instrument mounted on the 4.1\,m SOAR
telescope, and NIR photometry obtained with the Flamingos-2 instrument on the 8.1-m Gemini-South observatory.
The optical photometry extends from a few days before maximum light to $+418$\,days after maximum, and the NIR photometry
extends from $+38$\,days to $+600$\,days.

\subsubsection{Optical}

To discount any potential issue in the public ZTF photometry of SN\,2022jli, we downloaded the original ZTF images
from the NASA/IPAC Infrared Science Archive,\footnote{\url{https://irsa.ipac.caltech.edu/Missions/ztf.html}}
and performed host-galaxy template image subtraction using {\sc HOTPANTS} \citep{2015ascl.soft04004B}. Aperture
and PSF photometry was then computed from these template subtracted science images using {\sc Python} scripts. ATLAS
forced photometry from the ATLAS forced photometry server was also downloaded and included in our analysis.\footnote{\url{https://fallingstar-data.com/forcedphot/}}
We include $g$-band ASAN-SN photometry to complement the photometry during the decline from the first maximum and on
the rise to the second maximum. The ASAN-SN photometry correspond to photometry published by \citet{moore23} and \citet{chen24}.

The reduction of the LDSS-3 and Goodman images was performed in IRAF following standard procedures that
include bias subtraction and the normalisation of the science images using a normalised flat image,
the astrometric registration of the images was done using {\sc astrometry.net} \citep{lang10}.
The reduction of the EFOSC2 $V$-band acquisition image was done with the PESSTO pipeline \citep{smartt15}.
Aperture and PSF photometry was performed with {\sc Python} scripts \citep[see e.g.,][]{hueichapan22}.
The optical photometry of SN\,2022jli is presented in Fig. \ref{lightcurves_fig}. LDSS-3, EFOSC2
and Goodman photometry is reported in Table \ref{op_phot_tab} in the Appendix.

\subsubsection{Near-infrared}

NIR photometry of SN\,2022jli was obtained using the Flamingos-2 instrument \citep{eikenberry04, eikenberry12}
mounted on the Gemini-South telescope. Flamingos-2 images were reduced using custom {\sc IRAF} scripts.
The reduction steps include the creation of a clean sky image, and the subtraction of this image from the
science frames, flat fielding the science frames and the astrometric registration of the science images
using {\sc astrometry.net} \citep{lang10}. The sky images were created from frames obtained in a position
far from the host galaxy to avoid over subtraction of the host galaxy extended emission. 
The pre-explosion images of NGC\,157 in $JHK_{s}$ obtained with the
VIRCAM instrument \citep{dalton06} mounted on the VISTA telescope \citep{emerson06} at Paranal observatory
were employed to subtract the host galaxy background emission from the SN images. The NIR photometry of
SN\,2022jli is reported in Table~\ref{ir_phot_tab} in the Appendix, and shown in Fig.~\ref{lightcurves_fig}.

\begin{figure*}
\centering
\includegraphics[width=180mm]{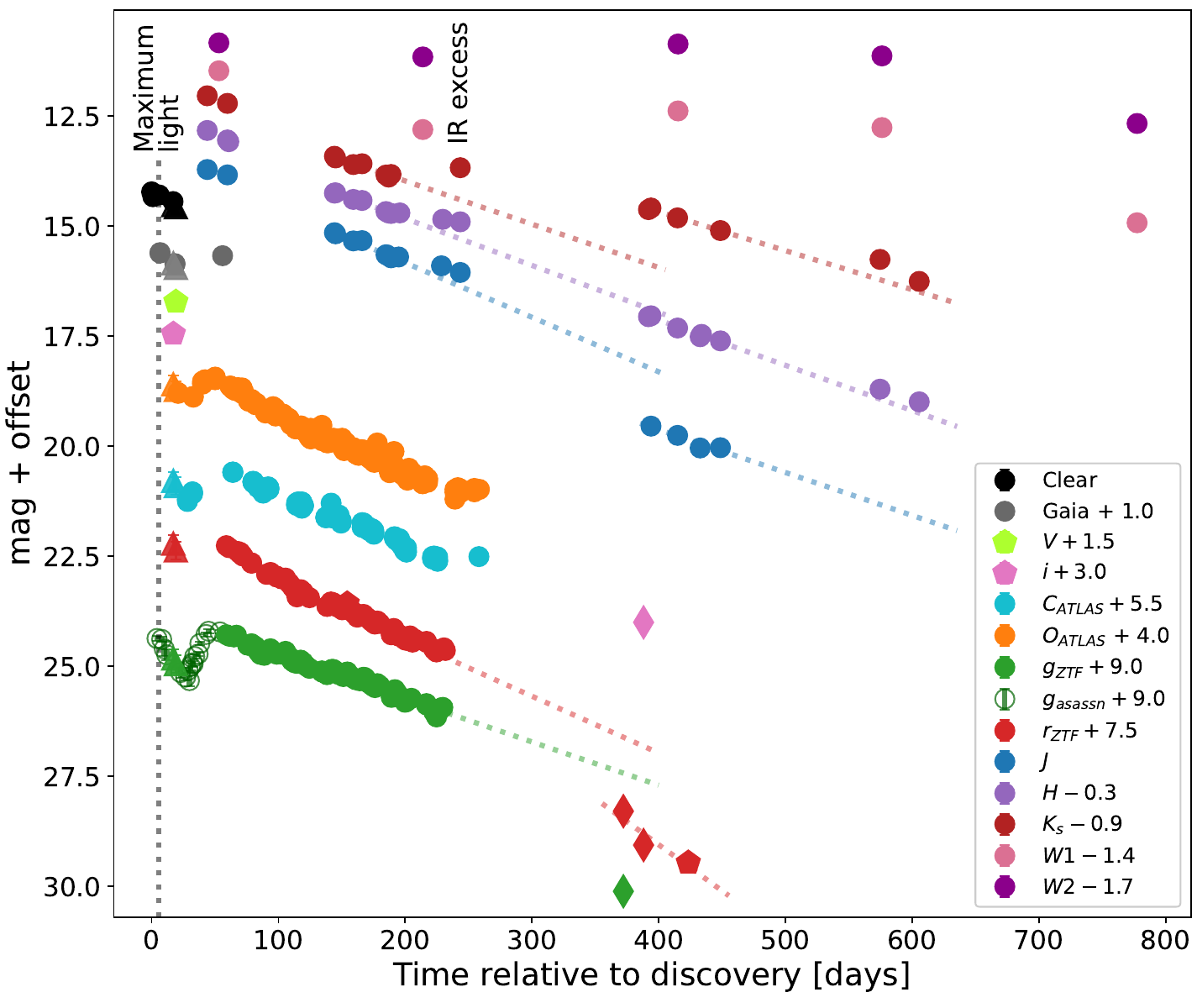} 
\caption{Optical and NIR light curves of SN\,2022jli. Circles correspond to survey photometry or NIR
  photometry repeatedly obtained using the same instrument or photometric system. Pentagons correspond to
  photometry computed from LDSS-3 and EFOSC2 spectroscopic acquisition images, diamonds are from late time
  Goodman photometry and triangles represent synthetic photometry computed from colour-matched spectra.
  The $g$-band photometry from ASAS-SN is shown using dark-green open circles to avoid confusion with
  other $g$-band photometry. The legend on the right specifies the colour code and the offsets employed
  to plot the different bands. Dotted lines are used to compare (extrapolate) the brightness decline from
  an early epoch to late times in a few relevant bands.}
\label{lightcurves_fig}
\end{figure*}

\subsubsection{Mid-infrared}

The field of the SN has been repeatedly visited by the Wide-field Infrared Survey Explorer (WISE)
satellite \citep{wright10} as part of the NEOWISE \citep{mainzer11} and NEOWISE-R \citep{mainzer14}
missions. The NEOWISE-R mission observed the SN five times during 2022, 2023 and 2024.
A careful inspection of the NEOWISE-R images before and after the SN explosion shows a clear mid-IR emission
from the SN in the $W1$ and $W2$ bands, and no variability pre-SN explosion; therefore, pre-explosion
images were used to subtract the host galaxy background emission from the images containing
the SN. We obtained aperture photometry from host subtracted images.
The NEOWISE $W1$ and $W2$ photometry is presented in Fig. \ref{lightcurves_fig} and summarised
in Table \ref{mir_phot_tab} in the Appendix.

\subsection{Spectroscopy}
\label{spectra_sec}

\subsubsection{Optical}

We obtained optical spectra of SN\,2022jli with the LDSS-3 spectrograph mounted at the Clay telescope
at Las Campanas Observatory, the Goodman spectrograph \citep{clemens04} mounted at the SOAR telescope
and the GMOS-S spectrograph \citep{hook04,gimeno16} at the Gemini-South telescope. The LDSS-3 and
Goodman data were reduced using {\sc IRAF} routines. The reduction steps include bias subtraction,
flat fielding, cosmic ray rejection using {\sc LACosmic} \citep{vandokkum01}, wavelength calibration,
flux calibration and telluric correction. We performed the flux calibration and the telluric
correction using a flux standard star observed on the same night of the SN spectra.

The GMOS-S spectra were reduced using the {\sc gemini} {\sc IRAF} package following standard
reduction procedures similar to the ones described before, but using the {\sc gmos} {\sc IRAF} task.
The flux calibration was performed using a spectrophotometric
standard star observed on a different night with the same instrument setup employed to observe the SN.
The GMOS-S instrument is composed of three detectors, each one covering a different portion of the wavelength
range. To obtain a continuous spectrum, almost free from the gaps between the detectors, two spectra
of the SN shifted in wavelength were obtained.\footnote{For the blue spectral range,
the B600 grating was employed. The B600 grating provides a combined spectra are free from gaps between
detectors. For the red spectra was employed the R400 grating with the GG455 second order blocking filter. The red
portion of the GMOS combined spectra have a small gap of about $100$\,\AA\ that affects the \ion{O}{I} $\lambda 7774$
line (from $+55$ to $+232$\,days), or the [\ion{Ca}{II}] line in the nebular spectrum at $+412$\,days.}

\begin{figure}
\centering
\includegraphics[width=90mm]{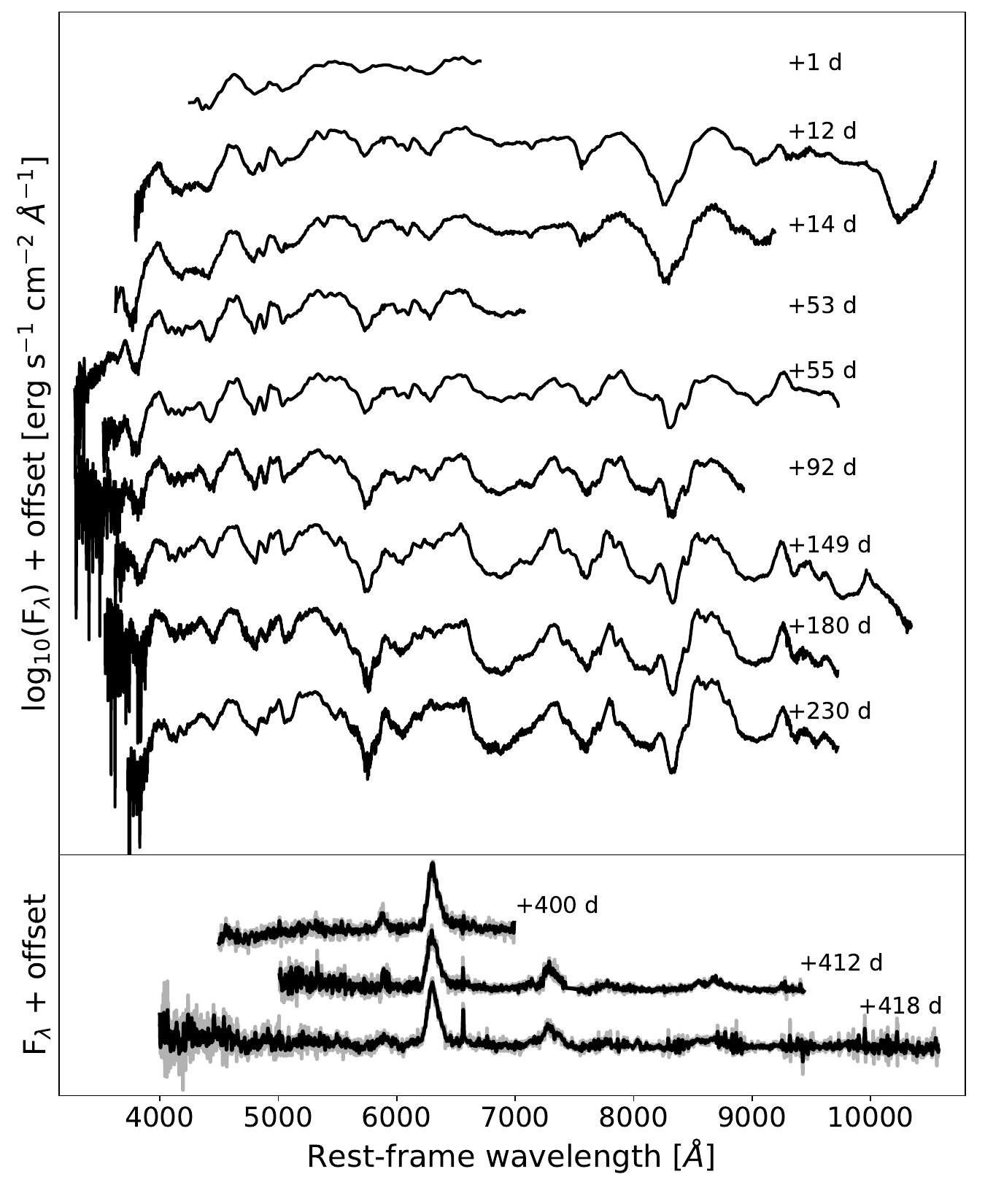} 
\caption{Optical spectral sequence of SN\,2022jli. The phase relative to the time of maximum
  light is indicated on the right. The spectra during the first $+250$\,days are shown
  in the top panel in logarithmic scale, the nebular phase spectra are presented in
  the bottom panel. The smoothed nebular spectra using the Savitzky–Golay filter are shown
  in black, the original spectra are shown in grey.}
\label{specseq_op_fig}
\end{figure}

We include in our analysis the classification spectra reported to the TNS by \citet{2022TNSCR1261....1G}
and by the ePESSTO+ collaboration. The ePESSTO+ classification spectrum is publicly available and
was observed using the EFOSC2 instrument \citep{buzzoni84,snodgrass08} mounted on the New-Technology
Telescope (NTT) at La Silla observatory. The ePESSTO+ classification was independently reduced by our team
using the PESSTO pipeline \citep{smartt15}. The optical spectra are presented in Fig. \ref{specseq_op_fig},
and a summary of the optical spectroscopic observations of SN\,2022jli is presented in Table \ref{op_spec_summary_tab}.

\begin{table*}
  \centering
  \caption{Summary of optical-wavelength spectroscopic observations of SN~2022jli.}
  \label{op_spec_summary_tab}
  \begin{tabular}{@{}lcccccc}
    \hline
    Date UTC   & MJD       & Phase  & Instrument/        & Wavelength       & Dispersion  & Resolution ($FWHM$) \\
               &           & (days) & Telescope          & Range (\AA)      & (\AA/pixel) & (\AA)       \\
    \hline
    22-05-2022 & $59721.4$ & $+11.7$ & LDSS-3/Clay       & $3800 - 10\,300$  & $2.0$ & $8.0$ \\
    24-05-2022 & $59723.4$ & $+13.7$ & EFOSC2/NTT        & $3660 - 9240$    & $5.5$ & $23.0$\\
    03-07-2022 & $59763.3$ & $+53.4$ & Goodman/SOAR      & $3300 - 7115$    & $2.0$ & $5.5$ \\
    04-07-2022 & $59764.3$ & $+54.4$ & GMOS-S/Gemini-S   & $4700 - 9750$    & $1.5$ & $6.7$ \\
    05-07-2022 & $59765.3$ & $+55.4$ & GMOS-S/Gemini-S   & $3550 - 7080$    & $1.0$ & $4.7$ \\
    11-08-2022 & $59802.3$ & $+92.2$ & Goodman/SOAR      & $3600 - 8950$    & $2.0$ & $5.5$ \\
    07-10-2022 & $59859.3$ & $+148.9$ & LDSS-3/Clay      & $3800 - 10\,300$  & $2.0$ & $8.0$ \\
    07-11-2022 & $59890.1$ & $+179.5$ & GMOS-S/Gemini-S  & $3560 - 7080$    & $1.0$ & $4.4$ \\
    07-11-2022 & $59890.1$ & $+179.5$ & GMOS-S/Gemini-S  & $4700 - 9750$    & $1.5$ & $6.8$ \\
    26-12-2022 & $59939.1$ & $+228.2$ & GMOS-S/Gemini-S  & $3550 - 7080$    & $1.0$ & $4.7$ \\
    30-12-2022 & $59943.1$ & $+232.2$ & GMOS-S/Gemini-S  & $4700 - 9750$    & $1.0$ & $6.8$ \\
    17-06-2023 & $60112.4$ & $+400.6$ & Goodman/SOAR     & $3600 - 7020$    & $2.0$ & $9.4$ \\
    29-06-2022 & $60124.4$ & $+412.5$ & GMOS-S/Gemini-S  & $4600 - 9460$    & $1.5$ & $6.8$ \\
    05-07-2023 & $60130.4$ & $+418.5$ & LDSS-3/Clay      & $3800 - 10\,300$  & $2.0$ & $8.0$ \\
    \hline
  \end{tabular}
  \tablefoot{The phase is measured relative to the time of maximum brightness.
   }
\end{table*}

\subsubsection{Near-infrared}

Seven epochs of NIR spectroscopy of SN\,2022jli were obtained\footnote{11 individual spectra.} using the Flamingos-2 instrument
\citep{eikenberry04, eikenberry12} mounted on the Gemini-South telescope. The observations
were performed following an ABBA pattern, we observed the $JH$ and $HK$ configurations
that cover from about $8750$ to $17000$ \AA\ and from $13300$ to $24700$ \AA, respectively.
We obtained a $JH$ and $HK$ pair as close in time as possible to obtain full NIR coverage.

One NIR spectrum of SN\,2022jli was obtained using the fourth generation of
the Triple-Spec spectrograph \citep{schlawin14} mounted on the NIR Nasmyth platform
of the SOAR telescope. Triple-Spec is a cross-dispersed, long-slit spectrograph
with a resolution of about 3500, covering the full NIR wavelength range in
one observation. The observations were performed following the ABBA pattern.
The Triple-Spec spectra were reduced using its custom reduction pipeline,
included in the {\sc Spextool} {\sc IDL} package \citep{cushing04}. We always observed
an A0V telluric star before or after observing the SN and at a similar air mass. After
the extraction of the individual spectra, we used the {\sc xtellcorr} task \citep{vacca03} included
in the {\sc Spextool} {\sc IDL} package \citep{cushing04} to perform the
telluric correction and flux calibration of the NIR spectra.

\begin{figure*}
\centering
\includegraphics[width=180mm]{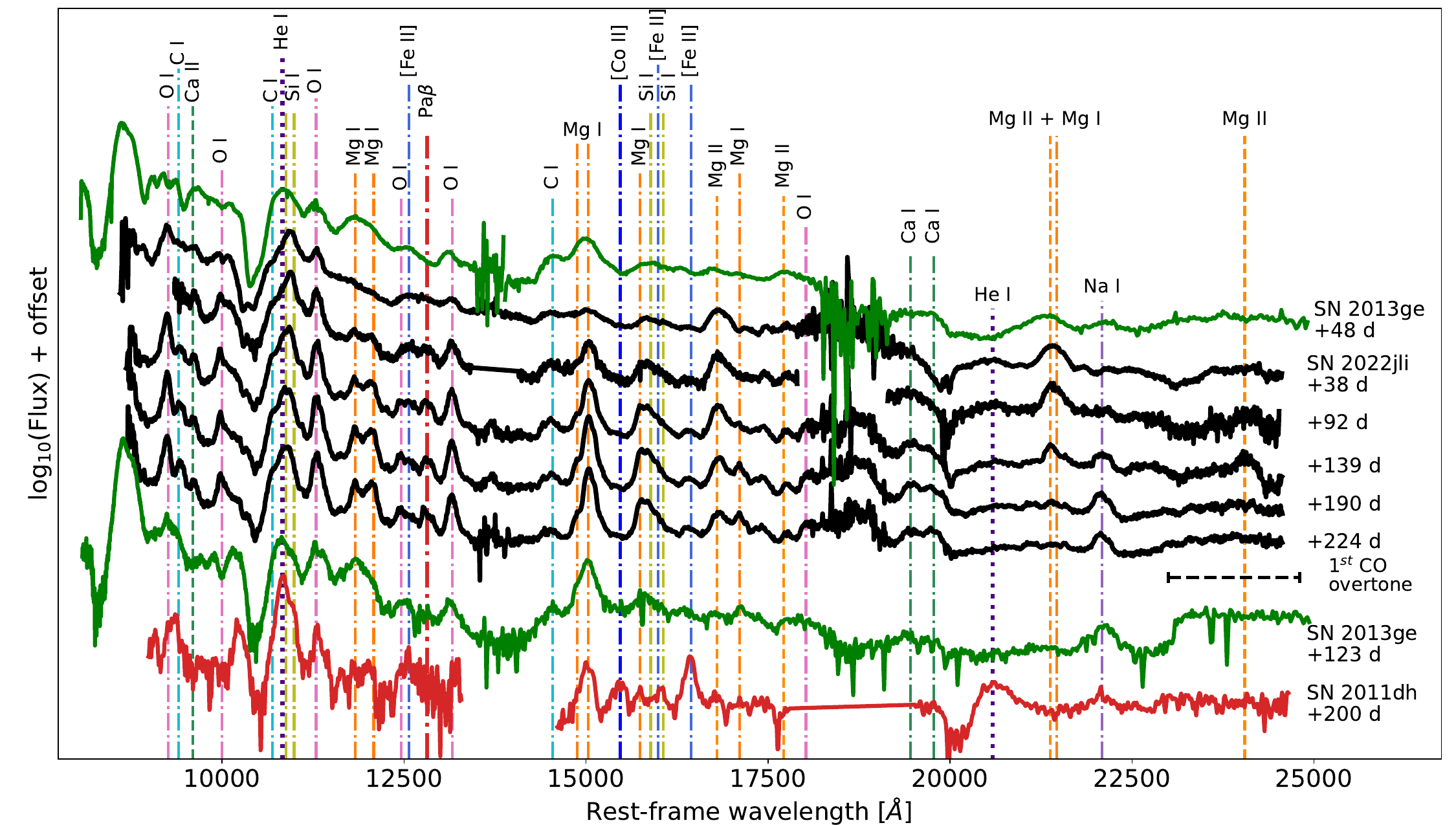} 
\caption{NIR spectral sequence of SN\,2022jli (in black) compared with SN\,2013ge \citep[in green;][]{drout16}
  and SN\,2011dh \citep[in red;][]{ergon15}. The position of several lines are indicated with vertical lines and
  the phase relative to the maximum is indicated on the right. Five selected epochs are presented in this
  figure.
}
\label{specseq_nir_fig}
\end{figure*}

\begin{table*}
  \centering
  \caption{Summary of NIR spectroscopic observations of SN\,2022jli.}
  \label{nir_spec_summary_tab}
  \begin{tabular}{@{}lcccccc}
    \hline
    Date UTC   & MJD       & Phase  & Instrument/          & Wavelength       & Dispersion  & Resolution ($FWHM$)\\
               &           & (days) & Telescope            & Range (\AA)      & (\AA/pixel) & (\AA)       \\
    \hline
    18-06-2022 & $59748.3$ & $+38.5$  & Flamingos-2/Gemini-S    & $8670 - 17\,000$   & $6.5$ & $18$ \\
    18-06-2022 & $59748.4$ & $+38.6$  & Flamingos-2/Gemini-S    & $13,300 - 24\,700$ & $7.6$ & $22$ \\
    11-08-2022 & $59802.2$ & $+92.1$  & Triple-Spec/SOAR        & $9400 - 24\,650$   & $2.2$ & $4.5$ \\
    27-09-2022 & $59849.2$ & $+138.8$ & Flamingos-2/Gemini-S    & $13,300 - 24\,700$ & $7.6$ & $24$ \\
    27-09-2022 & $59849.3$ & $+138.9$ & Flamingos-2/Gemini-S    & $8780 - 17\,000$   & $6.8$ & $18$ \\
    19-10-2022 & $59871.0$ & $+160.5$ & Flamingos-2/Gemini-S    & $8700 - 17\,000$   & $6.4$ & $18$\\
    17-11-2022 & $59900.0$ & $+189.4$ & Flamingos-2/Gemini-S    & $8740 - 17\,000$   & $6.5$ & $18$ \\
    18-11-2022 & $59901.1$ & $+190.4$ & Flamingos-2/Gemini-S    & $13,300 - 24\,700$ & $7.6$ & $22$ \\
    21-12-2022 & $59934.1$ & $+223.3$ & Flamingos-2/Gemini-S    & $8750 - 17\,000$   & $6.0$ & $18$ \\
    22-12-2022 & $59935.1$ & $+224.3$ & Flamingos-2/Gemini-S    & $13,300 - 24\,700$ & $7.6$ & $22$ \\
    08-07-2023 & $60133.4$ & $+421.5$ & Flamingos-2/Gemini-S    & $13,300 - 23\,800$ & $7.6$ & $24$ \\
    15-07-2023 & $60140.4$ & $+428.4$ & Flamingos-2/Gemini-S    & $8700 - 17\,000$   & $6.5$ & $21$ \\
    \hline
  \end{tabular}
  \tablefoot{The phase is measured relative to the time of unfiltered maximum brightness.
  }
\end{table*}

\section{Analysis}
\label{sec:analysis}

\subsection{Host galaxy}
\label{sec:host_galaxy}

NGC\,157 is an isolated SAB(rs)bc galaxy, with no close companions of comparable brightness detected
in POSS images \citep{ryder98}. The galaxy heliocentric redshift is $z=0.005500 \pm 0.000002$ \citep[from NED;][]{springob05}.
The SN\,2022jli is located in the north-west spiral arm of NGC\,157. In this paper we adopt a distance of $23.5 \pm 1.6$ Mpc,
corresponding to the distance derived from the host-galaxy recession velocity by NED considering a correction for Virgo,
Great Attractor, and Shapley Supercluster infall \citep{mould00}, assuming $H_{0} = 70.0$ km\,s$^{-1}$ Mpc$^{-1}$.

\subsection{Light curves}
\label{sec:light_curve_chara}

\subsubsection{Epoch of maximum brightness}

\begin{figure}
\centering
\includegraphics[width=85mm]{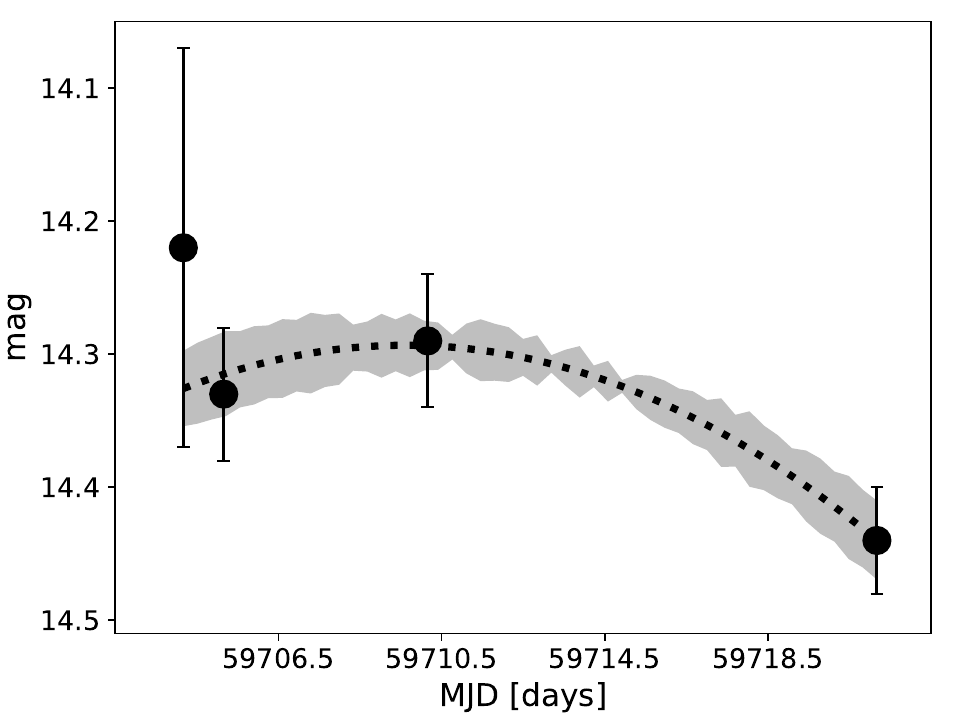} 
\caption{Unfiltered photometry reported by Monard to the TNS obtained at the Kleinkaroo Observatory
in South Africa (black points). The dotted black line corresponds to a second order polynomial fitted
to the photometry. The polynomial fit was performed weighting by the photometric uncertainties and
the shaded area corresponds to the $1 \sigma$ uncertainty computed from the covariance matrix of the
polynomial fit. From this fit we obtained that the epoch of maximum brightness was on
MJD$= 59709.6 \pm 1.2$\,days at an unfiltered brightness of $14.29 \pm 0.02$ mag; however, we
assume more conservative uncertainties for these parameters (see text).
}
\label{maximum_polyfit_fig}
\end{figure}

We used the unfiltered photometry obtained from the Kleinkaroo Observatory in South Africa and
reported by Monard to the TNS to estimate the time of maximum light. In Fig. \ref{maximum_polyfit_fig}
we show the reported photometry fitted using a second order polynomial to estimate the
SN\,2022jli maximum brightness. From the polynomial fit we obtained that the epoch of maximum
brightness ($t_{\mathrm{max}}$) was on MJD$= 59709.6 \pm 1.2$\,days at an unfiltered brightness
of $14.29 \pm 0.02$ mag, where the uncertainties are estimated using the covariance matrix
of the polynomial fit. Conservatively, we assume an error on $t_{\mathrm{max}}$ equal to the time
lapse between the discovery and the estimated epoch of maximum light, and an error on the brightness
equal to the uncertainty on the closest photometric epoch, this is $t_{\mathrm{max}}= 59709.6 \pm 5.4$\,days
and $14.29 \pm 0.05$ mag. Hereafter we will refer to this epoch as the maximum time unless
otherwise indicated. Our estimate of the epoch of maximum brightness is
consistent with the phase reported based on the spectroscopic classification of
\citet{2022TNSCR1261....1G}.

\subsubsection{Secondary maximum and decline rates}

SN\,2022jli shows double-peaked light curves, and we estimate the epoch of the second maximum
by fitting a low order polynomial to the photometry in the $g$, $c$ and $o$ bands. This procedure yields
that the time second maximum took place on MJD equal to $59763.3 \pm 8.8$\,days, $59752.4 \pm 2.5$\,days
and $59747.9 \pm 1.4$\,days for the $c$, $o$ and $g$ bands, respectively. The time between the first and the second maximum is
between $+43$ to $+53$\,days, the minimum brightness in the $g$ band was reached on $59724.4 \pm 1.9$\,days, coincident with
the minimum in the $c$ band at about $+20$\,days after the first maximum.

After the second maximum, the SN shows a linear decline which is well described by a linear polynomial fit. The average
decline rate of the optical light curves in the time range between the MJD equal to $59790$ and $59850$ was measured,
corresponding to a phase between $+80$ and $+140$ days relative to the first maximum. The uncertainties are estimated
using the Monte Carlo methodology described in \citet{cartier22}. The light curve decline rates are summarised in table 3.
The optical bands with bluer effective wavelengths ($g$ and $c$ bands), decline more slowly than the bands with redder
effective wavelengths ($r$ and $o$ bands), at a significance of $8 \sigma$ or higher. This is reflected in bluer optical
colours at later times (see Section \ref{colour_sec} below), meaning that the ejecta is continuously heated.

From about $+200$\,days the NIR light curves show a re-brightening (see Fig. \ref{lightcurves_fig}). The NIR photometric coverage
allow us to measure the light curve linear decline rate in two regions, in the MJD range between $59790$ and $59900$\,days
($+80$ to $+189$ days; this is before the NIR re-brightening), and in the MJD range between $60090$ and $60160$\,days ($+378$ and
$+448$ days) at late times. These linear decline rates are also summarised in table \ref{decl_rate_tab}. The $J$ band declines
faster than the $H$ and $K_{s}$ bands. In concrete, the $J$ band declines at a linear rate similar to the $o$ band in
the optical, while the $K_{s}$ band decline rate is similar to the $g$ band, this is before the onset of of the NIR re-brightening.
Thus the linear decline rates in the NIR follow the opposite trend than the optical bands, this is a consequence of two different
physical processes such as heating of the SN ejecta and dust emission.

At late times (\textgreater $+350$\,days) the decline rates in the NIR light curves
in all bands are about $1$\,mag/100\,days and are consistent with each other within
the uncertainties, while the decline in the $r$ band is about twice faster than
the decline measured in the NIR bands.

The linear extrapolation from $+80$ to $+189$\,days to late times shows that the
SN is significantly brighter than the extrapolation in $K_{s}$ ($\sim 1.2$\,mag brighter),
similarly the SN brightness in the $H$ band is slightly fainter than the linear
extrapolation ($\sim 0.15$\,mag fainter), and the $J$ band brightness is significantly fainter than
the linear extrapolation ($\sim 1.2$\,mag fainter).

\begin{table}
  \centering
  \caption{Summary of the linear decline rates of SN\,2022jli.}
  \label{decl_rate_tab}
  \begin{tabular}{@{}lcc}
    \hline
    Band    & Decl. rate      & Decl. rate ($+450$\,d \textgreater $t$ \textgreater $+350$\,d)\\
            & (mag/100\,days) & (mag/100\,days) \\
    \hline
    Pseudo-Bol & $1.12$($0.10$)   & $\cdots$ \\
    $g$        & $0.984$($0.004$) & $\cdots$ \\
    $c$        & $1.084$($0.012$) & $\cdots$ \\
    $r$        & $1.290$($0.004$) & $2.09$($0.25$) \\
    $o$        & $1.231$($0.012$) & $\cdots$ \\
    $J$        & $1.239$($0.022$) & $0.98$($0.21$) \\
    $H$        & $1.067$($0.066$) & $1.03$($0.11$) \\
    $K_{s}$    & $0.984$($0.065$) & $0.89$($0.11$) \\
    \hline
  \end{tabular}
\end{table}

\subsubsection{Host-galaxy reddening}
\label{sec:reddening}

\begin{figure}
\centering
\includegraphics[width=85mm]{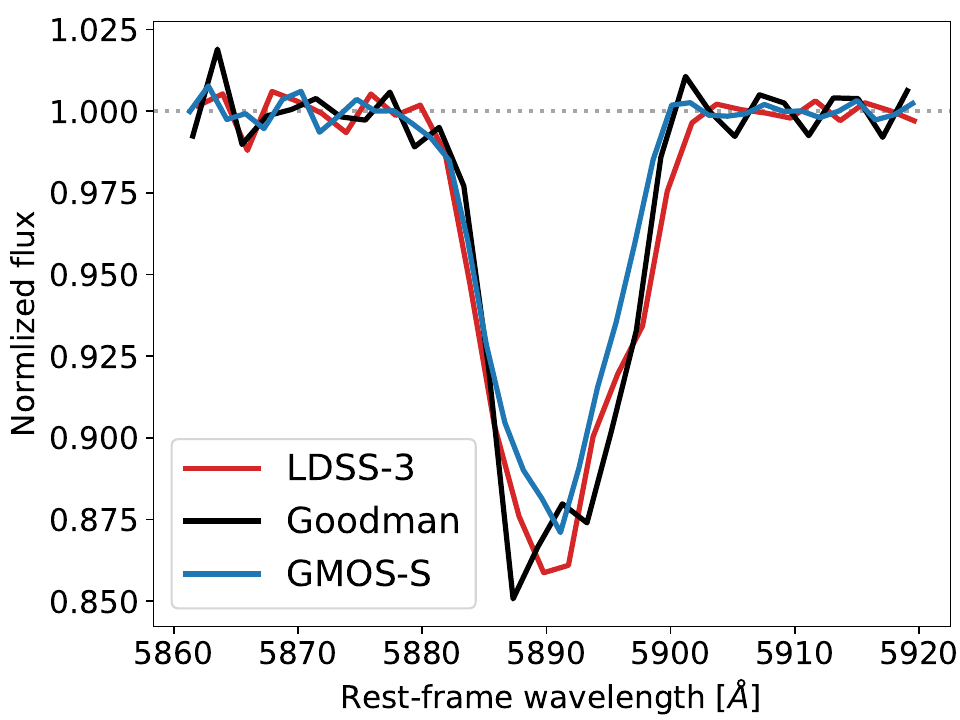} 
\caption{Normalised host-galaxy \ion{Na}{I} D line for SN\,2022jli at $+11.7$\,days (LDSS-3/Clay),
  $+53.4$\,days (Goodman/SOAR) and $+54.4$ (Gemini-S/GMOS-S). The average equivalent width of
  the \ion{Na}{I} D line for the three epochs is $EW = 1.54 \pm 0.12$\,\AA.}
\label{normlized_nai_fig}
\end{figure}

The Milky Way reddening in the direction of SN\,2022jli is $E(B-V)_{\mathrm{gal}} = 0.0383$ \citep{schlafly11}.
The spectra of SN\,2022jli are significantly reddened, showing a clear \ion{Na}{I}\,D
narrow absorption feature from the host-galaxy (see Fig. \ref{normlized_nai_fig}), implying that the SN
suffers significant host-galaxy extinction. The pseudo equivalent width (hereafter $EW$) of  \ion{Na}{I} was measured in
our highest signal-to-noise ratio spectra obtained at a phase before the host-galaxy \ion{Na}{I} narrow line
spectral region becomes too close to the edge of the underlying broad \ion{Na}{I} spectral feature to be properly
normalised. The spectra used to measure the \ion{Na}{I} $EW$ correspond to the LDSS-3, Goodman and GMOS-S spectra obtained
at $+11.7$\,days, $+53.4$\,days, and $+54.4$\,days, respectively (see Fig. \ref{normlized_nai_fig}). An average
value of $EW = 1.54 \pm 0.12$\,\AA\ was measured for the host \ion{Na}{I} D1+D2 feature.  Adopting the  \citet{poznanski12} relation between
the EW of the host \ion{Na}{I} and colour excess, implies a host-galaxy colour excess of $E(B-V)_{\mathrm{host}} = 0.9 \pm 0.3$~mag.
\citet{phillips13} note that although a relation between the \ion{Na}{I} $EW$ and the extinction $A_{V}$ value exists,
the uncertainty in $A_{V}$ from the \ion{Na}{I} $EW$ can be significant of the order of $0.3$ dex,
this is larger than the $0.08$\,dex quoted by \citet{poznanski12}. Assuming the uncertainty of \citet{phillips13} we obtain
$E(B-V)_{\mathrm{host}} = 0.9 \pm 0.7$~mag. Since the \citet{poznanski12} relations are calibrated using the original
\citet{shlegel98} extinction maps, we need to multiply the values derived from \citet{poznanski12} relations by
$0.86$ to place them in the re-calibration of \citet{schlafly11}, then the host galaxy reddening is
$E(B-V)_{\mathrm{host}} = 0.8 \pm 0.6$~mag from the \ion{Na}{I} D1+D2 absorption feature.

\citet{drout11} reported a small scatter of $\pm 0.06$\,mag in the $V-R$ colours of SE SNe (IIb/Ib/Ic)
around $+10$\,days relative to the $V$-band maximum. This property was latter confirmed by the radiative
transfer models of \citet{dessart15} and investigated in detail by \citet{stritzinger18} using multi-band
photometry of SE SNe \citep{stritzinger18b} from the Carnegie Supernova Project I \citep[CSP-I;][]{hamuy06}.
Using the CSP-I photometry, \citet{stritzinger18} created unreddened colour templates for SE SNe, for the
different sub-types (i.e.  IIb, Ib, Ic). According to \citet{stritzinger18} these templates can provide
a host reddening estimate with photometry obtained between $+0$ to $+20$\,days relative to $V$-band maximum.

We used the LDSS-3 and NTT (ePESSTO+) optical spectra of SN\,2022jli obtained at $+11.7$ and $+13.7$\,days,
respectively, to compute synthetic colours. These spectra combined with the CSP-I broad-band transmission
curves and zero points from \citet{krisciunas17}, provide the colour information needed to estimate the SN
reddening by means of computing synthetic photometry. As in \citet{stritzinger18} and throughout this paper,
the \citet{fitzpatrick99} reddening law is adopted, and in this work $R_{V}=3.1$ is assumed. We also assume
an uncertainty of $0.075$ mag for the synthetic magnitudes, which translates to an uncertainty of $0.10$ mag in colour.
The total reddening along the line of sight ($E(B-V)_{\mathrm{tot}}$), which is the host galaxy and the
Milky Way reddening, was inferred as the difference between the synthetic colours and the colour templates.

The reported colour excess value for each colour corresponds to the weighted average of the reddening values
obtained for the two spectra. The quoted uncertainties were computed by adding in quadrature the uncertainties
in the intrinsic colour-curve templates and in the synthetic colours. Table~\ref{reddening_tab} lists our total
and host-galaxy ($E(B-V_{\mathrm{host}})$) colour excess estimations for SN\,2022jli, inferred from nine colour
combinations. We used the \citet{fitzpatrick99} reddening law with $R_{V}=3.1$ to transform from a given
colour excess to $E(B-V)$ colour excess.

The host-galaxy reddening estimates derived from the \citet{stritzinger18} colour templates yield values
of $E(B-V)_{\mathrm{host}}$ between $0.14-0.62$~mag, with the $E(B-g)$ colour being the most discrepant value
of $E(B-V)_{\mathrm{host}} = 0.62 \pm 0.20$~mag. Not considering the latter value, which seems to be an outlier,
we obtain a weighted mean of $E(B-V)_{\mathrm{host}} = 0.23 \pm 0.06$~mag, where the uncertainty is equal to
standard deviation of the $E(B-V)_{\mathrm{host}}$ values. Throughout the paper we assume $E(B-V)_{\mathrm{host}} = 0.23 \pm 0.06$~mag
for SN\,2022jli, which when tacking on the MW reddening colour excess corresponds to a total value of
$E(B-V)_{\mathrm{tot}} = 0.27 \pm 0.06$~mag.

\begin{table*}
  \centering
  \caption{Summary of reddening determinations for SN\,2022jli.}
  \label{reddening_tab}
  \begin{tabular}{@{}lccc}
    \hline
    Method    & $E(B-V)_{\mathrm{tot}}$ & $E(B-V)_{\mathrm{host}}$ & Reference  \\
    \hline
    $E(B-g)$    & $0.66$($0.20$)          & $0.62$($0.20$) & \citet{stritzinger18} \\
    $E(B-V)$    & $0.32$($0.07$)          & $0.28$($0.07$) & \citet{stritzinger18} \\
    $E(B-r)$    & $0.31$($0.05$)          & $0.27$($0.05$) & \citet{stritzinger18} \\
    $E(B-i)$    & $0.30$($0.03$)          & $0.26$($0.03$) & \citet{stritzinger18} \\
    $E(g-V)$    & $0.17$($0.12$)          & $0.14$($0.12$) & \citet{stritzinger18} \\
    $E(g-r)$    & $0.18$($0.06$)          & $0.14$($0.06$) & \citet{stritzinger18} \\
    $E(g-i)$    & $0.22$($0.04$)          & $0.18$($0.04$) & \citet{stritzinger18} \\
    $E(V-r)$    & $0.18$($0.13$)          & $0.15$($0.13$) & \citet{stritzinger18} \\
    $E(V-i)$    & $0.27$($0.06$)          & $0.23$($0.06$) & \citet{stritzinger18} \\
    \ion{Na}{I} $\mathrm{EW}$ & $\cdots$   & $0.8$($0.6$)    & \citet{poznanski12, phillips13} \\
    \hline
  \end{tabular}
\end{table*}

\subsubsection{Periodicity in the light curves}
\label{sec:periodicity}

After an inspection of the light curves of SN\,2022jli, it was clear that in addition
to the main re-brightening that occurs at 25-60\,days after maximum light we could
distinguish several softer bumps or undulations, which are clearly detected in the ATLAS and ZTF
light curves. We fitted a fifth order polynomial to detrend the ZTF and ATLAS light curves
to study the structure of these undulations. The polynomial fits were performed between
  $+50$ and $+230$\,days relative to the first maximum. Polynomial fits using third
or fourth order yield very similar results. For the analysis of the ATLAS light curves
we consider observations with uncertainties smaller than $0.15$\,mag, and weighted averages were
computed for the observations obtained within the same night (i.e. observed within $8$\,hrs difference),
this procedure significantly reduces the noise of the ATLAS light curves.

\begin{figure}
\centering
\includegraphics[width=85mm]{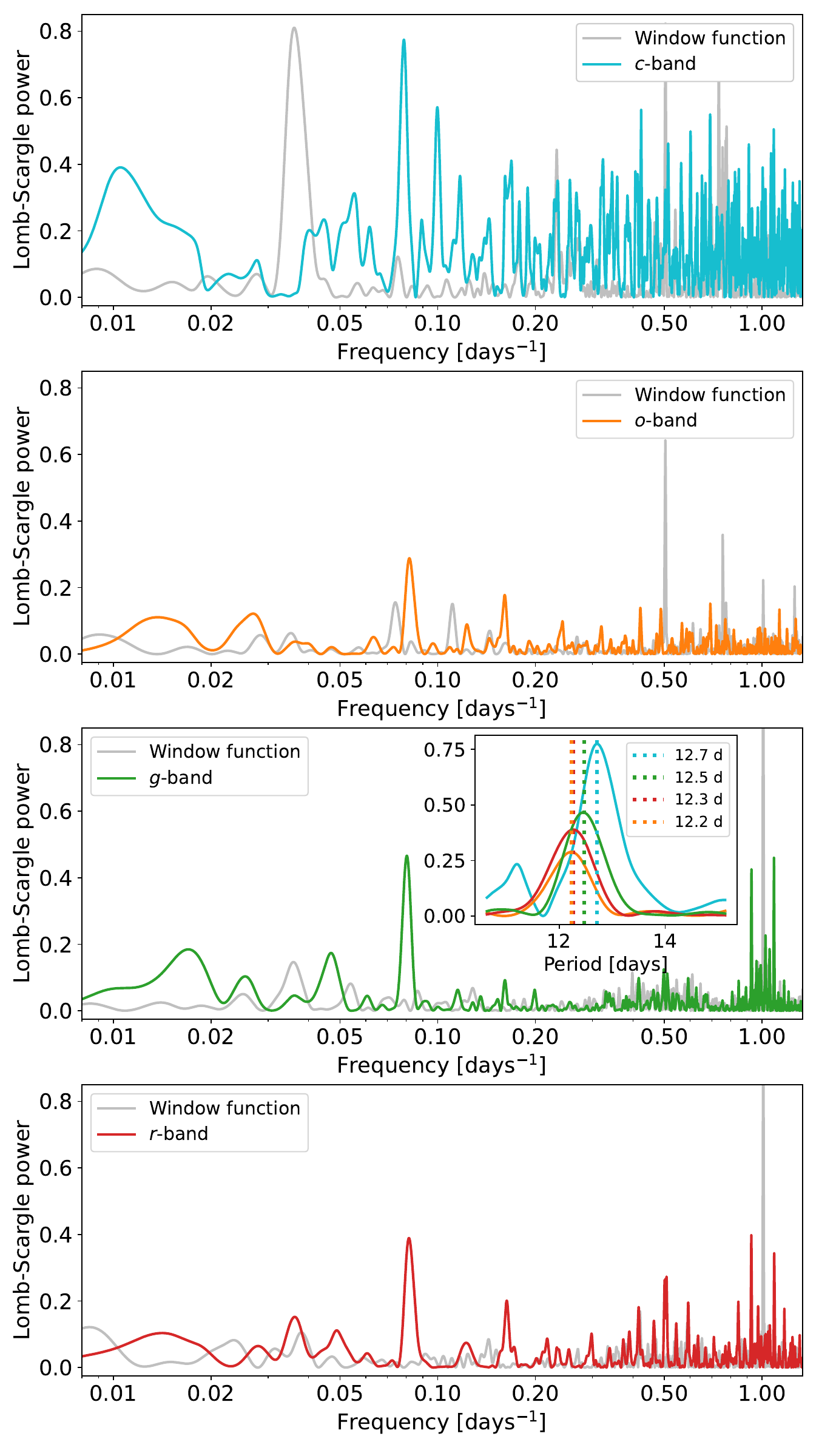} 
\caption{Lomb-Scargle periodogram of the detrended and normalised
$g$ (green), $r$ (red), $c$ (cyan), and $o$ (orange) light curves of
  SN\,2022jli. The window functions are shown in grey for the different
    bands, and the inset shows in detail the region around the
    peak frequency.
}
\label{periodogram_fig}
\end{figure}

\begin{figure}
\centering
\includegraphics[width=85mm]{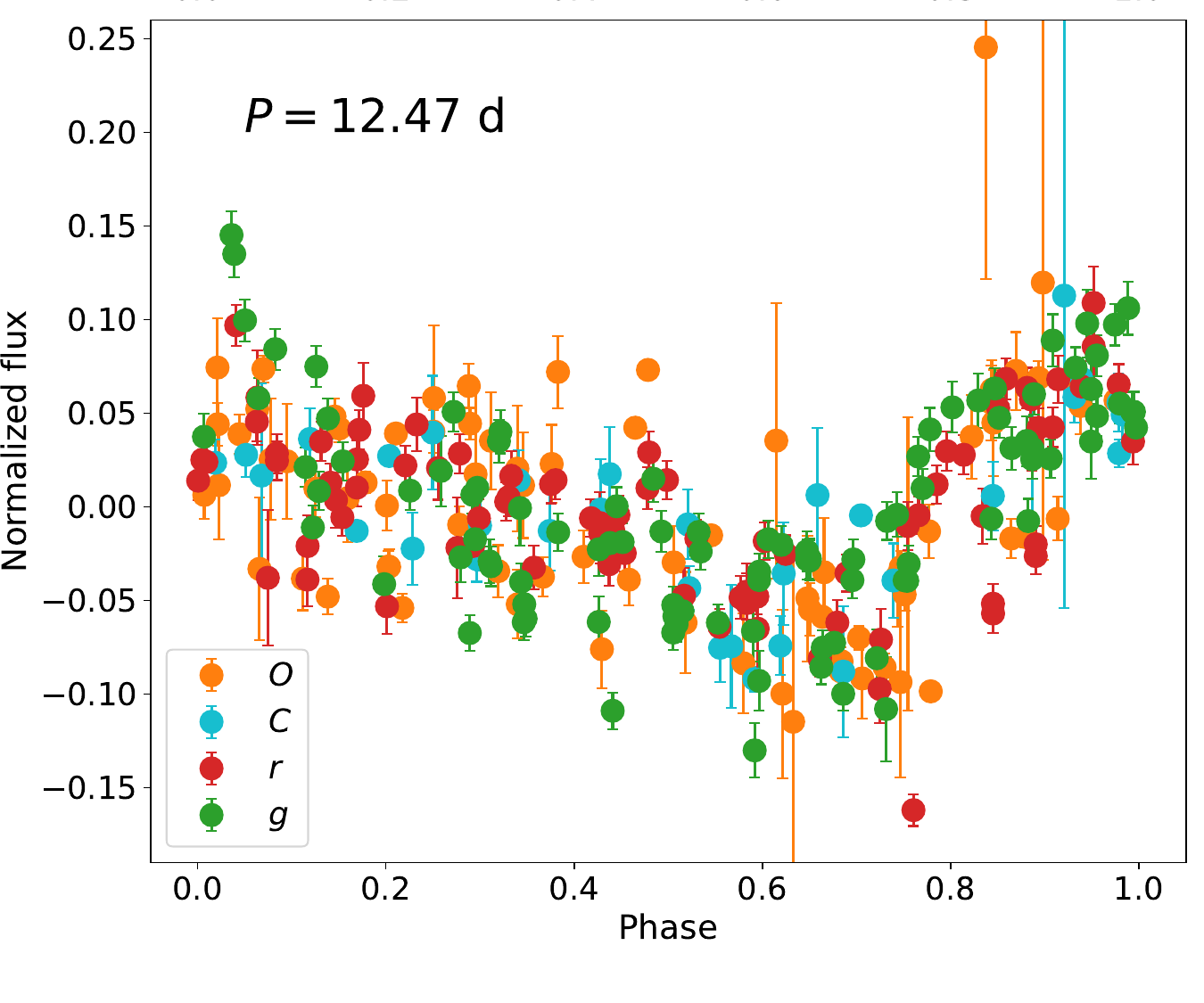} 
\caption{Folded ATLAS and ZTF light curves to a common period
of 12.47\,days.}
\label{folded_fig}
\end{figure}

The detrended residuals reveal periodic or quasi-periodic undulations, with the $g$ band showing
larger amplitudes than the $r$ band. We found that the root mean square of the
detrended $g$-band flux is $1.0 \times 10^{-16}$\,erg\,s$^{-1}$\,cm$^{-2}$\,\AA$^{-1}$ compared
with  $0.5 \times 10^{-16}$\,erg\,s$^{-1}$\,cm$^{-2}$\,\AA$^{-1}$ in the $r$ band,
confirming that the amplitude is larger in the bluer bands.
We analysed the normalised light curve residuals using a Lomb-Scargle periodogram
(see Fig. \ref{periodogram_fig}) to discover that the periodic signal
is stronger in $g$ than in the $r$ band, with a period of 12.47\,days and 12.27\,days
in the $g$ and $r$ bands, respectively. A similar analysis reveals a period of
12.71\,days and 12.23\,days in the $c$ and $o$ bands
of ATLAS, respectively. Despite the small differences in the period found for the
different bands, with bluer ($g$ and $c$) bands having slightly longer
periods than the redder bands ($r$ and $o$), folding the ATLAS and ZTF light curves
to a common period of 12.47\,days they show
a very good agreement in the periodic variability (see Fig. \ref{folded_fig}).
This is consistent with \citet{moore23} and \citet{chen24} results; thus,
we adopt this period hereafter in our analysis.

\subsubsection{Colour evolution}
\label{colour_sec}

\begin{figure}
\centering
\includegraphics[width=85mm]{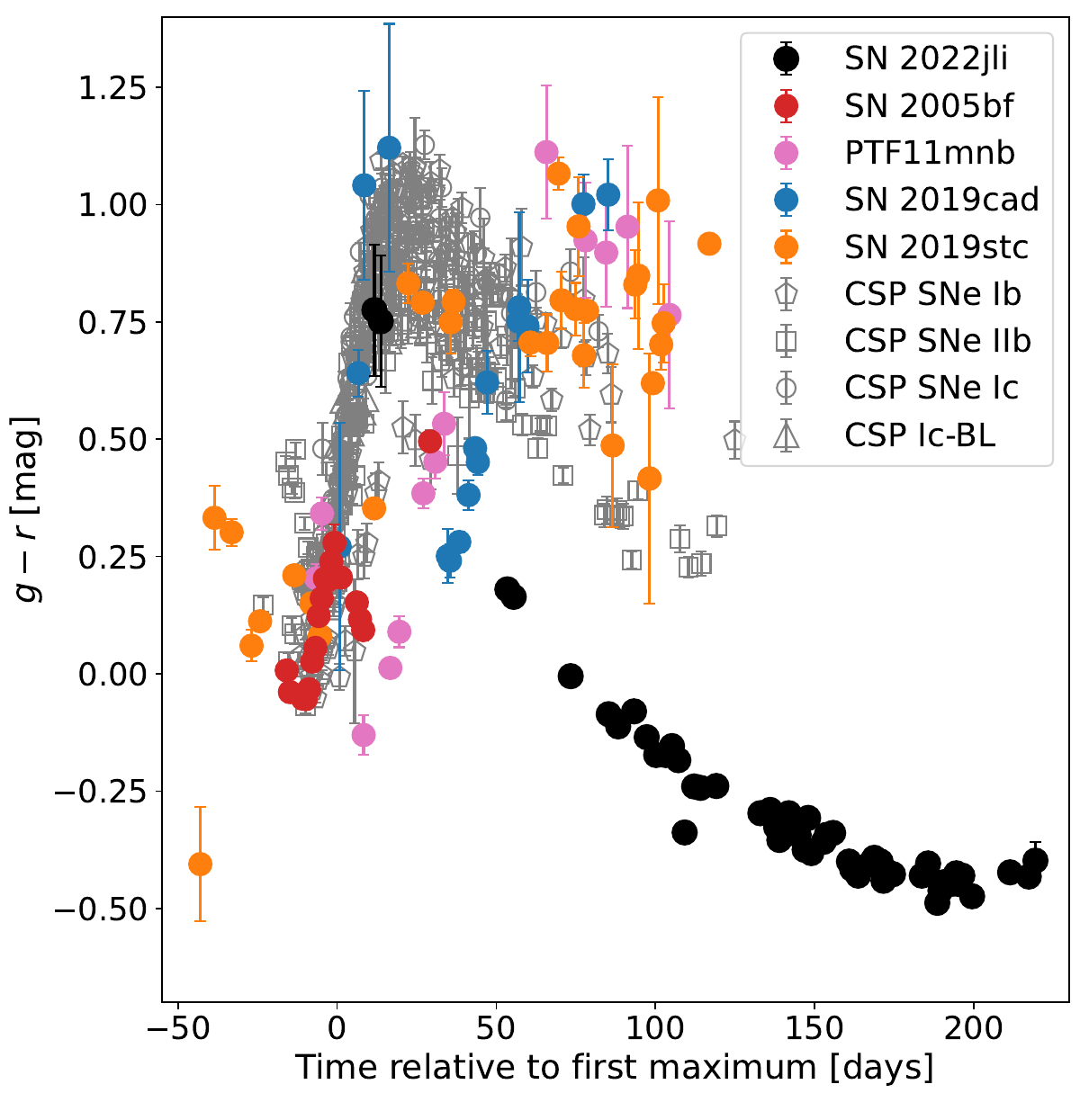} 
\caption{$g-r$ colours of SE SNe corrected by reddening using the \citet{fitzpatrick99}
  reddening law. SN\,2022jli is plotted in black, while in grey are the CSP-I SE SNe \citep{stritzinger18}
  with different symbols used for each SE SN sub-type as is indicated in the legend.
  For comparison are also presented the double-peaked SE SN\,2005bf \citep[Ib-pec; MJD$_{\mathrm{max}} = 53474.0$;][]{folatelli06},
  PTF11mnb \citep[Ic-pec; MJD$_{\mathrm{max}} = 55827.4$;][]{taddia18}, SN\,2019cad \citep[Ic-pec; MJD$_{\mathrm{max}} = 58558.6$;][]{gutierrez21}
  and SN\,2019stc \citep[Ic-pec; MJD$_{\mathrm{max}} = 58793.5;$][]{gomez21}.
}
\label{g_r_colour_fig}
\end{figure}

In Fig. \ref{g_r_colour_fig} we present the $g-r$ colours of SN\,2022jli compared with the colours of the
CSP-I SE SN sample \cite{stritzinger18b,stritzinger18}, the double-peaked SE SNe 2005bf \citep{folatelli06},
PTF11mnb \citep{taddia18}, 2019cad \citep{gutierrez21}, and 2019stc \citep{gomez21}.
As could be expected, the synthetic colours of SN\,2022jli at $+12$ and $+14$\,days corrected for reddening
(see Section~\ref{sec:reddening}) are located at the locus of the SE SNe intrinsic colour-curve evolution from
maximum to $+20$\,days.

As can be readily observed in Fig. \ref{g_r_colour_fig}, after the onset of the SN re-brightening
(at $20-35$\,days), the SN displays very blue colours compared to the bulk of SE SNe. The SN is
$\sim 0.65$\,mag bluer than the bulk of SE SNe at about $+50$\,days. The SN continues evolving to
bluer $g-r$ colours until about $+190$\,days after maximum, when a colour minimum is reached. Although
all normal SE SNe tend to evolve from red to blue colour after reaching a peak in
the $g-r$ colour at about 25-30\,days, the extreme colours of SN\,2022jli are remarkable. The blue colour
is presumably related with the extra power injection which produces the secondary maximum, increasing the
ejecta temperature as we discuss in Section \ref{bolometric_sec} below. If the SN\,2022jli is not corrected by Galactic
and host-galaxy reddening, the $g-r$ colour of the SN would be $\sim 0.25$\,mag bluer than the bulk of
the SE SNe. It is worth noting that SN\,2022jli has bluer colours and higher
ejecta temperature than the double-peaked SN\,2019stc which follows a colour evolution comparable
to the bulk of SE SNe. This suggests that the physical mechanism(s) producing the secondary maximum
is probably different in these two SNe.

We note that the strong \ion{Na}{I}\,D absorption line is readily noticeable throughout the SN evolution,
suggesting that the dust and gas is not destroyed by the SN radiation, or by the potential ejecta-CSM interaction. Unfortunately,
at late phases  the \ion{Na}{I} feature is located at the edge of an underlying broad spectral feature that prevents the most
reliable measurement of its $EW$ \citep[see e.g.,][]{santiago24}.

\begin{figure*}
\centering
\includegraphics[width=180mm]{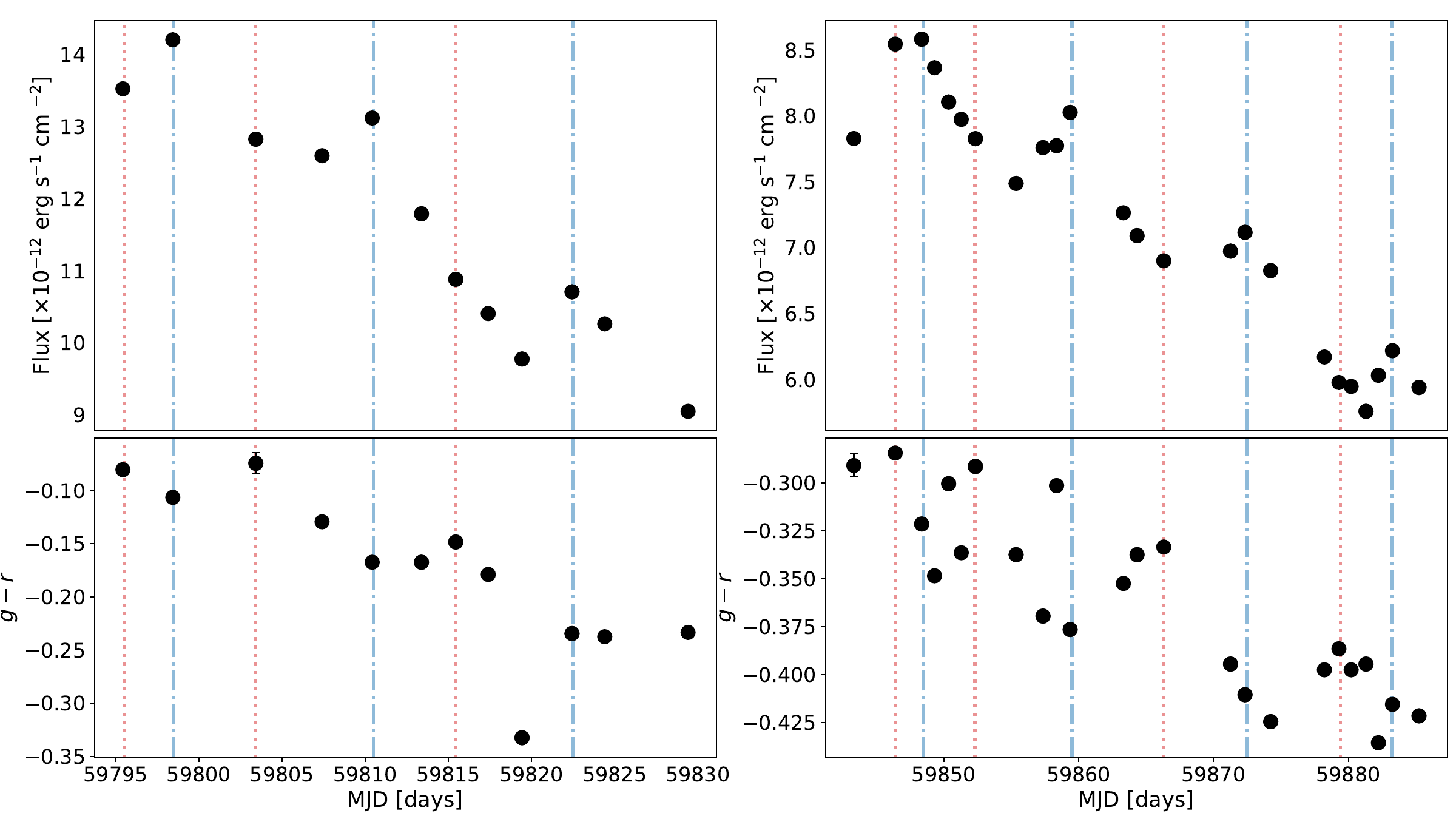} 
\caption{Top panels: Total ($g+r$) flux corrected by reddening ($E(B-V) = 0.27$) following the \citet{fitzpatrick99}
reddening law for two regions with well-sampled light curves. Bottom panels: $g-r$ colours (corrected by reddening)
for the same two regions with well-sampled light curves. The vertical dash-dotted blue lines mark the observed flux
maxima, and the vertical dashed red lines mark the colour maxima (redder colours). The flux and the colours
show periodic undulations.
}
\label{flux_color_var_fig}
\end{figure*}

In Fig. \ref{flux_color_var_fig} we investigate the colour behaviour during the light curve
undulations in detail. We present the total flux emission in the ($g+r$) bands corrected by reddening,
compared with the $g-r$ colours in two well-sampled regions of the ZTF light curves. We find that in
addition to becoming blue with time, there is large scatter in the $g-r$ colour showing periodic or
quasi-periodic evolution. The peak to trough variation in the $g-r$ colours is of the order of $0.07$ to $0.10$ mag,
which is $5-7\sigma$ larger than the $0.015$\,mag variation expected from the median of colour photometric uncertainty alone.
In Fig. \ref{flux_color_var_fig}, we mark the maxima of the ($g+r$) flux (dash-dotted blue line) and
the peaks of the SN colours (dashed red lines), finding that at the epochs of flux maxima the SN tend to be bluer,
while at the epochs of $g-r$ colour peaks (redder colours) the SN is (generally) declining in ($g+r$)
flux and close to a flux minima following a periodic or quasi-periodic cycle.

\subsubsection{Light curve comparison}

\begin{figure}
\centering
\includegraphics[width=85mm]{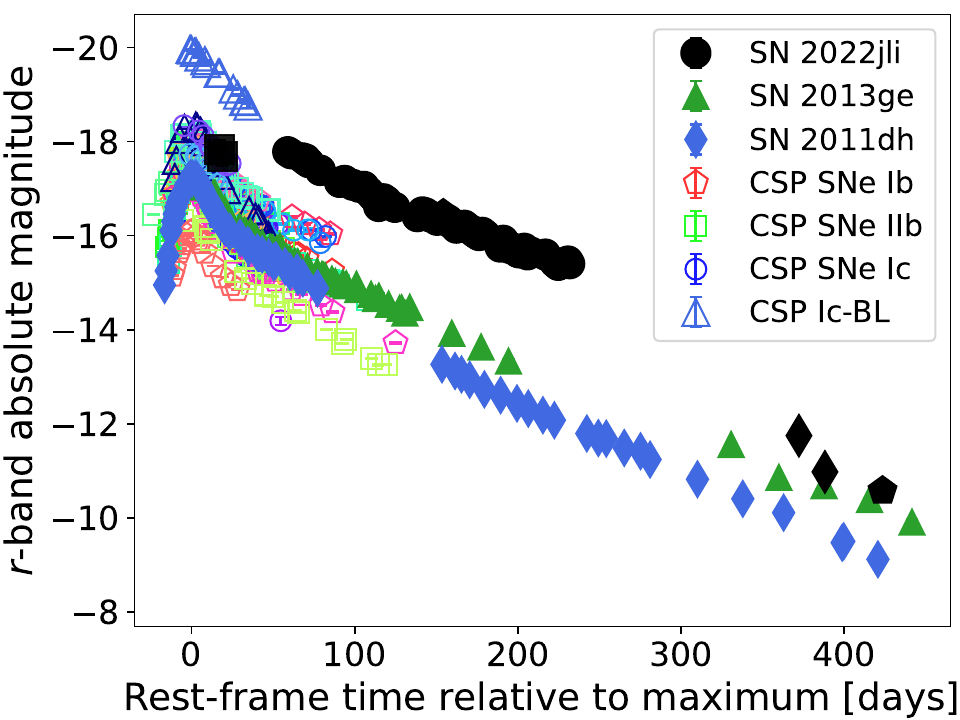} 
\caption{Comparison between the $r$-band absolute magnitude of SN\,2022jli (black symbols)
  and well-observed SE SNe from CSP-I \citep[open symbols;][]{stritzinger18}, SN\,2011dh
  \citep[blue diamonds;][]{ergon15, ergon14} and SN\,2013ge \citep[green triangles;][]{drout16}.
  SNe IIb, Ib, Ic and Ic-BL from the CSP-I are shown with open squares, pentagons,
  circles and triangles, respectively.
}
\label{Optical_lc_comp_fig}
\end{figure}

In Fig.~\ref{Optical_lc_comp_fig} the $r$-band absolute magnitude light curve of SN\,2022jli
is compared with the well observed SN\,2011dh \citep[Ib;][]{ergon14, ergon15}, SN\,2013ge
\citep[Ic;][]{drout16} and the SE SNe from the CSP-I \citep{stritzinger18b, stritzinger18}.
The absolute magnitude at the first maximum of SN\,2022jli is $\simeq -17.9$\,mag\footnote{Estimated from
the Gaia photometry of SN\,2022jli close to maximum ($G = 14.60$\,mag), after correcting by reddening in the line-of-sight.}
compared with the maxima absolute magnitude of $\simeq -17.2$\,mag of SNe 2011dh and 2013ge.
As can be observed, SN\,2022jli seems to be a bright SE SN, but within the range of normal luminosity
objects. After the second maximum, the light curve of SN\,2022jli is nearly parallel
to the comparison objects, but with SN\,2022jli is $\sim 2.3$\,mag brighter.
At later times, SN\,2022jli suffered a sudden luminosity drop between about $+250$ and $+375$\,days
\citep[see also][]{chen24}, and by $+400$\,days the SN spectrum (see Section \ref{nebular_spec_sec} below)
and brightness in the $r$ band are similar to SN\,2013ge.

\subsubsection{Bolometric light curve}
\label{bolometric_sec}

\begin{figure}
\centering
\includegraphics[width=80mm]{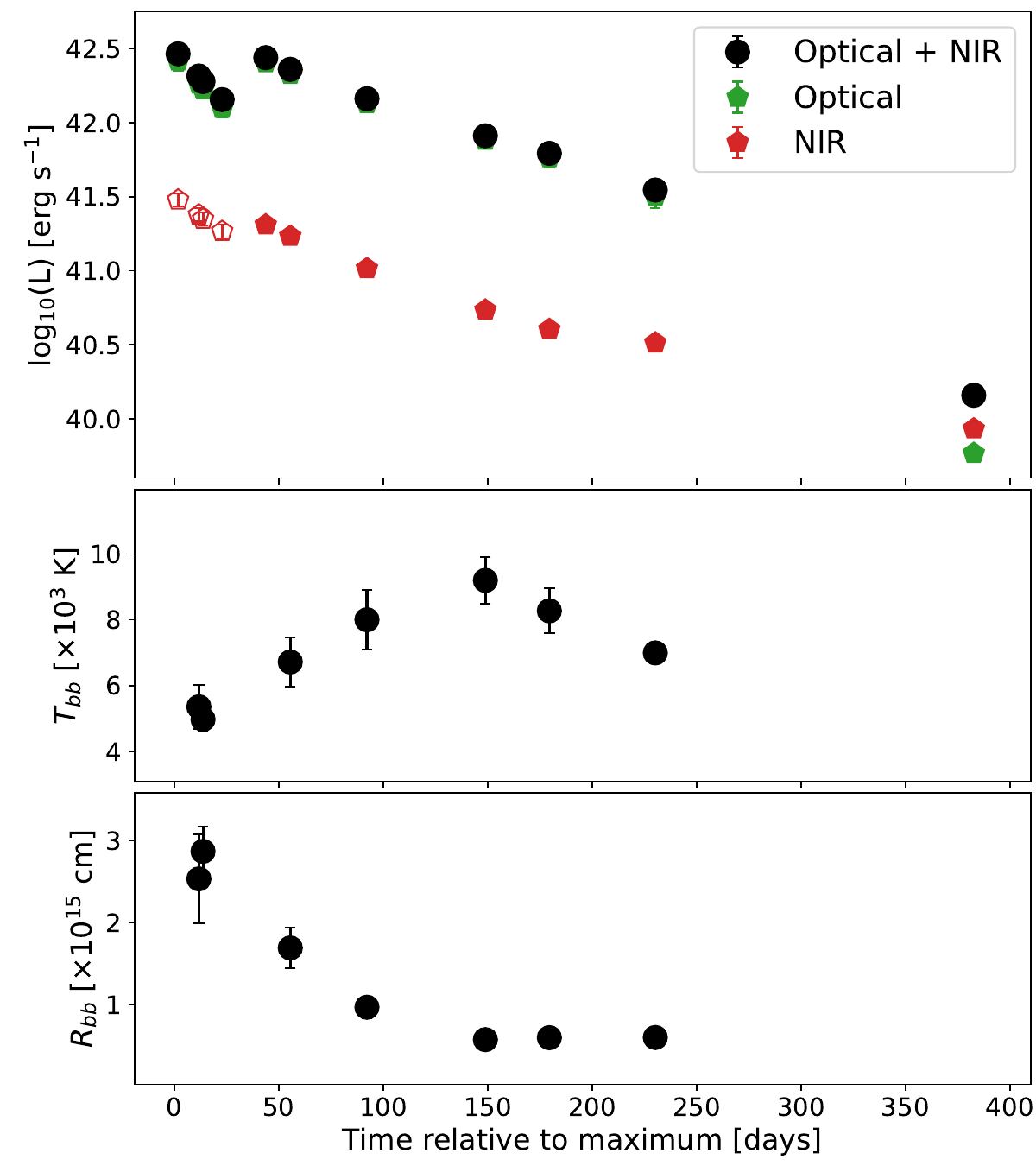} 
\caption{Top: Pseudo-bolometric light curve of SN\,2022jli (black circles) computed integrating
  the SN spectra in the optical region between $3800-9000$\,\AA\ (green pentagons) and taking in account
  the NIR contribution (red pentagons). The NIR emission during the decline from the first maximum was
  estimated from optical+NIR templates of SN\,2013ge as described in the text, and these estimates
  are shown using open pentagons. The pseudo-bolometric luminosity at $+383$\,days was estimated
  directly from the $griJHK_{s}$ photometry.
  Middle: $T_{\mathrm{bb}}$ computed from a black-body fit to the spectra of SN\,2022jli.
  Bottom: $R_{\mathrm{bb}}$ from a black-body fit to the spectra of SN\,2022jli.}
\label{bol_bb_params_fig}
\end{figure}

In Fig. \ref{bol_bb_params_fig} we present the pseudo-bolometric light curve of SN\,2022jli.
At early times during the first and secondary maximum, the scarcity of
multi-band photometric observations poses a problem for the construction
of a pseudo-bolometric light curve of SN\,2022jli. We circumvent this
problem by constructing spectral templates which are scaled in flux,
using contemporary photometry to achieve an accurate flux calibration
of the templates. The photometry used to scale the spectra was obtained
as close in time as possible to the observing epoch of the spectra,
usually in the same or within a couple of nights. In a few cases the
photometry used to scale the spectra comes from the acquisition images
of the spectra. At the minimum of the $c$ band and the maximum of the $o$ band
we scaled template spectra with a time difference of 9\,days and 12\,days relative to the
time of the photometry, respectively. We consider it important to have an estimate of the
SN luminosity at these phases. In Section \ref{sec:spec_analysis} we show that the spectra
vary slowly during this time (from $+14$ to $+55$\,days in Fig. \ref{specseq_op_fig}).
The uncertainty introduced by the use of spectra and photometry with a relevant
time difference is accounted in the bolometric error (see Appendix).
As is shown in Section \ref{sec:spec_analysis} and in the Appendix,
the spectrum of SN\,2022jli after maximum light is similar to the spectra of
SN\,2013ge at a similar phase. We exploited this similarity to estimate the
bolometric luminosity of SN\,2022jli close to maximum light by employing the
template spectrum of SN\,2013ge at maximum. The maximum light spectral template
of SN\,2013ge was carefully colour-matched to the photometry
of SN\,2013ge and corrected by reddening along the line-of-sight
to the SN. Then, the reddening free spectral template of SN\,2013ge was reddened by the
$E(B-V)$ in the line-of-sight to SN\,2022jli and scaled to the Gaia
photometry of SN\,2022jli close to maximum light (see the Appendix for more details).
The pseudo bolometric luminosity of SN\,2022jli is summarised in
Table \ref{pseudo_bol_tab}.

We estimate the NIR contribution at the first maximum and during the
decline from it using spectrophotometric templates of SN\,2013ge.
These optical-NIR templates were constructed by colour matching the optical spectra
to the photometry, and the NIR spectra were scaled to the optical spectra in their
overlapping regions (from $8000-9500$\,\AA\ approximately).
We estimate that the NIR contribution is of the order of $10-15$\% of
the total optical and NIR emission during this phase. From $+38$\,days
the NIR contribution was estimated directly from the NIR photometry
of SN\,2022jli, and is of the order of $7-9$\% until $\sim 230$\,days.
At about $+380$\,days the NIR corresponds to $59$\% of the SN emission.

The bolometric light curve of SN\,2022jli is double-peaked, with
both maxima reaching a similar bolometric luminosity of about
$3 \times 10^{42}$\,erg\,s$^{-1}$.\footnote{The SN reached a luminosity of
$2.9 \pm 0.4 \times 10^{42}$\,erg\,s$^{-1}$ in the first maximum and a luminosity of
about $2.8 \pm 0.3 \times 10^{42}$\,erg\,s$^{-1}$ in the second maximum.}
The estimated decline rate from the first maximum is $3.67 \pm 0.94$\,mag/100\,days,
and the decline rate after the second maximum is $1.18 \pm 0.10$\,mag/100\,days.
At later times the SN showed a significant drop in luminosity, and the pseudo-bolometric
luminosity becomes dominated by the NIR emission (corresponding to 60\%).

A black-body model was fitted to the optical spectra of SN\,2022jli
after scaling them and matching their colours to the photometry. The estimated black-body
temperature ($T_{bb}$) and radius ($R_{bb}$) are shown in Fig. \ref{bol_bb_params_fig}.
During the decline from the first maximum the ejecta temperature is $\sim 5000$\,K
and $R_{bb} \simeq 2.7 \times 10^{15}$\,cm ($\simeq 39000$\,$R_{\odot}$). After the second maximum
the ejecta temperature increases and the $R_{bb}$ shows a decrease with time
reaching a minimum of $R_{bb} \simeq 6 \times 10^{14}$\,cm ($\simeq 8600\,R_{\odot}$)
at about $+150$\,days and staying constant thereafter. The ejecta temperature
reaches a maximum temperature of $\simeq 9000$\,K at about $+150$\,days, and the ejecta
temperature decreases thereafter.

\subsubsection{$^{56}$Ni and ejecta mass}
\label{nickel_sec}

In order to estimate the mass of the ejecta ($M_{\mathrm{ej}}$) and of the
radioactive $^{56}$Ni ($M_{\mathrm{Ni}}$) produced by the SN, we used the \citet{arnett82}
analytical solution for type I SNe, assuming that the first maximum is powered by the
radioactive decay of $^{56}$Ni and employing the methodology described in \citet{cartier22}.
We fitted the bolometric light curve of SN\,2022jli from the first maximum to the minimum at about $+20$\,days,
assuming an ejecta velocity at maximum light of $v_{ej} = 8500$ km\,s$^{-1}$ (see Section \ref{exp_vel_sec}),
an optical opacity of $\kappa =0.07$\,cm$^{2}$\,gr$^{-1}$ and a rise time to maximum
light of $\tau_{\mathrm{rise}} = 14$\,days. This rise time provides the best fit to the bolometric
light curve measured using the reduced $\chi^{2}$. The rise time to maximum calculated for SN\,2022jli
is also consistent with the typical rise times measured in normal SNe\,Ic. For example,
\citet{taddia15} reports average rise times in the $r$ band of $22.9 \pm 0.8$\,days, $21.3 \pm 0.4$\,days,
$11.5 \pm 0.5$\,days and $14.7\pm 0.2$\,days for SNe IIb, Ib, Ic and Ic-BL from SDSS-II, respectively. Although
Arnett's analytical solution has limitations \citep[see e.g.,][]{khatami19, afsariardchi21, rodriguez23},
like the assumption of a constant optical opacity, Arnett’s model remains a widely used, simple, and powerful
analytical model to obtain reasonable estimates for $M_{\mathrm{ej}}$ and $M_{\mathrm{Ni}}$.
Other methodologies cannot be used to estimate the $M_{\mathrm{Ni}}$ because; i) the time of the explosion
is not well constrained from the observations, ii) the lack of multi-band photometry during the rise to the
first maximum, and iii) only one (pseudo) bolometric epoch is available at late times when the power
source that produced the second maximum had significantly decreased.
Following \citet{taddia18b}, it is assumed $\kappa =0.07$\,cm$^{2}$\,gr$^{-1}$, since it provides
consistent results for the $M_{\mathrm{ej}}$ and $M_{\mathrm{Ni}}$ parameters when Arnett's
analytical model is compared with the hydro-dynamical models of \citet{bersten11, bersten12}, both
fitted to the bolometric light curves of the CSP-I sample \citep[see][for a discussion]{taddia18b}.

\begin{figure}
\centering
\includegraphics[width=85mm]{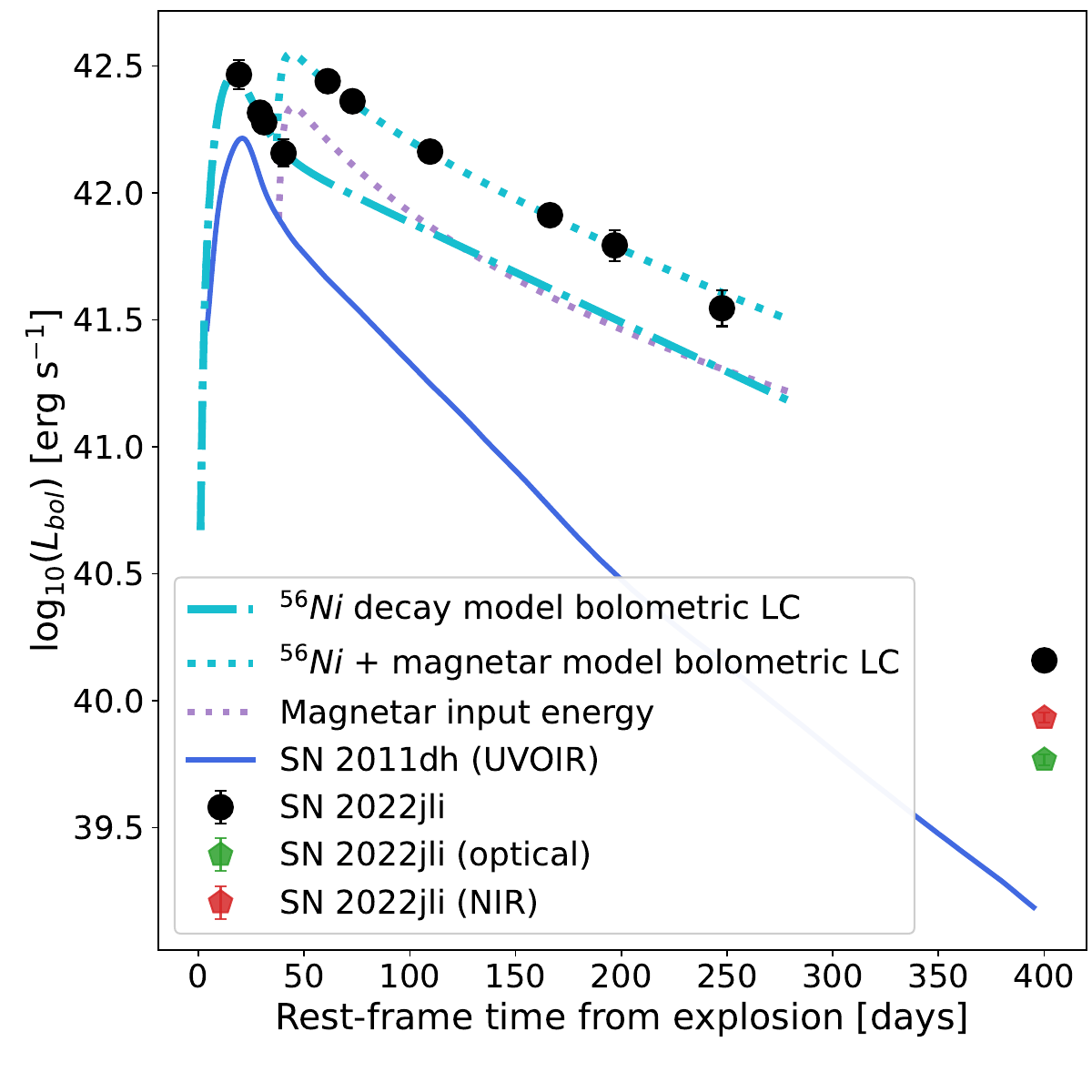} 
\caption{Fit of the radioactive decay of $^{56}$Ni to the first maximum of the bolometric light
curve of SN\,2022jli, using the \citet{arnett82} analytic approximation. The parameters of the
best-fit model are $\tau_{\mathrm{rise}} = 14$\,days, $M_{\mathrm{ej}} \sim 1.5$\,$M_{\odot}$,
$M_{\mathrm{Ni}} \sim 0.12$\,$M_{\odot}$, $v_{\mathrm{ej}}  = 8500$\,km\,s$^{-1}$ and
$\kappa = 0.07$\,cm$^2$\,gr$^{-1}$. This model illustrates that the first maximum of SN\,2022jli
is consistent with the typical rise time, $^{56}$Ni and ejecta mass of a normal SN\,Ic.
The estimated MJD of the explosion is $59695.4$ days. The secondary maximum
is modelled as the sum of the radioactive decay of $^{56}$Ni and the
spin-down energy from a magnetar \citep[dotted cyan line; see][for details about the magnetar model]{cartier22}.
To model the secondary maximum the magnetar birth is delayed 37\,days
relative to the assumed time of the SN explosion and the magnetar parameters are
$B \sim 8.5 \times 10^{14}$\,G and $P \sim 48$\,ms. The dotted purple line shows the
magnetar contribution to the bolometric light curve. The optical and NIR luminosity
at $+383$\,days are shown using green and red pentagons, respectively. The
pseudo-bolometric (UVOIR) light curve of the well-observed SE SN\,2011dh (IIb) is shown for comparison
\citep[$M_{\mathrm{Ni}} = 0.075$\,$M_{\odot}$;][]{ergon15}.
}
\label{Ni_bol_lc_fig}
\end{figure}

Figure \ref{Ni_bol_lc_fig} presents the analytical fit to the first bolometric peak of  SN\,2022jli.
Given that the rise to maximum of SN\,2022jli is not constrained, we explored a range of $\tau_{\mathrm{rise}}$
from $+10$ to $+23$\,days, which comprises the range of typical $\tau_{\mathrm{rise}}$ found
in SNe~Ibc \citep[see][]{taddia15}. It was found that 14\,days provides the best analytical fit
for SN\,2022jli, with good fits having $\tau_{\mathrm{rise}}$ in the range of 11 to 17\,days.
In this range of $\tau_{\mathrm{rise}}$ the analytic solution provides an ejecta mass ($M_{\mathrm{ej}}$)
and a radioactive $^{56}$Ni mass ($M_{\mathrm{Ni}}$) that do not significantly vary. The values
obtained for SN\,2022jli are $M_{\mathrm{ej}} \sim 1.5 \pm 0.4$\,$M_{\odot}$
$M_{\mathrm{Ni}} \sim 0.12 \pm 0.01$\,$M_{\odot}$, where the quoted errors reflect
the uncertainty in the assumed $\tau_{\mathrm{rise}}$ and $v_{\mathrm{ej}}$. The good quality of
the fit to the first maximum of the bolometric light curve, the fact that $M_{\mathrm{ej}}$ and
$M_{\mathrm{Ni}}$ are typical of a standard SN Ic and that SN\,2022jli shows spectral
features of a SN\,Ic, supports the idea that the first maximum of SN\,2022jli is a standard SN\,Ic
mainly powered by the decay of $^{56}$Ni.

As was noted previously, at later times the SN shows a drop in luminosity, which marks
a decrease in the power injection from the energy source producing the second maximum. This might be
a consequence of the well known drop in the ejecta opacity to $\gamma$-rays at late times, due to
the expansion of the SN ejecta. This is illustrated in Fig. \ref{Ni_bol_lc_fig} with the rapid
decline of the UVOIR light curve of SN\,2011dh \citep{ergon15} after maximum light.
SN\,2011dh is chosen for comparison because is a normal SE SN (type IIb) with a high-quality and well-sampled UVOIR
light curve, constructed from high-quality and well-sampled optical and IR observations
from the $U$ band to the mid-IR {\it Spitzer} $S_{2}$ band collected from $+4$\,days until about $+400$\,days.
SN\,2011dh has an estimated $M_{\mathrm{Ni}} = 0.075$\,$M_{\odot}$, and its decline rate after
maximum is significantly faster than the decline rate expected of full trapping from the $^{56}$Co decay,
the daughter product of $^{56}$Ni. As in the case of SN\,2011dh, a significant fraction of the power from the
$^{56}$Co radioactive decay probably escape after maximum in SN\,2021jli.\footnote{After 200\,days, is expected that
practically all $\gamma$-ray photons from the electron capture process will escape \citep[see e.g.,][]{milne99}.}

\subsection{Spectra}
\label{sec:spec_analysis}

\subsubsection{Spectral comparison}

\begin{figure*}
\centering
\includegraphics[width=180mm]{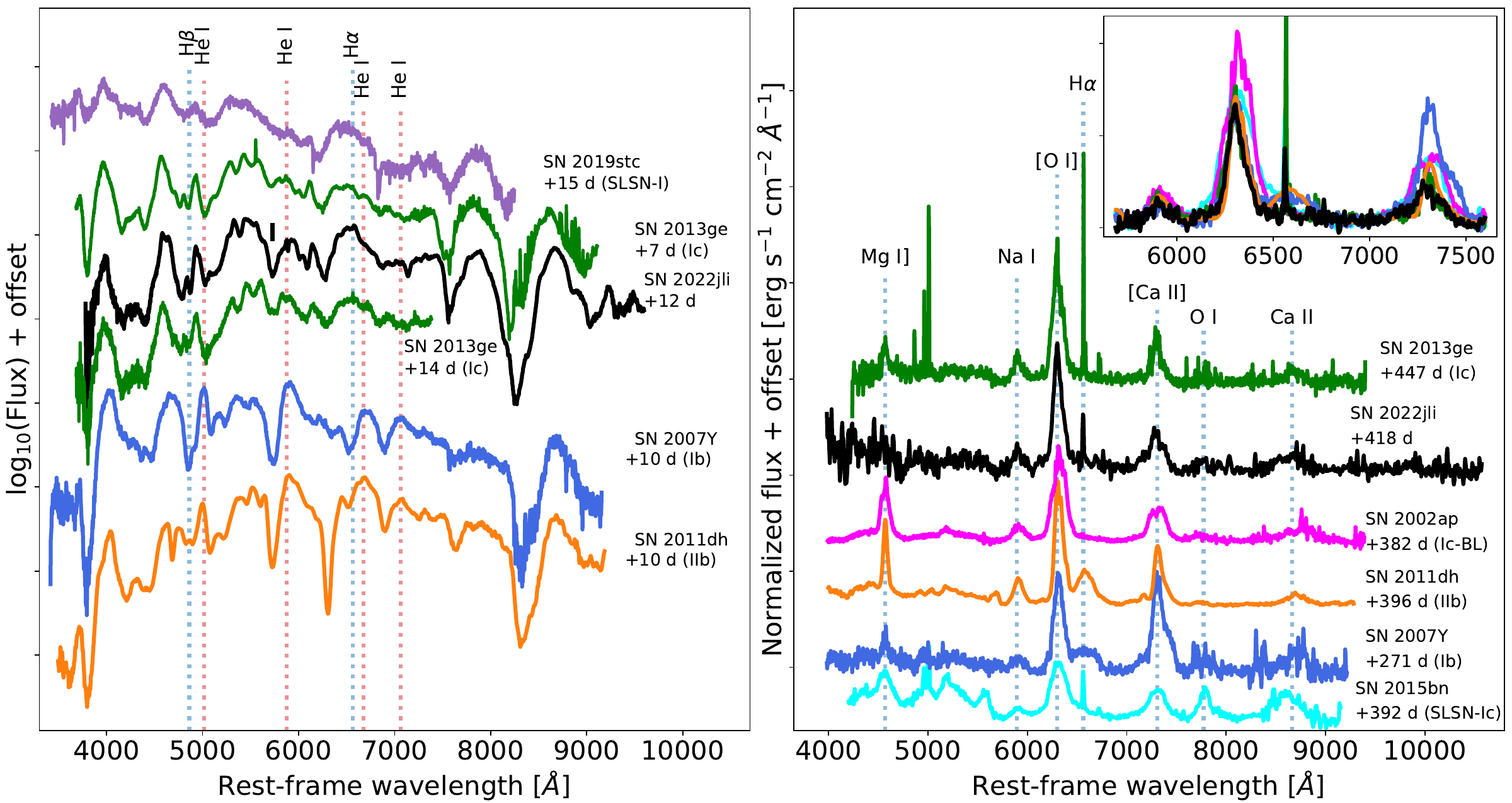} 
\caption{Comparison of SN\,2022jli with other SE SNe in decline from
  the first maximum and at a late nebular phase. On the left panel, SN\,2022jli is
compared at about $+12$\,days relative to the first maximum, during the
decline from maximum light, with the bumpy SLSNe-I SN\,2019stc \citep{gomez21},
the Ic SN\,2013ge \citep{drout16}, the Ib SN\,2007Y \citep{stritzinger09,stritzinger23}
and the IIb SN\,2011dh \citep{ergon14}. The black tick marks the \ion{Na}{I}
  D feature which is strong in SN\,2022jli possibly due to the \ion{He}{I} $\lambda 5876$
  line contribution. Similarly, on the right panel, SN\,2022jli
is compared with the SLSN-I SN\,2015bn \citep{nicholl16b}, the Ic-BL
SN\,2002ap \citep{shivvers19}, SN\,2013ge and SN\,2011dh \citep{ergon15}
at nebular phase. We indicate the rest-frame wavelength of the main
spectral features. All the spectra have been corrected for reddening assuming
$R_{V} = 3.1$ and values from the literature, in the case of SN\,2022jli
we assume $E(B-V)_{\mathrm{tot}} = 0.27$ mag.}
\label{spectral_comp1_fig}
\end{figure*}

In Fig. \ref{spectral_comp1_fig}, SN\,2022jli is compared with other SE SNe
at about $+12$\,days, during the decline from maximum brightness and before its re-brightening.
SE SNe are classified as Type IIb, Type Ib and Type Ic depending
on the presence of clear and unambiguously detectable amounts of hydrogen and helium
in the SN ejecta. At this phase, the spectral features of SN\,2022jli correspond to a SN\,Ic
very similar to SN\,2013ge during its decline from maximum light. The spectra of SN\,2022jli
do not show strong helium or Balmer absorption lines such
as in SN\,2007Y (Ib) and in SN\,2011dh (IIb). A comparison with the bumpy SLSN-I SN\,2019stc
shows that this SLSNe exhibit a bluer continuum and broader spectral lines yielding more blended
spectral features compared to their lower luminosity cousins like SN\,2022jli.

A manifestation that the envelope stripping is probably not complete is that some
SNe\,Ib exhibit small amounts of hydrogen material \citep[see e.g.,][]{holmbo23,dong23}
and some SNe Ic show evidence of helium in their ejecta \citep[see e.g.,][]{drout16,shahbandeh22,holmbo23}.
This is also the case of SN\,2013ge that although it is classified as a SN\,Ic,
it shows weak \ion{He}{I} features in the optical and in the NIR \citep[see][]{drout16}
and has an estimated ejecta mass of 2-3\,$M_{\odot}$, not far from the estimated ejecta
mass of SN\,2022jli. Similarly, SN\,2022jli also shows weak \ion{He}{I} features as
discussed below \cite[see also][]{moore23}. Additionally, a double-peaked, weak
and variable H$\alpha$ feature is robustly detected after the second maximum.
This spectral feature shows a periodic shift in its peak following
the light curve undulations and it seems to be a manifestation of the accretion of
hydrogen-rich material from a companion star in a binary system \citep{chen24}.
The Ic classification of SN\,2022jli is reinforced by its nebular spectrum
(see Fig. \ref{spectral_comp1_fig}); this is discussed in detail below.

\begin{figure*}
\centering
\includegraphics[width=180mm]{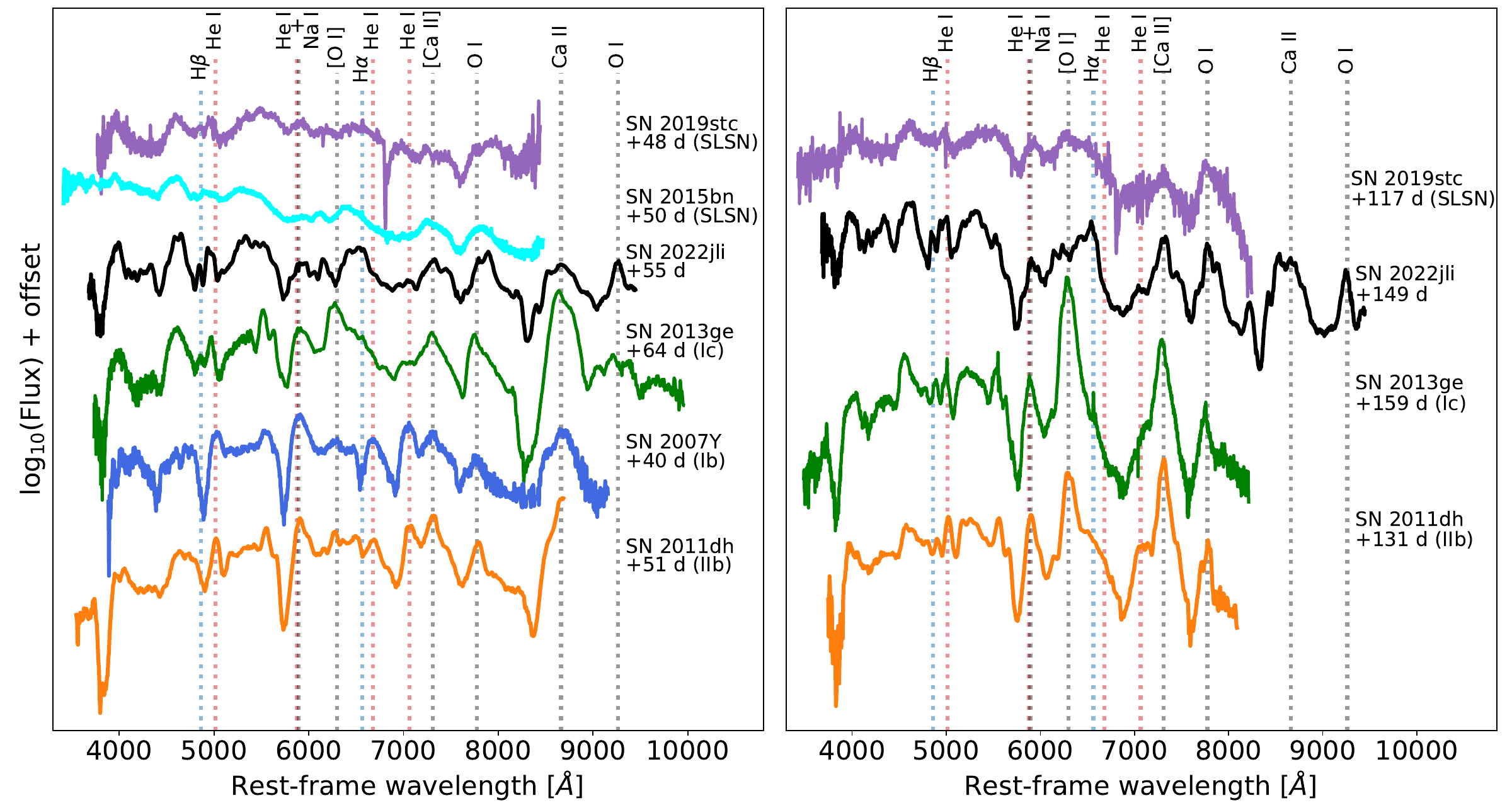} 
\caption{Comparison of SN\,2022jli with other SE SNe during the
SN re-brightening. On the left panel, the SN is compared at the time of its
second maximum with the SLSNe-I SN\,2019stc \citep{gomez21} and
SN\,2015bn \citep{nicholl16}, the Ic SN\,2013ge \citep{drout16},
the Ib SN\,2007Y \citep{stritzinger09,stritzinger23} and the IIb
SN\,2011dh \citep{ergon14} at a similar phase relative to the
first maximum. Similarly, on the right panel, SN\,2022jli
is compared with SN\,2019stc, SN\,2013ge and SN\,2011dh
\citep{ergon15} at about $+149$\,days relative to the first
maximum. We indicate the rest-frame wavelength of the main
spectral features. All the spectra have been corrected for reddening assuming
$R_{V} = 3.1$ and values from the literature, in the case of SN\,2022jli
we assume $E(B-V)_{\mathrm{tot}} = 0.27$ mag.
}
\label{spectral_comp2_fig}
\end{figure*}

Close to the time of the second maximum, the spectral features of SN\,2022jli
do not show a significant evolution (see Fig. \ref{spectral_comp2_fig}). At this phase, the spectrum of
SN\,2022jli still shows a clear pseudo-continuum, and a few (pseudo-)nebular features begin to
appear in the red part of the optical spectrum, such as [\ion{Ca}{II}] $\lambda\lambda 7291, 7323$
and \ion{O}{I} $\lambda 7774$. In turn, the spectral features in the blue part of the optical region
(\textless 7000 \AA) remain essentially the same, and remain almost unaltered for the next 100\,days,
evolving very slowly.

Between $+45$ and $+65$\,days, normal SNe Ic become pseudo-nebular,
the continuum emission decreases and becomes redder. Spectral features of [\ion{Ca}{II}] $\lambda\lambda 7291, 7323$,
and a strong \ion{Ca}{II} NIR P-Cygni emission are clearly present, and a weak
[\ion{O}{I}] $\lambda\lambda 6300, 6364$ feature can be distinguished. In turn, the \ion{Ca}{II} NIR triplet
in SN\,2022jli is shallower than in the comparison SE SNe, and even compared with the same SN during
its decline from the first maximum. All of this is a consequence of the extra energy input which makes
the pseudo continuum bluer compared with other SE SNe. The [\ion{O}{I}] $\lambda\lambda 6300, 6364$
doublet is absent in SN\,2022jli at this epoch, and it may weakly contribute to the SN spectrum after
$+150$\,days. However, the [\ion{O}{I}] doublet seems to appear after $+200$\,days in the
SN spectrum and it becomes the strongest optical spectral feature by $+400$\,days
(see Fig. \ref{spectral_comp1_fig}).

In the right panel of Fig. \ref{spectral_comp2_fig} we compare SN\,2022jli with other SE SNe at 150\,days.
At this phase SN\,2022jli is becoming pseudo-nebular, a decrease in the continuum can be appreciated
at wavelengths redder that 5500\,\AA, but the continuum remains bluer than that of normal luminosity
SE SNe, its continuum shape is more similar to SLSNe-Ic. The shape of the spectral features, on the other hand,
is more similar to SE SNe rather than to SLSNe. For comparison, at this phase SE SNe are nebular,
showing [\ion{O}{I}] $\lambda\lambda 6300, 6364$ and [\ion{Ca}{II}] $\lambda\lambda 7291, 7323$ lines
(see Fig. \ref{spectral_comp2_fig}).

In summary the evolution of the optical spectra of SN\,2022jli after the re-brightening resembles the
evolution of SLSNe-I in the sense that these SNe show a slow evolution and bluer continuum compared
to normal luminosity SE SNe. However, SN\,2022jli has stronger spectral features typical of a SN\,Ic,
whereas SLSNe-Ic have shallower and broader spectral features \citep[see Fig. \ref{spectral_comp2_fig}
  and ][for a discussion]{cartier22}. The spectral features displayed by SN\,2022jli and by all normal
SNe\,Ic show less blending than most SLSNe-Ic, pointing to a less energetic explosion of SN\,2022jli
than SLSNe-Ic.

Presented in Fig.\ref{specseq_nir_fig} is the NIR spectral sequence of SN\,2022jli starting at $+38$\,days
relative to the first maximum and continuing for nearly $+200$\,days. As in the optical,
SN\,2022jli evolves slowly in the NIR. We identify strong lines of \ion{O}{I}, \ion{C}{I}, \ion{Mg}{I},
\ion{Mg}{II}, \ion{Na}{I}, weak lines of \ion{He}{I} $\lambda 2.058$ and Pa$\beta$ as is discussed below,
and potential lines of \ion{Si}{I}. Weak lines of [\ion{Fe}{II}] seem to appear between 100 and 200\,days,
these lines are commonly observed in most core-collapse SNe. Initially, at $+38$\,days, we identify that
the strongest spectral features in the $H$ and $K_{s}$ bands correspond to \ion{Mg}{II}. As the \ion{Mg}{II}
lines fade, the \ion{Mg}{I} spectral features become stronger.

\subsubsection{Helium}

The optical spectra of SN\,2022jli do not show unambiguous \ion{He}{I} lines, but
the potential presence of a small amounts of helium is suggested by the strong absorption
feature at about $5730$\,\AA, which is usually associated with the \ion{Na}{I}\,D line in SNe Ic,
but in this case the absorption feature seems to be strengthened by presence of the
\ion{He}{I} $\lambda 5875$ line (see Fig. \ref{spectral_comp1_fig}). In addition, a weak
spectral feature coincident with \ion{He}{I} $\lambda 7065$ from about $\sim +50$\,days
(see Fig. \ref{spectral_comp2_fig}) also suggests the presence of small amounts of helium. The existence
of helium is the SN ejecta is further supported  by a very weak broad spectral feature barely detected at
$\sim 2.058\,\mu$m in the NIR spectra at $+36$ and $+92$ days \citep[see also][]{tinyanont24}. As is
discussed in \citet{dessart20} and \citet{shahbandeh22}, the \ion{He}{I} $\lambda 2.058$ line is
probably the best suited spectral feature for an unambiguous detection of helium in SNe Ic.

\subsubsection{Hydrogen}

\begin{figure*}
\centering
\includegraphics[width=180mm]{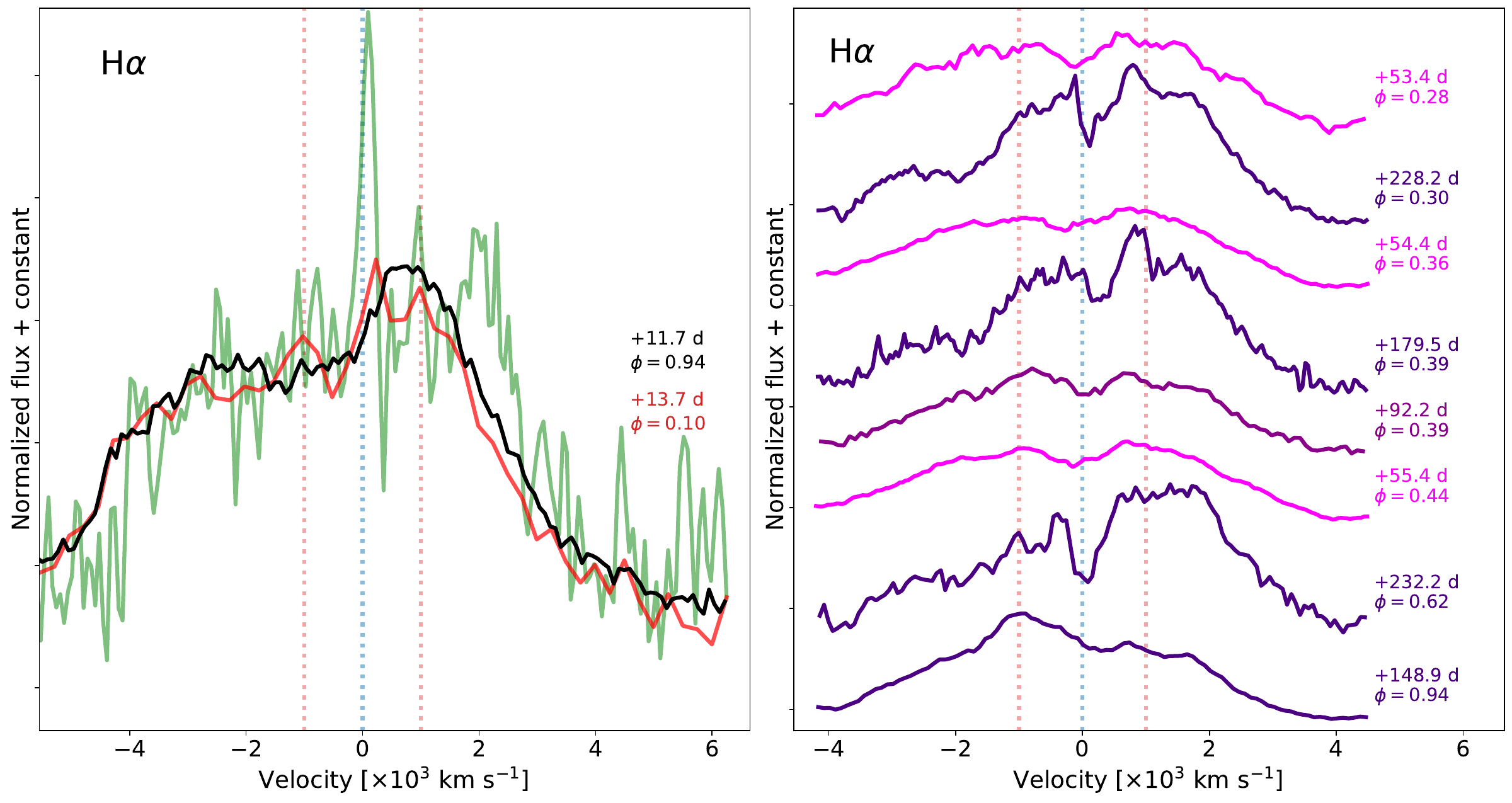} 
\caption{Velocity profile of the region around the H$\alpha$ spectral feature. A pseudo-continuum on each
side of the profile has been used to subtract the underlying continuum emission, and the result
was normalised by the median residual flux over the region. On the left panel,
the spectral region of SN\,2022jli during the decline from the first maximum is compared with SN\,2013ge
(in green) at a similar spectral phase of $\sim 12-14$\,days. On the right hand panel, the evolution of the
H$\alpha$ velocity profiles of SN\,2022jli from $+53$\,days to $+232$\,days are shown, during and after
the second maximum, ordered with respect to the phase folded period of $12.472$\,days ($\phi$).
The near-maximum spectra are shown in magenta, the intermediate-phase spectrum at $+92.1$\,days
spectrum is shown in dark magenta, and the late-phase spectra (\textgreater $+100$\,days) is shown in indigo.
The phase with respect to maximum and the periodic phase ($\phi$) are indicated on the right
of each spectrum. Vertical dotted red lines indicate a velocity of $\pm 1000$\,km\,s$^{-1}$ for comparison.
}
\label{Halpha_fig}
\end{figure*}

\begin{figure}
\centering
\includegraphics[width=90mm]{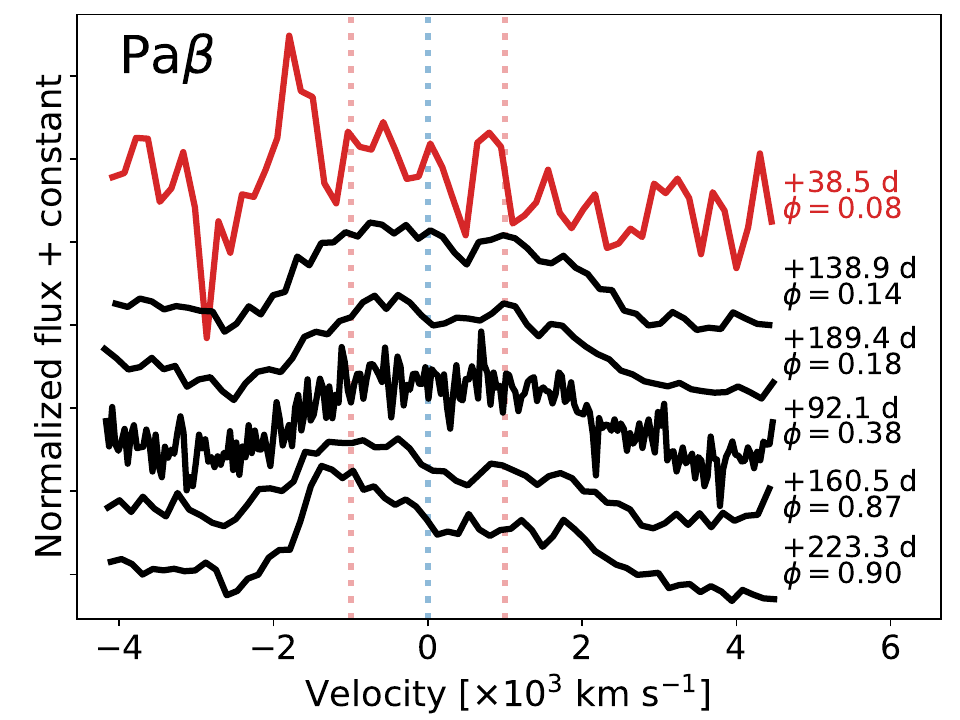} 
\caption{Velocity profile of the region around the Pa$\beta$ spectral feature in the NIR. As in the
case of H$\alpha$, a pseudo-continuum on each side of the profile has been used to subtract the
underlying continuum emission, and the result was normalised by the median residual
flux over the region. Vertical dotted red lines indicate a velocity of $\pm 1000$\,km\,s$^{-1}$
for comparison.
}
\label{PaBeta_fig}
\end{figure}

SN\,2022jli shows a broad and blended spectral feature at the location of H$\alpha$ (Fig. \ref{Halpha_fig}),
and a double-peaked profile at the location of the Pa$\beta$ (Fig. \ref{PaBeta_fig}). From about the
time of the second maximum the H$\alpha$ feature begins to show a double-peaked profile (see Fig. \ref{Halpha_fig}).
The peaks of the double horned profiles in the H$\alpha$ and Pa$\beta$ features are located at
about $\pm 1000$\,km\,s$^{-1}$ from the central rest-frame line position (see Figs. \ref{Halpha_fig}
and \ref{PaBeta_fig}).

The evolution of the H$\alpha$ profile is presented in Fig.\ref{Halpha_fig}. A pseudo-continuum on each
side of the profile has been used to subtract the underlying continuum emission, and the resulting residuals
were normalised by the median residual flux over the region. In the left panel, the spectra at $+12$ and $+14$\,days
are compared with the spectra of SN\,2013ge at a similar phase. There is a potential excess emission at
$\sim +800$\,km\,s$^{-1}$; however, the overall profile is very similar to the profile of SN\,2013ge and no
clear evolution in the shape of the spectral feature is detected between the two epochs. Therefore the presence
of H$\alpha$ emission is ambiguous during the decline from the first maximum.

In the right panel, the H$\alpha$ profiles from about the time of the second maximum until +232\,days are presented,
ordered by phase assuming a period of $12.47$\,days. The spectra obtained close to the secondary maximum
(53-55\,days) and at $+92.2$\,days ($\phi$ in the range of $0.28-0.44$) show a double-peaked profile similar to the one
first detected at +92.1\,days in the Pa$\beta$ line, but there is no apparent shift in the line profiles over
this phase. In the spectra at \textgreater $+100$\,days, the double-peaked feature becomes more conspicuous
and shows a clear shift in the peak position with phase. The latter is probably due to the larger phase
coverage in the spectra after 100\,days ($\phi$ from $0.30$ to $0.94$).

The Pa$\beta$ profile is not detected in the NIR spectrum at $+38.5$\,days, and only
becomes evident in the spectrum at $+92.2$\,days (see Fig. \ref{PaBeta_fig}), after the time of the secondary maximum.
The feature becomes more conspicuous with time, as can be observed in Fig. \ref{specseq_nir_fig}. The maxima
of the double-peaked profile seem to shift as a function of the phase of the periodic undulations and consistent with
the variation observed in the H$\alpha$ feature at \textgreater $+100$\,days (Fig. \ref{PaBeta_fig}). 
No other hydrogen lines are clearly detected in the NIR spectra of SN\,2022jli. Furthermore, we inspected
in detail the region of Br$\gamma$ and the line is not visible despite the high signal-to-noise ratio of
the NIR spectra. Some weak hydrogen lines may be blended with lines of other species. The H$\alpha$ and the Pa$\beta$
lines produce relatively weak spectral features indicative of a small amount of hydrogen, and the double-peaked
profiles could be a consequence of an accretion disc or of out-flowing material ejected in winds. At $+400$\,days
the SN spectrum is fully nebular and spectral feature associated with H$\alpha$ is no longer detected,
supporting the conclusion that the amount of hydrogen material mixed in the ejecta is very small, if any.

Narrow H$\alpha$ emission lines ($\mathrm{FWHM} = 3.5 - 6.5$\,\AA) are present in the nebular spectra obtained with the
GMOS-S and LDSS-3 spectrographs at late times, this narrow emission is not present in the nebular spectrum obtained a few days before with the Goodman
spectrograph. A careful inspection of the 2D spectra shows strong narrow H$\alpha$ and [\ion{N}{II}] emission lines from the host galaxy, where the H$\alpha$ line
is coincident with a night sky line. In addition, on the 2D spectra the H$\alpha$ emission has irregular structure in the spatial direction,
and small shifts (\textless $5$\,\AA) due to the galaxy rotation in the dispersion axis. We suspect that the narrow H$\alpha$ feature
is most likely due to an incomplete subtraction of the host galaxy H$\alpha$ background emission close to the SN. This is supported by
the small $\mathrm{FWHM}$ measured from the H$\alpha$ narrow feature in the GMOS-S spectrum, which is half the instrument resolution.
Thus, probably originates from an imperfect subtraction. However, we are cautious, and the potential appearance of late time
narrow hydrogen lines such as in SN\,2016adj \citep{stritzinger24} cannot be completely discounted. 

\subsubsection{Expansion velocity}
\label{exp_vel_sec}

As was previously discussed, SN\,2022jli shows spectral features of a SN\,Ic
after maximum light and throughout its evolution. Now we
pay attention to the ejecta expansion velocity. We estimate the
ejecta velocity from the minimum of the \ion{Fe}{II} $\lambda 4923.9$,
\ion{Fe}{II} $\lambda 5001.9$ and \ion{Fe}{II} $\lambda 5169.0$
absorption lines. In table \ref{feii_vel_tab} we summarise the
ejecta velocity estimated from each of these three lines, and provide
their weighted average. The uncertainty of the weighted average
corresponds to the maximum value between the error from the weighted
mean and the standard deviation from the three independent
measurements.

\begin{figure}
\centering
\includegraphics[width=85mm]{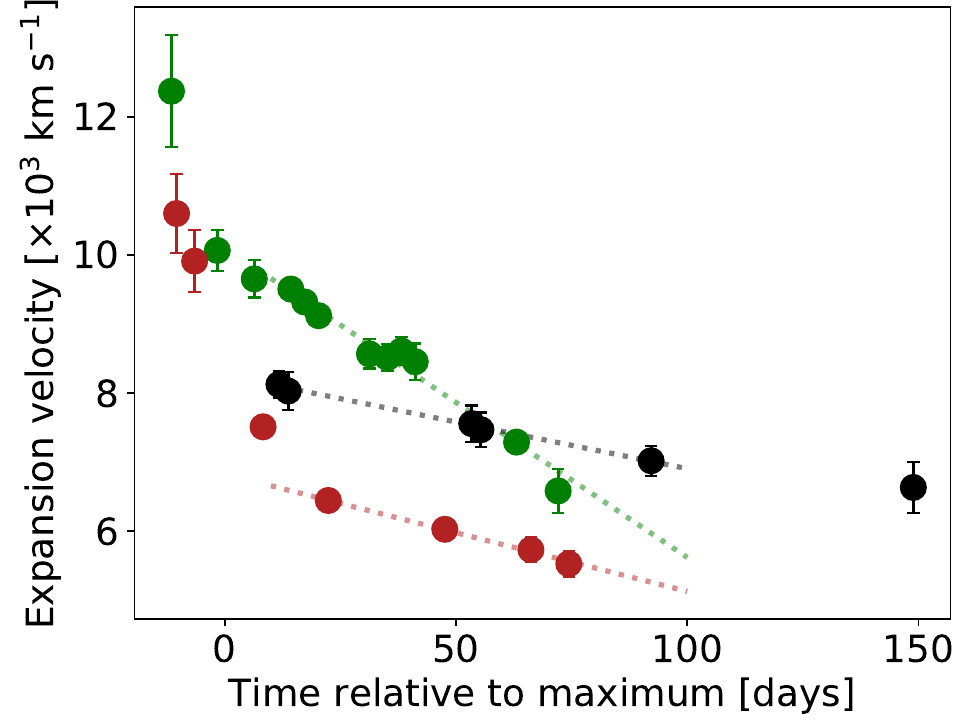} 
\caption{Expansion velocity estimated from the \ion{Fe}{II}
lines (see text) for SN\,2022jli (black), SN\,2013ge (green)
and SN\,2007gr (red). The dashed lines illustrate the
velocity decline rate, measured between +10 and +100\,d,
for the three SNe.}
\label{exp_vel_fig}
\end{figure}

Figure \ref{exp_vel_fig} compares the ejecta expansion
velocity of SN\,2022jli with the normal type Ic SN\,2007gr and
SN\,2013ge. For these three SNe, the ejecta velocity was estimated
using the \ion{Fe}{II} lines. \citet{taddia18b} shows that the
\ion{Fe}{II} velocity evolution from early times, before maximum light,
to late times can be described with a power law. Here we show that from about
$+10$\,days the ejecta velocity is well described by a linear decline, assuming
that the velocity decline rate for these three SNe was measured in
a time window between $+10$ and $+100$\,days. We found
that the velocity decline rates are  $13.5 \pm 0.8$ km\,s$^{-1}$\,day$^{-1}$ for
SN\,2022jli, $44.9 \pm 5.8$ km\,s$^{-1}$\,day$^{-1}$ for SN\,2013ge
and $17.0 \pm 0.6$\,km\,s$^{-1}$\,day$^{-1}$ for SN\,2007gr.
The expansion velocity of SN\,2022jli after maximum is comparable
to other SNe\,Ic, and extrapolating linearly its velocity to maximum light a value of
$v^{\mathrm{avg}}_{\ion{Fe}{II}} \simeq 8250$\,km\,s$^{-1}$ is obtained, the true value
must be slightly higher given the power-law behaviour of the ejecta velocity at early
times. The velocity decline rate post-maximum of SN\,2022jli
follows a linear evolution marginally slower than the comparison SNe.
The latter suggest that the physical mechanism producing the secondary maximum
and the undulations has a moderate effect on the expansion velocity
of the ejecta. Arguably the main effect is to inject energy to the ejecta
keeping a pseudo-photosphere for longer time at a higher ejecta velocity
compared with normal SNe Ic. The latter effect is evident at about 100\,d,
when the comparison SNe\,Ic are almost fully nebular, and the
non-nebular \ion{Fe}{II} lines are no longer detected.
The linear extrapolations of the ejecta velocities of the comparison
Ic's at $+100$\,days are slower ($\leq 6000$\,km\,s$^{-1}$) than the ejecta
velocity of SN\,2022jli at this phase ($\simeq 7000$\,km\,s$^{-1}$).

\begin{table*}
  \centering
  \caption{Summary of the \ion{Fe}{II} expansion velocity of SN\,2022jli.}
  \label{feii_vel_tab}
  \begin{tabular}{@{}lcllll}
    \hline
    MJD    & Phase  & $v_{\ion{Fe}{II} \lambda 4923.9}$ & $v_{\ion{Fe}{II} \lambda 5001.9}$ & $v_{\ion{Fe}{II} \lambda 5169.0}$ & $v^{\mathrm{avg}}_{\ion{Fe}{II}}$ \\
    (days) & (days) & (km\,s$^{-1}$)                    & (km\,s$^{-1}$)                    & (km\,s$^{-1}$)                    & (km\,s$^{-1}$) \\
    \hline
    $59721.4$ & $+11.7$  & $8332$($99$)  & $7888$($73$)  & $8250$($77$)  & $8121$($193$) \\
    $59723.4$ & $+13.7$  & $8296$($243$) & $7642$($371$) & $7886$($277$) & $8025$($270$) \\
    $59763.3$ & $+53.4$  & $7664$($47$)  & $7065$($98$)  & $7565$($92$)  & $7553$($262$) \\
    $59765.3$ & $+55.4$  & $7701$($22$)  & $7113$($25$)  & $7556$($52$)  & $7464$($250$) \\
    $59802.3$ & $+92.2$  & $7291$($238$) & $6766$($288$) & $6995$($101$) & $7015$($215$) \\
    $59859.3$ & $+148.9$ & $7279$($155$) & $6466$($148$) & $6516$($75$)  & $6628$($372$) \\
    \hline
  \end{tabular}
\end{table*}

\subsubsection{Nebular phase}
\label{nebular_spec_sec}

In Fig. \ref{spectral_comp1_fig} the spectrum of SN\,2022jli at $+418$\,days
is compared with the spectra of other SE SNe at nebular phase. SN\,2022jli
appears like a normal SN\,Ic similar to SN\,2013ge. The strongest spectral
features of SE SNe at this phase are the [\ion{O}{I}] $\lambda \lambda 6300,6364$
and [\ion{Ca}{II}] $\lambda \lambda 7291,7323$ lines.

The [\ion{O}{I}] and [\ion{Ca}{II}] emission lines are studied using Gaussian
profiles to fit the lines. The continuum emission regions were defined selecting the minimum
on each side of the spectral features, and a linear polynomial was fitted between these regions to
subtract the continuum emission. In SN\,2022jli, the [\ion{O}{I}] line does not show any contamination
from other broad features such as H$\alpha$ or [\ion{N}{II}], as is frequent in Type IIb and Ib SNe
  (see e.g., SNe 2011dh and 2007Y in Fig. \ref{spectral_comp1_fig}). The [\ion{O}{I}]
doublet was modelled using two Gaussian profiles fixing the ratio between the to components to 3:1, assuming
an optically thin nebular emission as in \citet{taubenberger09}.
The [\ion{Ca}{II}] lines show a weak but clear contamination from the [\ion{Fe}{II}] $\lambda 7155$
line on the blue part of the profile, and can also be contaminated by other relatively weak
[\ion{Fe}{II}] and [\ion{Ni}{II}] lines in this region. We assumed that the [\ion{Fe}{II}] $\lambda 7155$
emission can be well described by a Gaussian profile, the profile was fitted using blue
part of the [\ion{Fe}{II}] $\lambda 7155$ line, which does not seem to be affected by the [\ion{Ca}{II}]
emission. Making use of the Gaussian fit, the flux contribution from the [\ion{Fe}{II}] $\lambda 7155$ line
was subtracted from the [\ion{Ca}{II}] spectral feature.

The [\ion{O}{I}] emission line of SN\,2022jli does not require an additional component,
and it does not show a double-peaked profile found in some SE SNe \citep[see e.g.,][]{maeda08, taubenberger09, fang22, prentice22}.
In table \ref{neb_vel_tab} we summarise the central velocity of the [\ion{O}{I}] ($v_{\mathrm{[\ion{O}{I}}]}$)
and the [\ion{Ca}{II}] ($v_{\mathrm{[\ion{Ca}{II}}]}$ ) doublets, their $\mathrm{FWHM}$, and the [\ion{O}{I}]/[\ion{Ca}{II}]
flux ratio. The [\ion{Ca}{II}] line in the $+412$ days GMOS spectrum is affected by gap of $\sim 100$\,\AA\
on the red tail of the spectral feature (see Section \ref{spectra_sec}). We present the measurements of the
SN\,2013ge nebular spectrum for comparison.

\begin{table*}
  \centering
  \caption{Summary of nebular phase spectroscopic measurements.}
  \label{neb_vel_tab}
  \begin{tabular}{@{}llccccc}
    \hline
    SN    & Phase  & $v_{\mathrm{[\ion{O}{I}}]}$ & $\mathrm{FHWM}_{\mathrm{[\ion{O}{I}}]}$ & $v_{\mathrm{[\ion{Ca}{II}}]}$ & $\mathrm{FHWM}_{\mathrm{[\ion{Ca}{II}}]}$ & [\ion{O}{I}]/[\ion{Ca}{II}] \\
          & (days) & (km\,s$^{-1}$)              & (km\,s$^{-1}$)                          & (km\,s$^{-1}$)                & (km\,s$^{-1}$)                            &                              \\
    \hline
    SN\,2022jli & $+401$  & $-274$               & $3610$                                  & $\cdots$                      & $\cdots$                                  & $\cdots$ \\                     
    SN\,2022jli & $+412$  & $-294$               & $3700$                                  & $+250$                        & $4800$                                    & $2.1$ \\
    SN\,2022jli & $+418$  & $-336$               & $3776$                                  & $+430$                        & $6100$                                    & $2.2$ \\
    \hline
    SN\,2013ge  & $+447$  & $-495$               & $4400$                                  & $-979$                        & $4780$                                    & $2.7$ \\
    \hline
  \end{tabular}
\end{table*}

\citet{prentice22} studied the [\ion{O}{I}]/[\ion{Ca}{II}] ratio for a sample 86 SNe from the
literature, finding that this ratio does not evolve significantly between about $+200$\,days and $+400$\,days
in normal luminosity CC SNe. The different sub-types are found in a well-localised range
of values. For example, SNe\,II have [\ion{O}{I}]/[\ion{Ca}{II}] ratios in the range $0.3-1.4$ with
a median $0.5$, SNe IIb are found in the range of $0.5-3.3$, SNe Ibc in the range of $1.0-5.0$,
while SLSNe-I are found in a wide range of values overlapping with II and IIb SNe. The
[\ion{O}{I}]/[\ion{Ca}{II}]$= 2.2 \pm 0.1$ for SN\,2022jli confirms its SE SN nature, and places this
object within the Ibc range.

\subsection{Carbon monoxide emission and late-time near-infrared excess}
\label{carbon_and_dust_sec}

The formation of carbon monoxide (CO) is key in the process of molecule and dust formation in the
SN ejecta, playing an important role in cooling the SN ejecta \citep{liljegren20} to temperatures at
which dust can form, and key in the chemical reactions leading to complex molecule formation.
The CO molecule is readily detectable in late time NIR spectra of SNe from its first CO overtone emission in the spectral
region between $2.3$ and $2.5$\,$\mu$m
\citep[e.g.,][]{spyromilio88,elias88,gerardy02,hunter09,drout16,banerjee18,rho18,rho21,tinyanont19,stritzinger24, park25, hueichapan25b}.
The fundamental CO mode at about $4.6$\,$\mu$m is frequently detected in the small number of CC SNe
observed at late-times in the mid-IR
\citep[e.g., SN\,1987A, SN\,2004dj, SN\,2004et, SN\,2005af and SN\,2017eaw; see][]{catchpole88, meikle89, kotak05, kotak06, kotak09, szalai11, tinyanont19, dessart25, medler25, jacobsongalan25}.

\begin{figure}
\centering
\includegraphics[width=85mm]{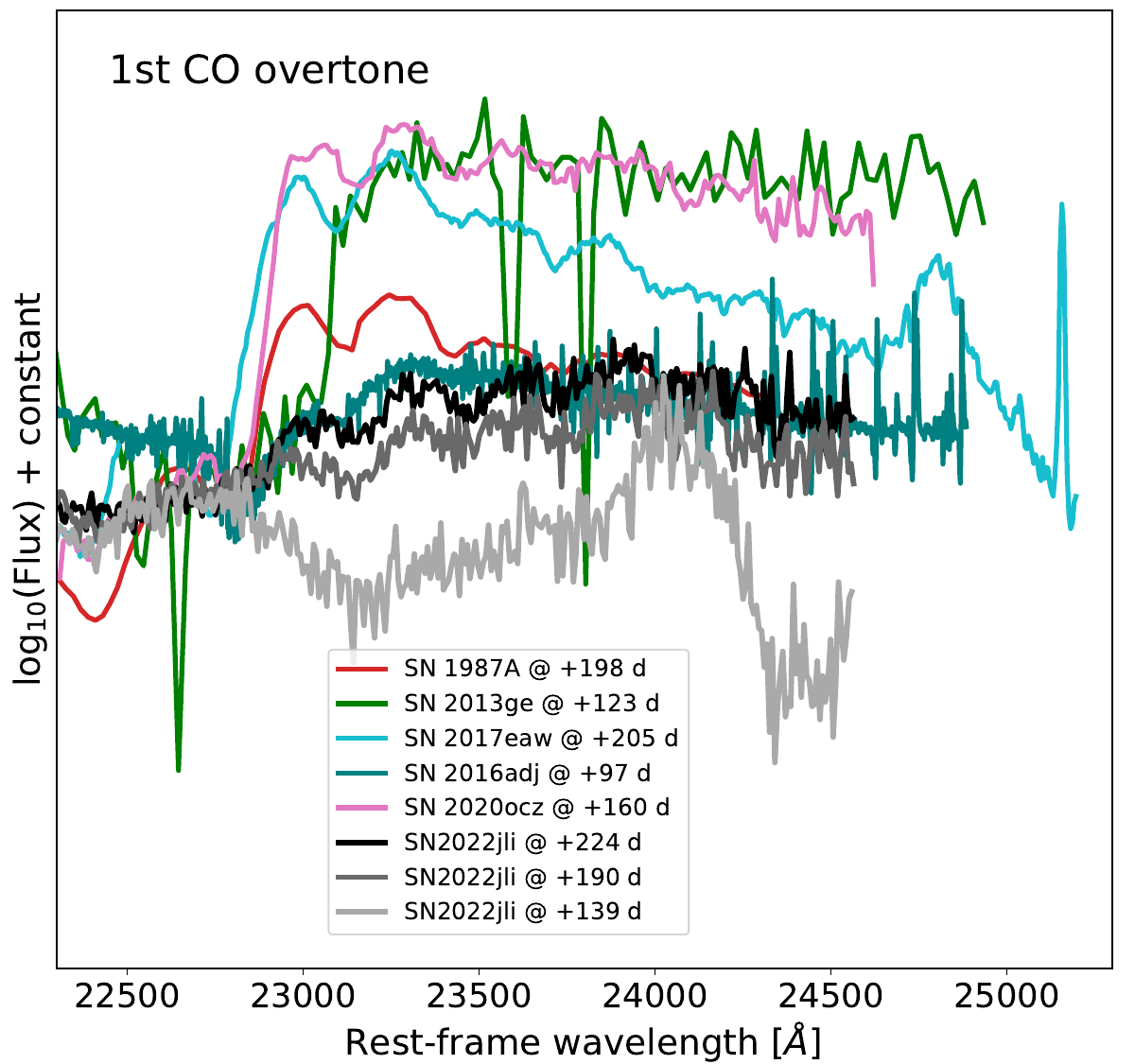} 
\caption{Comparison of the first CO overtone region of SN\,2022jli at $+139$, $+190$
  and $+224$\,days with the with nebular phase CO emission from the type Ic SNe\,2013ge \citep{drout16}
  and SN\,2016adj \citep{stritzinger24}, and the type II SNe 1987A \citep{bouchet91}, 2017eaw \citep{rho18}
  and 2020ocz \citep{cartier21}. The phase and the SN colour are indicated in the inset.}
\label{CO_fig}
\end{figure}

\begin{figure*}
\centering
\includegraphics[width=180mm]{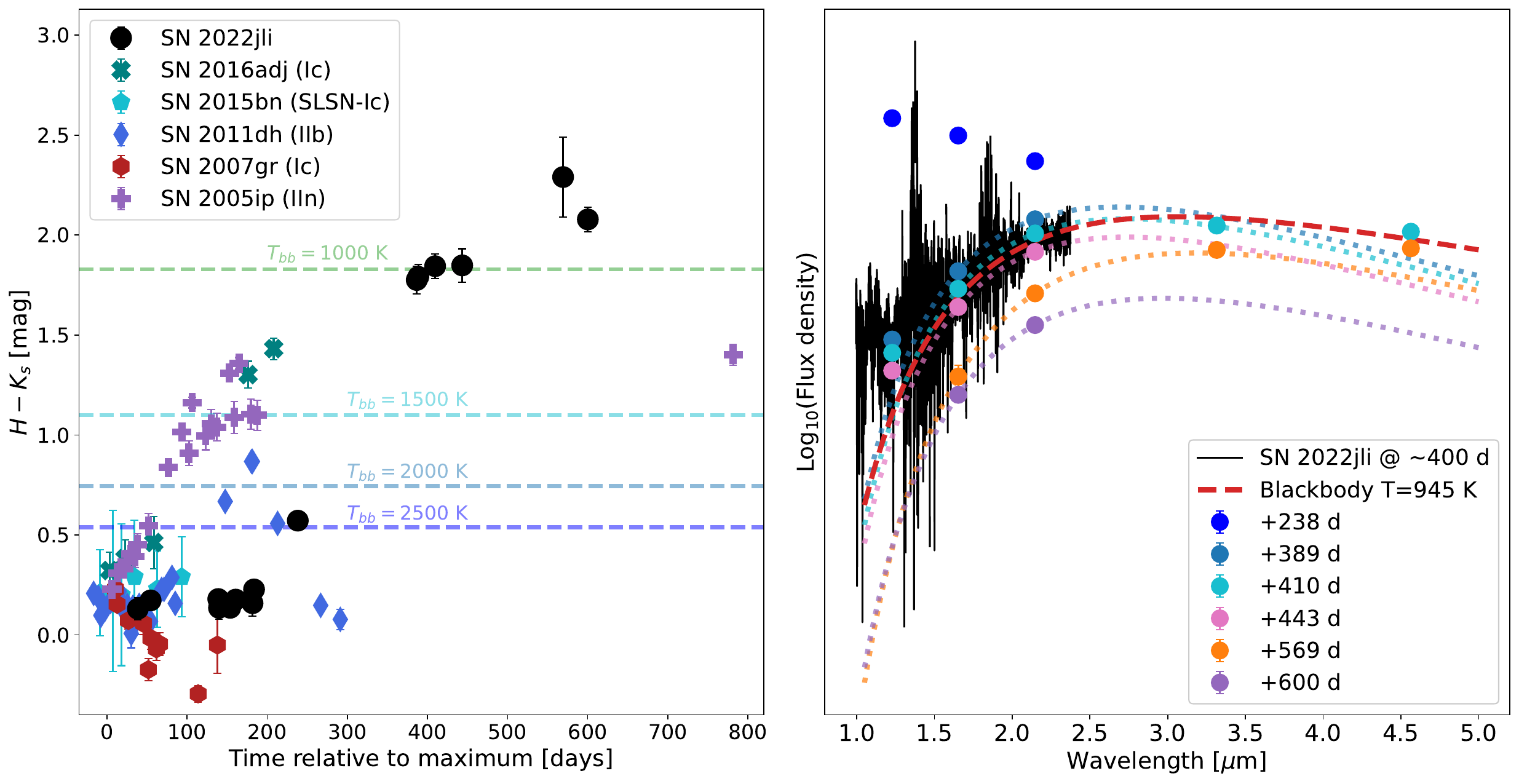} 
\caption{Left panel: $H-K_{s}$ colour evolution of SN\,2022jli (black circles) compared with the colour
  evolution of the type Ic SN\,2016adj \citep[teal X symbols;][]{stritzinger24},
  SLSN 2015bn \citep[cyan pentagons;][]{nicholl16}, type IIb SN\,2011dh \citep[blue diamonds;][]{ergon14, ergon15},
  the type Ic SN\,2007gr \citep[red hexagons;][]{hunter09} and the type IIn SN\,2005ip \citep[purple plus sign;][]{stritzinger12}.
  The horizontal dashed lines correspond to the $H-K_{s}$ colour of a given black-body temperature as
  indicated in the figure for comparison. Right panel: Evolution of the NIR SED
  of SN\,2022jli from $+238$ to $+600$\,days. A full NIR spectrum of SN\,2022jli at
  around $+422$\,days is presented, and for comparison a black-body SED having $T=945$\,K is presented
  for comparison (dashed red line). The dotted lines correspond to dust SED fitting results to the $H$,
  $K_{s}$ and $W1$ photometry of graphite dust with a radius of $0.1$\AA, using the $Q_{\mathrm{abs}}$
  from \citet{draine16}.
}
\label{latetime_ir_fig}
\end{figure*}

In Fig.\,\ref{CO_fig} the first CO overtone spectral feature of SN\,2022jli
is compared with other CC SNe with clear first CO overtone emission.
The first CO overtone feature is detected for first time in SN\,2022jli at $+190$\,days,
where the CO overtone band-heads can be clearly distinguished (see Fig. \ref{CO_fig}). Compared with other CC SNe,
the first CO overtone emission in SN\,2022jli is relatively weak, and it is significantly weaker than in the type Ic
SN\,2013ge, and similar to the first CO emission of SN\,2016adj. The potential detection of CO emission at earlier times
is ambiguous. For example, at $+139$\,days we can
distinguish a broad spectral feature at the expected location, but it seem to be dominated by the emission from the \ion{Mg}{II}
$\lambda 24048$ line and the characteristic CO band-heads are not distinguishable. At around $400$\,days the
first CO emission is no longer detected in SN\,2022jli, at this phase the $K_{s}$ band is dominated by dust thermal
emission (see Fig. \ref{latetime_ir_fig}).

The IR photometry of SN\,2022jli reveals a flattening in the $JHK_{s}W1W2$ emission at around 200\,days
(see Fig. \ref{lightcurves_fig}), which is more evident in the redder IR bands. This IR-excess in SN\,2022jli is
illustrated in the $H-K_{s}$ colour diagram of Fig. \ref{latetime_ir_fig},
where a jump in the $H-K_{s}$ colour of $\sim 0.35$ mag between $+180$ and $+238$ days marks the onset of strong
dust emission. The SN continues evolving to extremely red $H-K_{s}$ colours over the course of the following year,
which is a consequence of the dust cooling and of the predominance of the dust thermal emission over the
SN ejecta emission.

As is shown in the right panel of Fig. \ref{latetime_ir_fig}, despite the red SN colour, the SN
emission in the $J$ and $H$ bands at $+238$\,days is still dominated by the SN ejecta emission. Five
months later, at a phase of about $+380$\,days,
the ejecta emission has significantly faded and the NIR emission is now dominated by thermal dust emission
(see Fig. \ref{latetime_ir_fig}). A NIR spectrum at around $+420$\,days, reveals that the $J$ band still displays
broad nebular spectral features from the SN ejecta; however, the emission in the $H$ and $K_{s}$ bands is
dominated by dust thermal emission.

Assuming optically thin dust emission, we fitted the $H$, $K_{s}$, and $W1$ (when available) photometry
using the expression

\begin{equation}
 F_{\nu} =  \frac{M_{d} B_{\nu}(T_{d}) \kappa_{\nu}(a) }{d^{2}},
\end{equation}

\noindent where $\kappa_{\nu}(a)$, the dust mass absorption coefficient, is

\begin{equation}
 \kappa_{\nu}(a) = \left(\frac{3}{4 \pi \rho a^{3}} \right) (\pi a^2 Q_{\mathrm{a}}(a)).
\end{equation}

\begin{figure*}
\centering
\includegraphics[width=180mm]{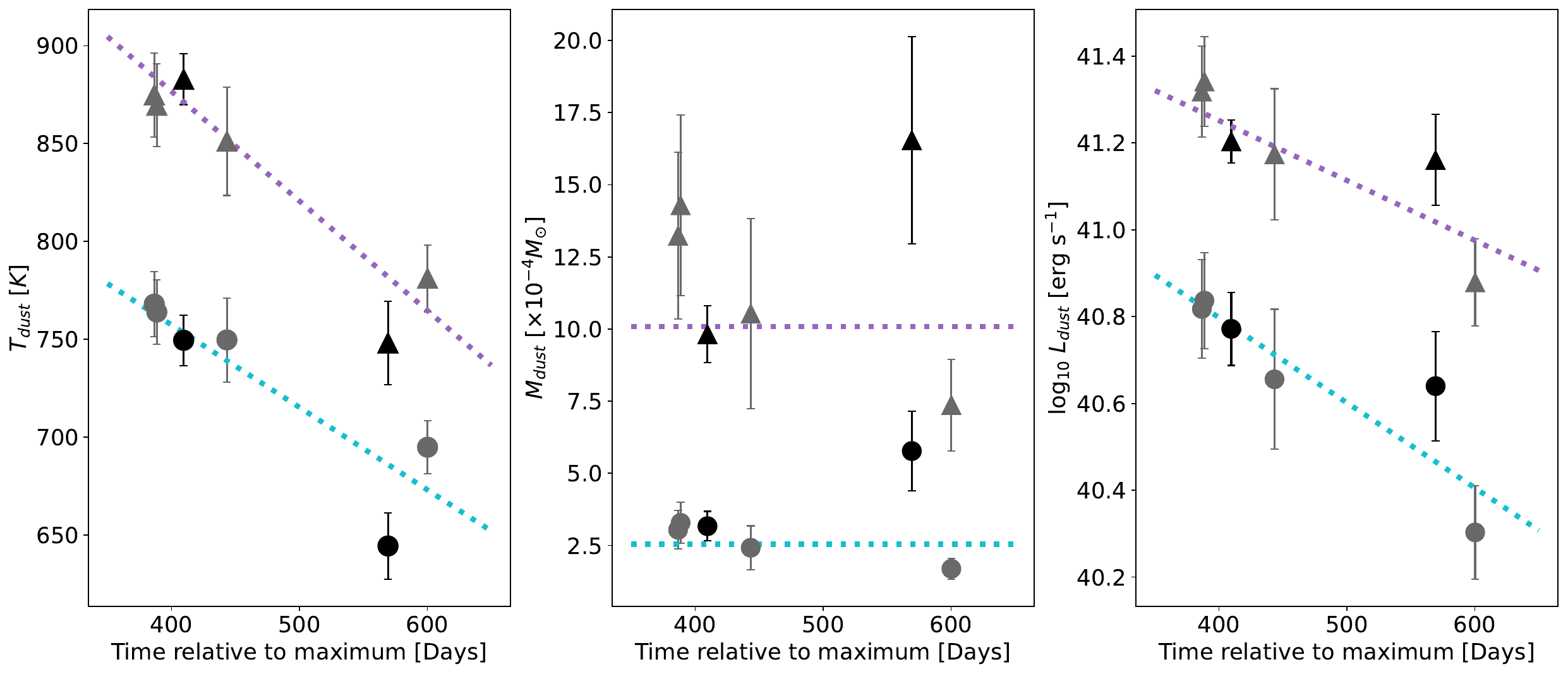} 
\caption{Evolution of the dust temperature, mass, and luminosity as a function of time relative to maximum
  light for silicate (triangles) and graphite (circles) dust, assuming a dust radius of $0.1$\,$\mu$m.
  Black symbols correspond to fits performed to $H$, $K_{s}$, and $W1$ photometry and grey
  symbols to fits performed using only $H$ and $K_{s}$ photometry. Dotted magenta and cyan lines
  correspond to linear fits to $T_{d}$ and $log_{10}(L_{d})$ for silicate and graphite dust, respectively.
  The horizontal dotted cyan and magenta lines indicate the weighted means for $M_{d}$ of $2.5 \pm 1.3 \times 10^{-4}$\,$M_{\odot}$
  and $10.1 \pm 3.0 \times 10^{-4}$\,$M_{\odot}$ for graphite and silicate dust, respectively.
}
\label{dust_evol_fig}
\end{figure*}

\noindent Here $a$ is the dust radius, $\rho$ is the dust density, and $Q_{\mathrm{a}}$ is the
dust absorption efficiency.
Here we study two cases: a single-crystal size graphite dust composition
for which the $Q_{\mathrm{a}}$ was computed by \citet{draine16} using the discrete dipole approximation,
and a single-size silicate dust composition (forsterite) with the $Q_{\mathrm{a}}$ computed
using the 1/3-2/3 approximation \citep{draine84,laor93}. The $W2$ band shows an excess of emission
compared with the dust models for both compositions, this band is probably affected by the emission
from the fundamental CO band which is coincident with the effective wavelength of the $W2$ band ($4.6$\,$\mu$m)
and is not included in the fits.\footnote{This IR excess at about $4.6$\,$\mu$m due to emission
from the fundamental CO band has been previously observed in several SNe
\citep[e.g., SN\,1987A, SN\,2004dj, SN\,2004et and SN\,2017eaw; see][]{catchpole88, kotak05, kotak09, szalai11, tinyanont19}.}
The evolution of the graphite and silicate dust temperature  ($T_{d}$),
mass ($M_{d}$) and luminosity ($L_{d}$) for SN\,2022jli assuming a dust radius of $a = 0.1$\,\AA\ are presented
in Fig. \ref{dust_evol_fig}. In this figure, a decrease in $T_{d}$ and $L_{d}$ parameters can be observed with time.
This trend in the evolution is independent of the dust size and composition assumed.
The weighted mean for $M_{d}$ is $2.5 \pm 1.3 \times 10^{-4}$\,$M_{\odot}$ for graphite dust
and $10.1 \pm 3.0 \times 10^{-4}$\,$M_{\odot}$ for forsterite silicate dust, both for single-size crystals
of a radius of $a=0.1$\,$\mu$m. In the fitting of the photometry we consider a significant range in the dust radius,
finding that for dust particles with a radius $a \leq 0.2$\,$\mu$m variations of \textless $20$\% in $M_{d}$ are
observed. When larger dust radii are considered $a \geq 0.3$\,$\mu$m the $M_{d}$ is scaled down by a factor of
$\sim 0.5$ compared with the reference radius of $a=0.1$\,$\mu$m for both dust compositions,
  graphite and silicate.

We considered the $M_{d}$ evolution in more detail. In the case of dust formation in the SN ejecta $M_{d}$
usually increases with time. An increase in $M_{d}$ with time is observed in the fits that include $W1$-band
photometry (black symbols in Fig. \ref{dust_evol_fig}), while a decreasing trend in $M_{d}$ is observed in the fits performed
using only $H$ and $K_{s}$ photometry. The inclusion of $W1$ photometry practically does not alter our estimates,
but reduce significantly the uncertainty, in particular in the fits at $+570$\,days. The observed decrease in $M_{d}$ in
the fits performed to only the $H$ and $K_{s}$ bands is largely due to the last measurement, with a difference between
the first and the last (minimum) value of $1.8 \sigma$, independent of the dust radius and composition
assumed. Future IR observations of SN\,2022jli with the {\it James Webb Space Telescope} ({\it JWST}) are key
to constrain the dust composition of the emitting dust, to reveal the presence of more than one dust component,
for example dust emission at a cooler temperature \citep[e.g.,][]{kotak09,szalai11,shahbandeh23} and to provide a
precise evolution of the total dust mass emitting in the IR. {\it JWST} observations are fundamental to fully
understand the nature and the composition of the dust emission in SN\,2022jli revealed by the IR observations presented here.

Fitting a linear relation to the $log_{10}(L_{d})$ vs. time shown in Fig.\,\ref{dust_evol_fig}, and
integrating the luminosity evolution of dust over the time range between 387 and 600\,days, we place a lower
limit of $8.0 \times 10^{47}$\,erg to the energy radiated by graphite dust and of $2.5 \times 10^{48}$\,erg
for the energy radiated by silicate dust. The decline rate luminosity for graphite and silicate dust are
$0.5$\,mag/100\,days and $0.3$\,mag/100\,days, respectively.

\begin{figure}
\centering
\includegraphics[width=90mm]{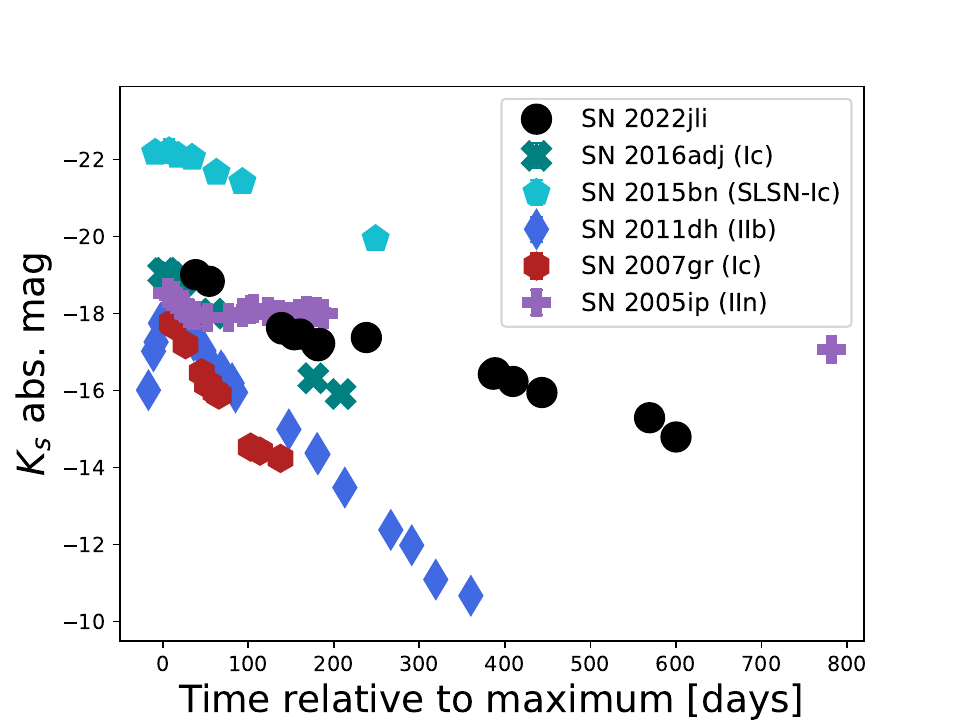} 
\caption{Comparison of the $M_{K}$ between SN\,2022jli and well observed core-collapse
  SNe in the NIR bands.
}
\label{Ks_abs_mag_fig}
\end{figure}

In Fig. \ref{Ks_abs_mag_fig}, we compare the $K_{s}$-band absolute magnitude ($M_{K}$) evolution of different local SNe with
well observed photometry in the NIR bands. With the exception of SLSNe which are intrinsically brighter and
have broad light curves in all bands, bright emission in the $K_{s}$ band ($M_{K} \textless -16$\,mag) at
late times ($\textgreater +200$\,days) can be considered a signature of strong dust emission. Normal
SE SNe decline faster, while SN\,2022jli has a broad $M_{K}$ light curve with the second maximum
powered by an extra energy source. SN\,2016adj is a peculiar and bright SE SN showing relatively broad
NIR light curve ($M_{K} \geq -16$\,mag at $+208$\,days) and narrow hydrogen emission lines at late times,
the latter is a potential signature of ejecta-CSM interaction. SN\,2022jli displays a broad light curve somewhat
similar to SLSNe-Ic, and shows a decline steeper than the strongly interacting SN\,2005ip, which remains bright
at an almost constant $M_{K}$ for several hundred days.

The process of dust formation and the heating mechanism of the dust detected in SN\,2022jli is not
obvious, the dust may be newly formed in the ejecta or pre-existing dust in the CSM. The first scenario
is emission from hot newly condensed dust in the SN ejecta or in a cool dense shell (CDS) produced
by the interaction of the ejecta with a dense CSM shell. Alternatively pre-existing CSM
dust may be radiatively heated by the SN luminosity, by the emission from the first and/or second
maxima, producing dust thermal emission, denominated IR echo.

\section{Discussion}
\label{sec:discussion}

A diverse set of mechanisms have been proposed to explain bumps and double-peaked light
curves in SNe. Until recently, two main scenarios were usually invoked to explain double-peaked
light curves. One is the power injection from a compact object such as a fast-spinning neutron star
with a strong magnetic field \citep[see e.g.,][]{maeda07} or fall-back accretion onto a compact object of stellar mass,
such as a black hole or a neutron star. Alternatively, double-peaked light curves and bumps have been frequently considered smoking gun signatures
of interaction with a dense CSM. In the following we analyse the potential scenarios
to explain the strong bump displayed by SN\,2022jli and its subsequent undulations.

\subsection{Super-Eddington accretion}

Recently, \citet{chen24} proposed that the second maximum and the subsequent periodic undulations
observed in SN\,2022jli are the result of the mass accretion from a companion star onto a compact
object in a binary system at their periastron encounters. In this paper, the
estimated (pseudo) bolometric luminosity of the second maximum is $\sim 1.7 \times 10^{42}$\,erg\,s$^{-1}$,\footnote{Discounting
the energy from the radioactive decay of $0.12$\,$M_{\odot}$ of $^{56}$Ni and assuming full trapping
of positrons and $\gamma$-rays. The full trapping assumption most likely overestimates the power input
to the light curve from the $^{56}$Ni $\rightarrow$ $^{56}$Co $\rightarrow$ $^{56}$Fe decay chain after
maximum light.} this corresponds to $\sim 10^{3}$ and $\sim 10^{4}$ times the Eddington ratio ($R_{\mathrm{Edd}}$)
for a black-hole mass ($M_{\mathrm{bh}}$) of $10$\,$M_{\odot}$ and a neutron star mass ($M_{\mathrm{ns}}$) of
$1.4$\,$M_{\odot}$, respectively.

Super-Eddington accretion into a compact object has been considered as the most likely possibility to explain
ultra-luminous X-rays sources \citep[ULXs; see][for a review]{kaaret17},
such as NGC\,5907 ULX \citep{israel17}. These sources can reach X-ray luminosities in excess of a few
$10^{39}$\,erg\,s$^{-1}$, which is the Eddington luminosity ($L_{\mathrm{Edd}}$)
for a $10$\,$M_{\odot}$ black hole. NGC\,5907 ULX is one of the brightest examples of this class, reaching a
luminosity of $\sim 2 \times 10^{41}$\,erg\,s$^{-1}$, and it might harbour a binary system composed of a
neutron star accreting at $\sim 10^{3}$ times the $L_{\mathrm{Edd}}$ from the companion star, assuming isotropic
luminosity \citep{israel17}. However, significant collimation of the disc radiation and strong multi-polar magnetic
fields are required to explain the extreme luminosity of NGC\,5907 ULX \citep{israel17}.
In the presence of a strong magnetic field ($\textgreater 10^{13}$\,G) the electron scattering cross sections in
the extraordinary mode are reduced for photon energies below the cyclotron energy
\citep[$E_{c} \sim 12 \frac{B}{10^{12}\,\mathrm{G}}$\,keV;][]{herold79}; thus, radiation can escape without halting the accretion flow \citep{israel17}. The assumption of isotropic emission in
ULXs strongly overestimates the total radiation output. For example
Cygnus X-3 in our galaxy, seems to be an ULX viewed side-on suffering significant geometrical beaming \citep{veledina24}.

\citet{chen24} reported the detection of an H$\alpha$ feature showing a shift in the peak emission and where its luminosity
follows the periodic undulations of the light curve of SN\,2022jli. This variability in the  H$\alpha$ feature
would be the result of the accretion of hydrogen-rich material from a companion star at periastron encounters
as was discussed previously. \citet{chen24}
also report a non-detection of the SN in X-rays placing an upper limit of $2.5 \times 10^{40}$\,erg\,s$^{-1}$ in the $30-60$
keV energy range at about $+230$\,days from NuSTAR observations, and of $1.32 \times 10^{38}$\,erg\,s$^{-1}$ in the
$0.5-7$ keV energy range at about $+260$\,days from Chandra observations.

The soft X-ray ($0.5-2$\,keV) opacity of CC SNe can remain very large for decades, even in SE SNe that are expected to have
a factor of 10 lower opacities compared to Type II SNe due to their faster ejecta \citep{alp18}. Assuming
that the opacity of SN\,2022jli is similar to the SN\,IIb models of \citet{alp18}, we estimate that
at the time of the X-ray observations the opacity at $2$\,keV, $10$\,keV, and $50$\,keV are
$\sim 30^{4}$, $10^{3}$, and $40$, respectively. Thus all the potential X-ray emission from a central source
is absorbed and reprocessed by the SN ejecta, note that the ejecta opacity is larger for the soft X-rays
than for more energetic emission. The reprocessing scenario is consistent with the blue optical SN colours,
the increase in the ejecta temperature during the re-brightening,\footnote{$R_{\mathrm{bb}}$ moves inwards during the
SN re-brightening (see Fig. \ref{bol_bb_params_fig}), a hint of a central heating source.} and the unusual detection of
\ion{Mg}{II} lines in the NIR at more than 150\,days, implying that the SN emission in the optical and in the NIR bands
is from a heated SN ejecta and not direct emission
from a central source. At later times (\textgreater 350\,days), the ejecta becomes fully nebular showing strong [\ion{O}{I}]
lines (Fig. \ref{spectral_comp1_fig}).

Following the \citet{chen24} hypothesis, \citet{king24} proposed an scenario in which the second maximum of SN\,2022jli
is produced by a highly beamed ULX, with a beaming factor  of $b \sim 10^{-3}$ or $b \sim 10^{-4}$ for a 10$M_\odot$ black hole or a neutron
star, respectively. Additionally, \citet{king24} propose that the second maximum corresponds to an ultra-soft ULX
emitting mostly in the UVOIR region of the electromagnetic spectrum. They propose that the orbital plane must be misaligned
with respect to the axis of the beamed emission to detect the highly beamed ULX emission and simultaneously observe the shift
in the H$\alpha$ emission produced by the orbital motion and accretion from the (low-mass) companion star.
The fact that the emission during the second maximum seems
to correspond to emission from heated ejecta and not to direct UVOIR emission from an ultra-soft ULX, in addition to
the extreme beaming required to explain the high luminosity of SN\,2022jli, and the fine tuned geometry to detect both
the highly beamed emission from the ULX and the H$\alpha$ periodic variability cast some doubts on the feasibility
of the scenario proposed by \citet{king24}.

Recent hydrodynamical simulations of \citep{hirai25} show that the luminosity
and the undulations of SN\,2022jli can be reproduced by the accretion from a shock-inflated
companion star into a neutron star. The model of \citep{hirai25} also predicts a temperature
increase (blue colours) at periastron encounters (higher luminosity), as is discussed in Section \ref{colour_sec},
but not the increase in the ejecta temperature observed in Fig. \ref{Ni_bol_lc_fig}. Similarly,
\citep{lu25} presented an analytical model for the super-Eddington accretion from a shock-inflated
companion star that reproduces the main characteristics of the bolometric light curve. This model
predicts that, in addition to the hydrogen-rich material accreted into the neutron star, a significant
amount of material from the companion star is ejected \citep[$\sim 10^{-3} - 10^{-2}$\,$M_\odot$;][]{hirai25, lu25}
at a velocity of a few $\times 10^{3}$\,km\,s$^{-1}$ \citep{lu25}. In principle, the detection of emission
lines from the ejected material at the nebular phase could be a smoking gun signature of this model.
Such hydrogen-rich material is not detected in our nebular phase spectra, which requires further
modelling and investigation.

\subsection{Power injection from a magnetar}

Neutron stars are one possible outcome of core-collapse SNe, and usually possess strong magnetic fields.
Magnetars are a type of neutron star and possess strong magnetic fields in the range of $10^{13} - 10^{15}$\,G.
The physics associated with the formation of magnetars is poorly understood.
The magnetars in our galaxy exhibit the most variable behaviour among neutron stars, emitting high
energy electromagnetic radiation mostly in the hard X-rays and soft $\gamma-$rays
\citep[$L_{X} \sim 5 \times 10^{30} - 10^{36}$\,erg\,s$^{-1}$;][]{kaspi17},
and are faint at optical wavelengths. If a fast-spinning neutron star with
a strong magnetic field is formed, the rotational spin-down energy of the magnetar can contribute to power the SN light curve.
However, the energy injection and the ejecta thermalisation process of the
magnetar wind are a complex phenomena \citep[see e.g.,][]{kotera13, vurm21} that can vary with time, producing humps,
bumps, or even periodic undulations in the SN light curves \citep{vurm21, moriya22}. 

Since the discovery of the double-peaked SN\,2005bf, the power injection from a magnetar has been
recognised as a viable mechanism to produce double-peaked light curves \citep{maeda07}. As is discussed
in \citet{israel17}, the magnetic field configuration required for the neutron star in NGC\,5907 ULX
to accrete at a super-Eddington rate is similar to the one envisaged for the so-called magnetars, and
as was discussed in \citet{medvedev13}, young pulsars or magnetars can be potential ULXs.
Therefore, a natural possibility is that the compact object in SN\,2022jli is a magnetar
accreting material from a companion star. This is similar to the scenario proposed by \citet{chen24},
but where a significant fraction of the power supply could
come from the magnetar, avoiding the need for super-Eddington accretion to power the
second maximum or a privileged viewing angle to an extremely beamed ULX. In this scenario,
the accretion at periastron encounters could explain the undulations
and the H$\alpha$ periodic variability.

\citet{moriya22} studied the effect of variable thermal energy injection in the SN ejecta
from a magnetar as the potential cause of bumps in SN light curves. They found that changes in the
thermal power injection produced in the SN ejecta from a magnetar can reproduce the light curve bumps
such as the ones observed in some SLSNe (e.g., SN\,2015bn). One of \citet{moriya22} predictions is that an
observable consequence of the increase in the power input from a magnetar is an increase in the SN ejecta
temperature, in addition to the luminosity increase. The temperature increase should translate in
bluer SN colours during the bump, such as the bluer $g-r$ colours observed in SN\,2022jli during
the maxima of the second bump and the undulations (see Figs. \ref{g_r_colour_fig} and \ref{flux_color_var_fig}).
Their model also predicts a non-noticeable change in the ejecta radius or velocity during the
luminosity increase. The latter effect seems to discount a magnetar power supply to explain, for example,
the second maximum in the case of SN\,2019stc \citep{moriya22}.

Potential causes of the change in the thermal energy injection are: i) conversion efficiency,
inefficient thermalisation at early times \citep[e.g.,][]{kasen16}, ii) changes in the fraction
of high-energy photons in the pulsar wind nebulae can lead to a change
in the thermalisation of the energy injected, since the opacity depends on the energy of the
photons injected, iii) a change in the angle between the dipole magnetic
field and the rotation axis of the neutron star changes the temporal energy injection, and iv)
fallback accretion on the magnetar can increase the mass and the angular momentum of the
neutron star, decreasing the magnetar spin period. A periodic variation in the angle between the magnetic
field axis and the rotation axis of the magnetar should yield periodic variability in the
light curves. However, this cannot explain the periodic variation observed in the \ion{H}{I} lines.

To explore the magnetar scenario, it is assumed that the second maximum is powered by a magnetar.
As is discussed in Section \ref{nickel_sec}, the first maximum of SN\,2022jli is consistent with a standard
SE SN event with an ejecta mass of about 1.5\,$M_{\odot}$. Following a methodology similar to the one described in
\citet{cartier22}, and assuming the same ejecta parameters used to fit the first maximum (Section \ref{nickel_sec}),
the second maximum of SN\,2022jli is reproduced by the combination of the radioactive decay of $0.12$ $M_{\odot}$
of $^{56}$Ni and the power input from a magnetar (see Fig. \ref{Ni_bol_lc_fig}).
We assume full trapping over the period analysed (no power leakage), with $\kappa = 0.07$\,cm$^{2}$\,gr$^{-1}$
and $v_{\mathrm{ej}}  = 8500$\,km\,s$^{-1}$, obtaining an initial spin period of $P \sim 48.5$\,ms
and a magnetic field of $B \sim 8.5 \times 10^{14}$\,G. We had to arbitrarily delay the {\it ``birth''}
of the magnetar by 37\,days relative to the estimated time of the explosion to reproduce the (pseudo)
bolometric light curve of SN\,2022jli. In this scenario the late time $\gamma$-ray emission reported
by \citet{chen24} could be interpreted as the emergence of a magnetar wind nebulae \citep[see e.g.,][]{kotera13},
bringing as a result a drop in the light curve due to leakage of the magnetar power supply. The drop
below the luminosity of the radioactive decay chain of $0.12\,M_{\odot}$ of $^{56}$Ni at
$+383$ requires the escape of a significant amount of $\gamma$-rays from the radioactive decay of $^{56}$Co.
The assumption of full trapping and fixed opacity are probably not realistic, but our modelling
provides evidence that the secondary maximum could be at least partially, if not completely,
powered by a magnetar.

Recently, \citep{orellana25} presented a hydrodynamic modelling of the bolometric light
curve of SN\,2022jli of \citet{chen24}. They obtain a good fit of the double-peaked light
curve using a hybrid model that combines the radioactive decay of $^{56}$Ni and a magnetar.
Assuming that magnetar power ceases to be efficiently thermalised at $\sim 270$ days, they
reproduce the rapid light curve drop. This model does not produce the small periodic
undulations of SN\,2022jli that may arise from the interaction between a neutron star and a
companion star in a binary system. \citet{orellana25} obtain a similar $^{56}$Ni mass to our
analytical model, and consistent parameters for the magnetic field and the spin of the magnetar
The main difference is in the ejecta mass, where they obtain $M_{\mathrm{ej}} = 9.55$\,$M_{\odot}$.
This significant difference may arise from the limitations of our simple analytical model,
in addition to the differences between our pseudo-bolometric light curve and that of \citet{chen24}.

\subsection{Ejecta circumstellar interaction}

The interaction between the SN ejecta and a dense CSM can efficiently transform the kinetic
energy of the SN ejecta in thermal energy, powering the SN brightness. This is mechanism behind
interacting SLSNe-IIn such as SN\,2006gy \citep{smith07}, and it has been proposed to explain
the large maximum luminosities of SLSNe-Ic \citep{sorokina16, tolstov17a, tolstov17b}.
There is evidence that the SN ejecta-CSM interaction may be responsible for the bumps or the light curve
undulations observed in some SLSNe after maximum light \citep{yan15, yan17}. Some SE SNe such
as SN\,2014C \citep{milisavljevic15}, SN\,2017ens \citep{chen18}, SN\,2017dio \citep{kuncarayakti18},
and SN\,2018ijp \citep{tartaglia21} interact with abundant hydrogen-rich CSM, becoming Type IIn SNe.
In fact, the presence of dense CSM is a natural consequence of the stellar stripping and bumps
in the light curves are often considered a signature of the ejecta-CSM interaction.

As is discussed in \citet{chen24}, it is unlikely that the undulations and the light curve of
SN\,2022jli is powered by ejecta-CSM interaction because the interaction on the rear side would smear
any periodic signature from the near side. Additionally, we consider that the spectroscopic evolution
of SN\,2022jli lack clear signatures of ejecta-CSM interaction.

The evolution of $R_{bb}$ can provide insights into the power mechanism. The radiative shock produced
by the interaction propagates throughout the dense CSM increasing the optical depth and as a consequence
the photospheric radius at the same time \citep{moriya18}. Such an increase in $R_{bb}$ has been observed
in the double-peaked superluminous SN2019stc \citep{gomez21,moriya22} and in the double-peaked luminous
type II ASASSN-13dn \citep{hueichapan25}, for example. Both SNe are double-peaked but show no clear
signs of interaction such as narrow CSM emission features. The $R_{bb}$ evolution in interacting type IIn
SNe is diverse, but generally show an increase at early times, in particular before maximum light, a
flat evolution or a slow decrease in their $R_{bb}$ after maximum \citep[see e.g.,][]{taddia13, ofek14}.
A very rapid increase in $R_{bb}$ in type IIn SNe has been also reported, which has been interpreted as a
signature of aspherical CSM \citep[see][]{soumagnac19,soumagnac20}. On the other hand, SN\,2022jli shows
the opposite, a decrease in $R_{bb}$ during the SN re-brightening (see Fig. \ref{bol_bb_params_fig}).
The second maximum of SN\,2022jli is produced by an energy source that heats the ejecta, this is
manifested by the increase in the temperature, but not in the radius. Thus we consider the ejecta-CSM
interaction an unlikely scenario to explain the second maximum of SN\,2022jli.


\section{Summary}
\label{sec:summary}

We have presented and analysed observations of the SE SN\,2022jli, which exhibits a double-peaked light
curve with two maxima separated by about 50\,days. After the second maximum, the well-sampled
ZTF and ATLAS light curves reveal periodic undulations with a period of about 12.5\,days
\citep[see also][]{moore23, chen24}. Using synthetic photometry inferred from spectra
obtained during the decline from the first maximum, a host galaxy reddening of
$E(B-V)_{\mathrm{host}} = 0.23 \pm  0.06$ was estimated. The SN shows unusually
blue $g-r$ colours during its re-brightening, reaching its bluest colour ($g-r \simeq -0.45$\,mag)
at $+190$\,days relative to maximum. We found that in addition to the blue colours
the SN shows clear colour variations during its periodic undulations, becoming blue when
brighter and redder when becomes fainter, showing more
conspicuous periodic undulations in the bluer bands.

The first maximum of SN\,2022jli is consistent with a SN\,Ic
with an estimated expansion velocity of $8500$\,km\,s$^{-1}$, an ejecta mass of $\sim 1.5$\,$M_\odot$,
and a radioactive $^{56}$Ni mass of $\sim 0.12$\,$M_{\odot}$ \citep[see also][]{orellana25}. The available photometry seems
to show a smooth evolution and does not show undulations during the first maximum,
although this cannot be completely discounted due to the sparse coverage in terms of
wavelength and sampling of the first maximum. The late optical spectra (\textgreater
$+400$\,days) confirm the SN\,Ic nature of SN\,2022jli.

We report the detection of relatively weak and broad H$\alpha$ and Pa$\beta$ double-peaked spectral
features. The main peak of H$\alpha$ and Pa$\beta$ features seems to shift following the periodic variation
of the light curve undulations, as has been reported by \citet{chen24} for H$\alpha$.
We notice that these features become more conspicuous with time, particularly after $+100$\,days. The shift in the \ion{H}{I} spectral
features is probably a signature of the accretion of hydrogen-rich material from a companion star in a
binary system. When the spectra of the SN become nebular ($\textgreater +350$\,days), the hydrogen
signatures are not longer detectable in the optical or in the NIR spectra; however, narrow unresolved H$\alpha$ emission most likely from the host galaxy is detected at late times ($FWHM \textless 400$\,km\,s$^{-1}$).

The second maximum was reproduced using the analytical solutions of \citet{arnett82} with
a hybrid model that combines the radioactive decay of $0.12\,M_{\odot}$ of $^{56}$Ni and the
power injection from a magnetar with $P\sim 48$\,ms and $B \sim 8 \times 10^{14}$\,G.
During the revision of this article, a similar hydrodynamical model
of the bolometric luminosity of SN\,2022jli was presented by \citet{orellana25}.
This demonstrates that the double-peaked light curve of SN\,2022jli could be powered
by a hybrid model that includes a magnetar.
In this scenario, the magnetar contributes to the second maximum, and moderate accretion from a
companion star is necessary to produce the small light curve undulations
and the weak hydrogen lines observed after the second maximum.
The power injection from a magnetar seems a viable scenario, since magnetars are the most
variable neutron stars displaying different types of episodic variability
\citep[see][]{mereghetti15,kaspi17}.
The magnetar contributing to power the second maximum of SN\,2022jli could fit into the more general picture whereby
magnetars are the power source of SLSNe-Ic and could produce their frequent bumps and undulations, and whereby pulsars could produce the late time excess observed in
some SE SNe \citep[see e.g.,][]{taddia16, ravi23}. The possibility of a magnetar as an additional power
source is not limited to 'exotic' SE SNe. \citet{rodriguez24} analysed 54 SNe IIb/Ib/Ic and found evidence
of a central engine in most of them. If these engines are magnetars, then their initial magnetic fields and
rotation periods align well with those expected for magnetars at birth.

Alternatively, the second maximum could be produced by the significant accretion of material from the
companion star into the compact object left from the explosion (a black hole or a neutron star),
at an extremely large super-Eddington accretion rate ($\sim 10^{3}$ for a black hole or $\sim10^{4}$ for a neutron star),
or could be a ULX displaying extremely beamed emission ($b \sim 10^{-3}$ or $10^{-4}$) viewed just
at the right angle, as has been proposed by \citet{king24}. A somewhat different scenario is the accretion from
a shock-inflated companion star into the neutron star. This model seems to reproduce the bolometric
luminosity \citep{hirai25,lu25} and the undulations of SN\,2022jli \citep{hirai25}, and predicts a significant
mass ejection of hydrogen-rich material from the inflated envelope of the companion star
\citep[$\sim 10^{-3} - 10^{-2}$\,$M_\odot$;][]{hirai25,lu25}. It is not clear whether this ejected material should
be detectable in the nebular phase or not. This is an interesting model that requires further comparison with
observations.

From $+190$\,days until about $+224$\,days, unambiguous first CO overtone emission is detected in the
NIR spectra of SN\,2022jli. At later times (\textgreater $+385$\,days), the first CO overtone is not
longer detected. From about $+200$\,days, a strong IR excess is detected in the NIR photometry
of SN\,2022jli, which is particularly strong in the $K_{s}$ band. Later NIR observations ($\textgreater +385$\,days)
demonstrate that this IR excess is produced by thermal emission from hot dust, with a temperature
of 900-650\,K. Depending on the assumptions of the dust composition, the estimated dust mass is
$2 - 16 \times 10^{-4}$\,$M_{\odot}$. This emission could be produced either by a strong IR echo
from pre-existing dust surrounding the SN, or by newly formed dust in the SN ejecta. The detection
of CO and of dust potentially formed in the ejecta of SN\,2022jli is important for understanding the formation
of molecules and dust formation in the ejecta of SE SNe. Future IR observations with {\it JWST}
should complement the observations presented here, providing an unprecedented view of the molecule and dust
emission in SN\,2022jli over several years, or even over a decade-long timescale.

\section{Data availability}

The spectra are available on request to the corresponding author and also available from the
WISeREP\footnote{\url{https://www.wiserep.org/}} archive \citep{yaron12}.

\begin{acknowledgements}

  We thank to \'Osmar Rodr\'iguez and the anonymous referee for useful comments that helped to improve this manuscript.
  R.C. acknowledges support from Gemini ANID ASTRO21-0036. M.D.S. is funded by the Independent Research Fund Denmark
  (IRFD, grant number  10.46540/2032-00022B). J.L.P and M.G. acknowledges support from ANID, Millennium Science Initiative,
  AIM23-0001. R.C. thanks Maria Elisa Ugarte for her support. Based on observations obtained at the international Gemini Observatory
  under proposals GS-2022A-Q-415, GS-2022B-Q-407 and GS-2022B-FT-206 (PI R. Cartier), a program of NSF’s NOIRLab, which
  is managed by the Association of Universities for Research in Astronomy (AURA) under a cooperative agreement with
  the National Science Foundation on behalf of the Gemini Observatory partnership: the National Science Foundation
  (United States), National Research Council (Canada), Agencia Nacional de Investigaci\'{o}n y Desarrollo (Chile),
  Ministerio de Ciencia, Tecnolog\'{i}a e Innovaci\'{o}n (Argentina), Minist\'{e}rio da Ci\^{e}ncia, Tecnologia, Inova\c{c}\~{o}es
  e Comunica\c{c}\~{o}es (Brazil), and Korea Astronomy and Space Science Institute (Republic of Korea).
  Also based on observations obtained at the Southern Astrophysical Research (SOAR) telescope, which is a joint project
  of the Minist\'{e}rio da Ci\^{e}ncia, Tecnologia e Inova\c{c}\~{o}es (MCTI/LNA) do Brasil, the US National
  Science Foundation’s NOIRLab, the University of North Carolina at Chapel Hill (UNC), and Michigan State University (MSU).
  Based on observations collected at the European Southern Observatory under ESO programmes 382.B-0331(A) and 108.220C.014.

  \smallskip

  \\

  This research has made use of the NASA/IPAC Extragalactic Database (NED), which is operated by the Jet Propulsion Laboratory,
  California Institute of Technology, under contract with the National Aeronautics and Space Administration. This work has also
  made use of data from the Asteroid Terrestrial-impact Last Alert System (ATLAS) project. The Asteroid Terrestrial-impact Last
  Alert System (ATLAS) project is primarily funded to search for near earth asteroids through NASA grants NN12AR55G, 80NSSC18K0284,
  and 80NSSC18K1575; byproducts of the NEO search include images and catalogs from the survey area. This work was partially
  funded by Kepler/K2 grant J1944/80NSSC19K0112 and HST GO-15889, and STFC grants ST/T000198/1 and ST/S006109/1. The ATLAS
  science products have been made possible through the contributions of the University of Hawaii Institute for Astronomy,
  the Queen’s University Belfast, the Space Telescope Science Institute, the South African Astronomical Observatory, and
  The Millennium Institute of Astrophysics (MAS), Chile. This publication makes use of data products from the Wide-field
  Infrared Survey Explorer, which is a joint project of the University of California, Los Angeles, and the Jet Propulsion
  Laboratory/California Institute of Technology, funded by the National Aeronautics and Space Administration.
  This work has made use of data from the European Space Agency (ESA) mission {\it Gaia} (\url{https://www.cosmos.esa.int/gaia}),
  processed by the {\it Gaia} Data Processing and Analysis Consortium (DPAC, \url{https://www.cosmos.esa.int/web/gaia/dpac/consortium}).
  Funding for the DPAC has been provided by national institutions, in particular the institutions participating in the {\it Gaia}
  Multilateral Agreement. 
  
\end{acknowledgements}

\bibliographystyle{aa}
\bibliography{references}

\begin{appendix}
  
  \section{Synthetic photometry}

  In Table \ref{syn_phot_tab} is presented the synthetic photometry performed to LDSS-3 and EFOSC2 spectra obtained
  on May 22 and 24, 2022, respectively.

  \begin{table*}
    \centering
    \caption{Summary of synthetic photometry for SN\,2022jli.}
    \label{syn_phot_tab}
    \begin{tabular}{@{}lccccccccc}
      \hline
      Template          & MJD phot & Scaling   & Scaling phot.      & $g_{\mathrm{ZTF}}$ & $r_{\mathrm{ZTF}}$ & $c_{\mathrm{ATLAS}}$ & $o_{\mathrm{ATLAS}}$ & $G$   & $\mathrm{Clear}$\\
      spectrum          & (days)   & band      & (mag)              & (mag)              & (mag)              & (mag)                & (mag)                & (mag) & (mag) \\
      \hline
      LDSS-3\_22-05-2022 & $59721.4$ & $i$ band & $14.44$($0.03$)  & $15.8$($0.1$)    & $14.7$($0.1$)    & $15.3$($0.1$)      & $14.6$($0.1$)      & $14.8$($0.1$) & $14.6$($0.1$) \\
      EFOSC2\_24-05-2022 & $59723.4$ & $V$ band & $15.22$($0.02$)  & $15.9$($0.1$)    & $14.9$($0.1$)    & $15.4$($0.1$)      & $14.7$($0.1$)      & $14.9$($0.1$) & $14.6$($0.1$) \\
      \hline
    \end{tabular}
  \end{table*}

  \section{Pseudo-bolometric light curves}

  To construct the pseudo-bolometric light curve of SN\,2022jli, we followed the approach of constructing
  optical spectral templates. During the first maximum and on the rise to the second one, a few epochs of multi-band
  photometry are available; therefore, we used the optical spectra of SN\,2022jli that provided the shape of
  the spectral energy distribution (SED) of the SN. The spectra of SN\,2022jli were scaled using optical photometry
  obtained as close in time to the spectra as possible. These spectrophotometric templates were integrated to derive
  the total luminosity emitted by SN\,2022jli in the optical regime ($3800-9000$\,\AA). When no spectra of SN\,2022jli
  are available, for example at maximum, we used the optical spectra of SN\,2013ge, which shows a SED and spectral
  features similar to SN\,2022jli during the normal Ic phase of SN\,2022jli, around maximum light.
  The spectra of SN\,2013ge were colour matched to the SN optical photometry of SN\,2013ge, dereddened using the reddening in
  the line-of-sight to SN\,2013ge \citep[$E(B-V) = 0.067$;][]{drout16}, and then the spectral template was reddened
  using the total colour excess in the direction of SN\,2022jli (i.e. $E(B-V)_{\mathrm{tot}} = 0.27$\,mag; see Section~\ref{sec:reddening}).
  Finally, the spectral template was scaled using Gaia photometry of SN\,2022jli close to maximum light.

  We have identified three sources of potential systematic errors in using spectral templates to derive bolometric light curves.
  The first source is a poor calibration of the spectrum due to slit losses or poor flux calibration. 
  When multi-band photometry is available this can be solved by matching the SN colours of the spectra to the photometry
  as described in \citet{cartier22}. This colour-matching procedure was applied to the spectra templates of SN\,2013ge, and
  correspond to minimal corrections. We have found comparing synthetic photometry with actual observations that this effect
  is usually small (\textless 5\%); however, conservatively we assume 10\% uncertainty for spectra that have been scaled and not
  colour-matched to multi-band photometry. The second potential source of systematic error is using a spectral template from a substitute
  object, as is the case of SN\,2013ge at maximum light. Given the similarity between SN\,2022jli at $+12$\,days and SN\,2013ge at
  $+7$\,days, to quantify this effect we used the spectral template of SN\,2013ge at $+7$\,days to compute the bolometric luminosity
  of SN\,2022jli at $+12$\,days, instead of the actual spectrum. This procedure shows that the use of a substitute object as
  spectral template, having a similar spectral shape, it can introduce a systematic error of the order of $\simeq 13$\%
  in the computation of the bolometric luminosity.

  Finally, we evaluate the effect of using a spectral template from a different phase, for example, this can be the case of using
  a spectral template obtained at the time of $V$-band  maximum to estimate the luminosity at the $i$-band maximum, even though
  both are close in time, the shape of the spectra is different at both epochs introducing a systematic error. We estimate this
  effect by comparing the pseudo-bolometric luminosity obtained for SN\,2022jli at maximum, from using the spectral templates of
  SN\,2013ge at $+1$\,days and at $+7$\,days relative to the $B$-band maximum. We find that this can introduce a systematic difference
  of $\simeq 10$\% in the pseudo-bolometric optical luminosity. We add in quadrature all these uncertainties in addition to the actual
  median of the errors from the photometry used to scale and colour match the spectra.

  \section{Photometry}

  \begin{table*}
    \centering
    \caption{Optical photometry of SN\,2022jli}
    \label{op_phot_tab}
    \begin{tabular}{lccccccl}
      \hline
      Date UTC & MJD       & Phase    & $V$             & $g$             & $r$             & $i$             & Instrument/ \\
                 & (days)    & (days)   & (mag)           & (mag)           & (mag)           & (mag)           & Telescope   \\
      \hline
      22-05-2022 & $59721.4$ & $+11.7$  & $\cdots$        & $\cdots$        & $\cdots$        & $14.44$($0.03$) & LDSS-3/Clay \\
      24-05-2022 & $59723.4$ & $+13.7$  & $15.22$($0.03$) & $\cdots$        & $\cdots$        & $\cdots$        & EFOSC2/NTT \\
      07-10-2022 & $59859.3$ & $+148.9$ & $\cdots$        & $\cdots$        & $16.07$($0.03$) & $\cdots$        & LDSS-3/Clay \\
      14-05-2023 & $60078.4$ & $+366.8$ & $\cdots$        & $21.11$($0.08$) & $20.79$($0.05$) & $\cdots$        & Goodman/SOAR \\
      30-05-2023 & $60094.4$ & $+382.7$ & $\cdots$        & $\cdots$        & $21.56$($0.08$) & $21.00$($0.07$) & Goodman/SOAR \\
      04-07-2023 & $60129.9$ & $+418.5$ & $\cdots$        & $\cdots$        & $21.96$($0.11$) & $\cdots$        & LDSS-3/Clay \\
      \hline
      \end{tabular}
    \tablefoot{Numbers in parentheses correspond to 1\,$\sigma$ statistical uncertainties. The phase is measured relative
      to the time of unfiltered maximum brightness.
    }
  \end{table*}

  \begin{table*}
    \centering
    \caption{Near-infrared photometry of SN\,2022jli}
    \label{ir_phot_tab}
    \begin{tabular}{lccccccl}
      \hline
      Date UTC & MJD     & Phase  & $J$ & $H$ & $K_{s}$  & Instrument/ \\
               & (days)  & (days) &     &     &          & Telescope   \\
      \hline
      18-06-2022 & $59748.3$ & $+38.5$  & $13.71$($0.03$) & $13.12$($0.03$) & $12.94$($0.04$) & Flamingos-2/Gemini-S \\
      04-07-2022 & $59764.3$ & $+54.4$  & $13.83$($0.01$) & $13.34$($0.01$) & $13.11$($0.02$) & Flamingos-2/Gemini-S \\
      05-07-2022 & $59765.4$ & $+55.5$  & $\cdots$        & $13.38$($0.01$) & $\cdots$        & Flamingos-2/Gemini-S \\
      27-09-2022 & $59849.2$ & $+138.8$ & $15.15$($0.01$) & $14.55$($0.01$) & $14.31$($0.02$) & Flamingos-2/Gemini-S \\
      28-09-2022 & $59850.1$ & $+139.7$ & $15.17$($0.01$) & $14.54$($0.05$) & $14.35$($0.03$) & Flamingos-2/Gemini-S \\
      12-10-2022 & $59864.2$ & $+153.8$ & $15.33$($0.01$) & $14.69$($0.01$) & $14.49$($0.02$) & Flamingos-2/Gemini-S \\
      19-10-2022 & $59871.1$ & $+160.6$ & $15.33$($0.01$) & $14.72$($0.02$) & $14.48$($0.02$) & Flamingos-2/Gemini-S \\
      07-11-2022 & $59890.1$ & $+179.5$ & $15.65$($0.01$) & $14.97$($0.01$) & $14.74$($0.03$) & Flamingos-2/Gemini-S \\
      08-11-2022 & $59891.1$ & $+180.5$ & $15.66$($0.02$) & $\cdots$        & $\cdots$        & Flamingos-2/Gemini-S \\
      09-11-2022 & $59892.2$ & $+181.6$ & $\cdots$        & $15.00$($0.03$) & $14.78$($0.06$) & Flamingos-2/Gemini-S \\
      11-11-2022 & $59894.1$ & $+183.5$ & $15.72$($0.01$) & $15.01$($0.03$) & $14.72$($0.01$) & Flamingos-2/Gemini-S \\
      17-11-2022 & $59900.1$ & $+189.5$ & $15.69$($0.01$) & $\cdots$        & $\cdots$        & Flamingos-2/Gemini-S \\
      18-11-2022 & $59901.1$ & $+223.3$ & $\cdots$        & $14.99$($0.01$) & $\cdots$        & Flamingos-2/Gemini-S \\
      21-12-2022 & $59934.1$ & $+223.3$ & $15.90$($0.01$) & $\cdots$        & $\cdots$        & Flamingos-2/Gemini-S \\
      22-12-2022 & $59935.1$ & $+224.3$ & $\cdots$        & $15.14$($0.01$) & $\cdots$        & Flamingos-2/Gemini-S \\
      05-01-2023 & $59949.0$ & $+238.1$ & $16.05$($0.01$) & $15.20$($0.01$) & $14.57$($0.02$) & Flamingos-2/Gemini-S \\
      03-06-2023 & $60098.3$ & $+386.6$ & $\cdots$        & $17.36$($0.06$) & $15.52$($0.03$) & Flamingos-2/Gemini-S \\
      05-06-2023 & $60100.3$ & $+388.6$ & $19.54$($0.06$) & $17.34$($0.05$) & $15.49$($0.04$) & Flamingos-2/Gemini-S \\
      26-06-2023 & $60121.4$ & $+409.6$ & $19.75$($0.08$) & $17.61$($0.04$) & $15.71$($0.04$) & Flamingos-2/Gemini-S \\
      14-07-2023 & $60139.3$ & $+427.4$ & $20.04$($0.06$) & $17.81$($0.05$) & $\cdots$        & Flamingos-2/Gemini-S \\
      15-07-2023 & $60140.4$ & $+428.4$ & $\cdots$        & $17.76$($0.08$) & $\cdots$        & Flamingos-2/Gemini-S \\
      30-07-2023 & $60155.4$ & $+443.4$ & $20.03$($0.11$) & $17.90$($0.06$) & $15.99$($0.06$) & Flamingos-2/Gemini-S \\
      04-12-2023 & $60282.0$ & $+569.3$ & $\cdots$        & $19.00$($0.18$) & $16.65$($0.09$) & Flamingos-2/Gemini-S \\
      04-01-2024 & $60313.1$ & $+600.2$ & $\cdots$        & $19.29$($0.05$) & $17.15$($0.04$) & Flamingos-2/Gemini-S \\
      \hline
    \end{tabular}
    \tablefoot{Numbers in parentheses correspond to 1\,$\sigma$ statistical uncertainties. The phase is measured relative
      to the time of unfiltered maximum brightness.
    }
  \end{table*}

  \begin{table*}
  \centering
  \caption{NEOWISE photometry of SN\,2022jli}
  \label{mir_phot_tab}
  \begin{tabular}{lcccc}
    \hline
    Date UTC & MJD     & Phase  & $W1$ & $W2$ \\
             & (days)  & (days) &      &     \\
    \hline
    27-06-2022 & $59757.5$ & $+47.6$  & $12.87$($0.06$) & $12.53$($0.07$) \\
    05-12-2022 & $59919.3$ & $+208.6$ & $14.20$($0.08$) & $12.85$($0.07$) \\
    26-06-2023 & $60121.7$ & $+409.8$ & $13.78$($0.07$) & $12.56$($0.07$) \\
    05-12-2023 & $60283.5$ & $+570.8$ & $14.16$($0.08$) & $12.83$($0.07$) \\
    24-06-2024 & $60485.9$ & $+772.0$ & $16.32$($0.17$) & $14.37$($0.11$) \\
    \hline
  \end{tabular}
  \tablefoot{Numbers in parentheses correspond to 1\,$\sigma$ statistical uncertainties.
  The phase is measured relative to the time of maximum brightness.
  }
\end{table*}

\begin{table*}
  \centering
  \caption{Optical and NIR pseudo-bolometric luminosity of SN\,2022jli}
  \label{pseudo_bol_tab}
  \begin{tabular}{lcrr}
    \hline
    MJD     & Phase   & $L_{\mathrm{Op}}$              & $L_{\mathrm{NIR}}$\\
            & (days)  & $\times 10^{40}$ erg\,s$^{-1}$ & $\times 10^{40}$ erg\,s$^{-1}$ \\
    \hline
    $59711.4$ & $+1.8$   & $261.9$($39.3$)              & $30.1$($3.2$) \\
    $59721.4$ & $+11.7$  & $182.9$($18.3$)              & $23.9$($2.5$) \\
    $59723.4$ & $+13.7$  & $167.3$($16.7$)              & $22.4$($2.4$) \\
    $59732.6$ & $+22.9$  & $124.8$($17.5$)              & $18.5$($1.9$) \\
    $59753.6$ & $+43.8$  & $254.7$($25.5$)              & $20.4$($1.4$) \\
    $59765.4$ & $+55.5$  & $212.3$($14.9$)              & $17.1$($1.2$) \\
    $59802.3$ & $+92.2$  & $134.9$($9.4$)               & $10.3$($0.7$) \\
    $59859.3$ & $+148.9$ & $76.2$($5.3$)                & $5.4$($0.4$) \\
    $59890.1$ & $+179.5$ & $58.1$($8.7$)                & $4.0$($0.3$) \\
    $59941.1$ & $+230.2$ & $31.8$($5.7$)                & $3.26$($0.06$) \\
    $60094.4$ & $+382.7$ & $0.58$($0.03$)               & $0.86$($0.04$) \\
    \hline
  \end{tabular}
  \tablefoot{Numbers in parentheses correspond to 1\,$\sigma$ statistical uncertainties.
  The phase is measured relative to the time of maximum brightness.
  }
\end{table*}

\end{appendix}

%
%

\end{document}